\documentclass[11pt,letterpaper]{article}

\usepackage[margin=1in]{geometry}
\usepackage{algorithm}
\usepackage{algorithmic}
\usepackage{natbib}
\usepackage{hyperref,enumitem}
\hypersetup{
	colorlinks=true,
	citecolor = blue 
}
\usepackage{url}

\usepackage{thmtools}
\usepackage{thm-restate}
\usepackage{amsmath,amssymb,amsthm}

\usepackage{xspace}
\usepackage{graphicx}
\usepackage{booktabs} 
\usepackage{xcolor}
\usepackage[font=small,labelfont=bf]{caption}
\usepackage{subcaption}
\usepackage{multirow}
\usepackage{authblk}

\usepackage{comment}

\usepackage{amsfonts} 
\usepackage{nicefrac} 
\usepackage{microtype} 

\usepackage[utf8]{inputenc} 
\usepackage[T1]{fontenc} 

\usepackage{tikz}
\usetikzlibrary{shapes}
\usetikzlibrary{arrows}
\usetikzlibrary{calc}
\usetikzlibrary{decorations.pathreplacing}

\newtheorem{theorem}{Theorem}
\newtheorem{lemma}[theorem]{Lemma}

\newtheorem{proposition}[theorem]{Proposition}

\theoremstyle{definition}

\newtheorem{example}[theorem]{Example}

\begin{document}

\title{Data Market Design through Deep Learning\textsuperscript{\textdagger}}

\author{%
  Sai Srivatsa Ravindranath\textsuperscript{*} \qquad Yanchen Jiang\textsuperscript{*} \qquad  David C. Parkes\\
  Harvard John A. Paulson School of Engineering and Applied Sciences\\
  \texttt{\{saisr, yanchen\_jiang, parkes\} @g.harvard.edu} \\
}

\maketitle 
\begingroup
\renewcommand{\thefootnote}{\textdagger}
\footnotetext{A preliminary conference version appeared in the proceedings of the Conference on Neural Information Processing Systems (NeurIPS 2023).}
\renewcommand{\thefootnote}{*}
\footnotetext{These authors contributed equally.}
\endgroup

\begin{abstract}%
The {\em data market design} problem is the problem of finding a set of signaling schemes (statistical experiments) to maximize expected revenue to an information seller, where each experiment reveals some information and has a corresponding price~\citep{Berg2018}. Each buyer chooses a downstream action based on their prior and signal, and their expected value for the information associated with a particular experiment comes from the improvement in deciding this action. In a setting with multiple buyers, a buyer's expected value for an experiment may also depend on the information sold to others~\citep{Bonatti2022}. We introduce the application of deep learning for the design of revenue-optimal data markets, looking to expand the frontiers of what can be understood and achieved. Relative to earlier work on deep learning for auction design~\citep{dutting2019}, we must learn signaling schemes (rather than allocation rules), and handle {\em obedience constraints}---these arising from modeling the downstream actions of buyers---in addition to incentive constraints on bids. Our experiments demonstrate that this deep learning framework can almost precisely replicate all known solutions from theory, expand to more complex settings, and be used to identify new designs for data markets that can be proved to be optimal as well as make conjectures in regard to the structure of optimal designs.
\end{abstract}%

\section{Introduction}\label{sec:intro}

Many characterize the current era as the Information Age. Companies such as Google and Meta (search and social media), Experian and TransUnion (credit agencies), and Amazon and American Express (commerce) hold vast quantities of data about individuals. In turn, this has led to data markets, where information about an individual can be purchased in real-time to guide decision-making (e.g., LiveRamp, Segment, Bloomreach). In this paper, we advance the design of market rules that govern the structuring and sale of this kind of information.
Using tools from machine learning, we leverage existing theoretical frameworks~\citep{Berg2018,Bonatti2022} to replicate all known solutions from theory, expand to more complex settings, generate new candidate designs, and make conjectures in regard to the structure of optimal designs. Wherever possible, we prove the optimality of these learned mechanisms through theoretical analysis. As such, this framework provides a tool for guiding intuition and generating insights that can support the development of new theory.
Although this is not our focus, there are also important ethical questions in regard to data markets, in regard to privacy, ownership, informed consent, and the use of data~\citep{acquisti2016economics, bonatti2022platform, socialdata, athey2017digital, choi2019privacy, fainmesser2022digital, acemoglu2022too}.

In settings with a single buyer, \citet{Berg2018} introduce a framework in which there is a {\em data buyer} who faces a decision problem under uncertainty, and has a payoff depending on their choice of action and the underlying state of the world. There is a {\em data seller} who knows the world state, and can disclose information through a menu of statistical experiments, each experiment offering a stochastic signal that reveals some (or all) of the seller's information and each of which is associated with a price. The buyer's willingness to pay for information is determined by their {\em type}, which defines their prior belief and their value for different outcomes in the world
(these arising from the action that the buyer will choose to take).

The {\em optimal data market design problem} is to find a set of experiments and associated prices with which to maximize expected revenue, given a distribution over the buyer's type. \citet{Berg2018} characterize the optimal design when there is a binary world state and a buyer with a binary action. In some settings there may be multiple buyers and buyers may compete downstream, and the more informed one buyer is, the lower the payoff may become for others. \citet{Bonatti2022} take up this problem, characterizing the optimal design for multiple buyers in the binary state/binary action setting, but for the restricted setting that buyers have a common prior on the world state. It remains an open problem to obtain theoretical results for richer, multi-buyer settings, and this motivates the need for gaining new understanding and making progress through computational approaches.

\subsection{Contributions}
Inspired by the recent advances in {\em differentiable economics}~\citep{dutting2019}, we initiate the use of deep learning for the automated design of data markets. This market sells information rather than allocates resources, and the value of information depends on the way in which a buyer will use the information---and this downstream action by a buyer needs to be modeled. Following the {\em revelation principle for dynamic games}~\citep[Section~6.3]{myerson1991game}, it is without loss of generality to model an experiment as generating a recommended action and insisting on designs in which the buyer will prefer to follow this action. This brings about new challenges, most notably to extend the differentiable economics framework to handle {\em obedience} (choosing to follow a recommended action). In particular, and in addition to achieving incentive alignment where a buyer will choose to truthfully report their prior beliefs and values for different outcomes, we also need to ensure there are no useful {\em double deviations}, where a buyer simultaneously misreports their type and acts contrary to the seller's recommended action.

In settings with a single buyer, we learn an explicit, parameterized representation of a menu of priced experiments to offer the buyer and with which to model the action choices available to the buyer. This extends the {\em RochetNet architecture}~\citep{dutting2019}, which has been used successfully for optimal auction design with a single buyer. This enables us to obtain exact incentive compatibility for the single buyer setting: a buyer has no useful deviation from a recommended action, no useful deviation in reporting their type, and no useful double deviation. In settings with multiple buyers, we learn revenue-maximizing designs while also minimizing deviations in disobeying action recommendations and misreporting types and including double deviations. This extends the {\em RegretNet framework}~\citep{dutting2019}, which has been used successfully for optimal auction design with multiple buyers and gives approximate incentive alignment.

Our first experimental result is to show through extensive experiments that these new neural network architectures and learning frameworks are able to almost exactly recover all known optimal solutions from~\citet{Berg2018} and \citet{Bonatti2022}. We adopt the notion of Bayesian incentive compatibility (BIC), using samples of a buyer's type to compute an {\em interim} experiment and {\em interim} payment, averaging over samples drawn from the type distribution (this builds on differentiable economics with BIC methods developed for budget-constrained auctions~\citep{feng2018deep}). We develop a new training method that enables the efficient reuse of computed {\em interim} experiments and {\em interim} payments from other samples to calculate the {\em interim} utility of misreports, dramatically speeding up training.

Whereas analytical results are only available for the BIC setting, we are able to study through computational techniques the design of data markets in the {\em ex post IC setting}, which is a setting without existing theory. A second result, in the setting of multiple buyers, is to use our framework to conjecture the structure of an optimal design and prove its optimality (for this, we make use of {\em virtual values} analogous to Myerson's framework~\citep{12}). We see this as an important contribution, as {\em ex post IC }is a stronger notion of IC than BIC.

A third result is to demonstrate how our framework extends its utility beyond empirical results and serves as a toolbox to guide economic theory. To illustrate this, we study how revenue varies with competition in the multi-buyer setting where the prior information is uncertain. Despite the absence of existing theoretical results for this setting, our framework enables us to derive trends in revenue with economically-relevant aspects of the problem setup.
We also conjecture the structure of solutions for problems in the single buyer setting with an enlarged type, where both the buyer payoffs and priors are uncertain. For this case, we again derive empirical results using our proposed framework and use these to conjecture the properties of the underlying theoretical solution.

\subsection{Related work}
\citet{ConitzerS02,ConitzerS04a} introduced the use of {\em automated mechanism design (AMD)} for economic design and framed the problem as an integer linear program (or just a linear program for Bayesian design). Responding to challenges with the enumerative representations used in these early approaches, ~\citet{dutting2023jacm} proposed {\em differentiable economics}, which leverages deep neural networks to achieve greater representational flexibility in auction design.

This approach has been applied to settings such as budget-constrained bidders~\citep{feng2018deep}, fairness–revenue trade-offs~\citep{kuo20}, sequential auctions~\citep{ravindranath2024deepreinforcementlearningsequential}, and combinatorial auctions~\citep{wang2025bundleflow}. Parallel advances in neural architectures include permutation-equivariant layers~\citep{rahme2020permutationequivariant} and transformer-based models~\citep{pmlr-v162-duan22a,ivanov2022optimal}, as well as designs that enforce dominant-strategy incentive compatibility via menu-based and affine maximizer mechanisms~\citep{shen2018automated,Tacchetti19,curry2022differentiable,duan2023a,curry2024dynamic,gemnet}. Other work explores adversarial training~\citep{rahme21a} and regret certification for robustness~\citep{curry2020certifying}, with recent theory explaining empirical performance despite non-convexity~\citep{NEURIPS2023_a5e4907a}.

Beyond auctions, differentiable economics based approaches have been applied to facility location~\citep{golowich2018deep}, two-sided matching~\citep{ravindranath2021deep}, contract design~\citep{wang2023deep}, and automated market makers~\citep{curry2024amm}; see~\citet{curry_automated} for a comprehensive overview. To the best of our knowledge, none of this work considers optimal data market design, which introduces additional challenges related to agent actions (obedience) and negative externalities.

In regard to the data market problem,~\citet{Berg2018} build upon the decision-theoretic model pioneered by~\citet{blackwell1953equivalent} and study a setting of selling information to a single buyer.~\citet{cai_et_al:LIPIcs.ITCS.2021.81} give computational results in this model, making use of linear programs to compute the optimal menu for discrete type distributions. They also investigate a generalization of this model that allows multiple agents to compete for useful information.
In this setting, at most, one agent receives the information. While this approach can be used in principle for continuous type distribution, by applying the LP to a discretized valuation space, solving it can be prohibitively expensive. Further advancing economic theory,~\citet{bergemann2022selling} consider more general type distributions and investigate both the cardinality of the optimal menu and the revenue achievable when selling complete information.~\citet{Bonatti2022} also study the multi-buyer setting, modelling competition through negative externality.

\citet{babaioff2012} give a related framework, with the key distinction that the seller is not required to commit to a mechanism before the realization of the world state. As a result, the experiments and prices of the data market can be tailored to the realized state of the world. \citet{9} extend their setting, considering budget-constrained buyers, formulating a linear program to find the solution of the problem for a discrete type space.

There exist several other models that study revenue-optimal mechanisms for selling data. For example,~\citet{LiuOptimal} characterize the revenue-optimal mechanism for various
numbers of states and actions, and
consider general payoff functions. However, in their setup, the state only impacts the payoff of the active action taken by a buyer, which provides considerable simplification and may not be realistic.
\citet{endo} investigates a setting where the buyer can conduct their own costly experiment at a cost after receiving the signal. Different from our model, after receiving the signal from the data broker, the agent can subsequently acquire additional information with costs. \citet{endo} also assumed that the valuation function of the agent is separable and that the private type of the agent represents her value of acquiring more information, which is different from our single buyer model, where the prior belief is also drawn from a distribution and constitutes a part of the buyer's type.~\citet{agarwal2019marketplace} explore data marketplaces
 where each of multiple sellers sells data to buyers who aim to utilize the data for machine learning tasks,
and~\citet{pmlr-v162-chen22m} consider scenarios where neither the seller nor the buyer
knows the true quality of the data. \citet{dataset} and \citet{agarwal2020towards}
also incorporate buyer externalities into the study of data marketplaces.

Another line of research studies problems of
{\em information design}, for example, the problem of {\em Bayesian Persuasion}~\citep{kamenica2011bayesian}. There, the model is different in that the sender of information has preferences on the
 action taken by the receiver, setting up a game-theoretic problem of strategic misrepresentation by the sender.
\citet{dughmi2016algorithmic} studied this from an algorithmic perspective, and \citet{castiglioni2022bayesian} also brought in considerations from mechanism design by introducing hidden types; see also \citep{dughmi2014hardness,daskalakis2016does,cai2020third} for more work on information design and Bayesian persuasion.

\section{Preliminaries}
\label{sec:prelim}
We consider a setting with $n$ data buyers, $N = \{1,\dots,n\}$. Each buyer faces a decision problem under uncertainty. The state of the world, $\omega$, is unknown and drawn from a finite state space $\Omega = \{\omega_1,\dots,\omega_m\}$. Each buyer $i$ selects an action, $a_i$, from a finite action set $\mathcal{A}_i$. Let $\mathcal{A} = \prod_{i=1}^{n}\mathcal{A}_i$ denote the set of joint actions.

Each buyer aims to choose an action that matches the true state of the world. Specifically, buyer $i$ receives a payoff $v_i>0$ if the action matches the realized state (i.e., $a_i = \omega$), and zero otherwise.\footnote{We focus on this special case of so-called {\em matching payoffs}, where $|\Omega| = |\mathcal{A}_i|$. However, our computational framework readily extends to settings where the \emph{ex post} payoff to buyer $i$ from choosing action $a_i \in \mathcal{A}_i$ given state $\omega \in \Omega$ is a general utility function $\mathcal{U}_i(\omega, a_i)$.} The valuation $v_i$ for each buyer is drawn independently according to the distribution $v_i \sim \mathcal{V}_i$. Additionally, each buyer $i$ holds an {\em interim belief} $\theta_i \in \triangle(\Omega)$ about the unknown state. The beliefs are drawn independently from a distribution $\theta_i \sim \Theta_i$. We denote the joint distribution on buyers' interim beliefs as $\Theta = \prod_{i=1}^{n}\Theta_i$. Further, let $\mathcal{P}_i=\mathcal{V}_i \times \Theta_i$ and $\mathcal{P}=\prod_{i=1}^n\mathcal{P}_i$.

A buyer's {\em type} is defined by the tuple $\rho_i := (v_i, \theta_i)$. We denote buyer $i$'s {\em type space} as $\mathcal{T}_i =\mathbb{R} \times \triangle(\Omega)$, and the joint type space across all buyers as $\mathcal{T} = \prod_{i=1}^{n}\mathcal{T}_i$. (If either valuations or beliefs are fixed, the types simplify to just $\rho_i = \theta_i$, or $\rho_i = v_i$, respectively).

The payoff function for a buyer $i$ is given by $u_i: \mathcal{A} \times \Omega \times \mathcal{T}_i \to \mathbb{R}$. The payoff for a buyer depends not only on their own action and type but also includes a negative externality caused by other buyers' actions. Following~\citet{Bonatti2022}, we assume a separable externality structure, leading to the following payoff function:
\begin{align}
u_i(a,\omega,\rho_i) &= v_i \left(\mathbf{1}\{a_i = \omega\} - \bar{\alpha}\sum_{j \in N \setminus \{i\}} \mathbf{1}\{a_j = \omega\} \right),
\end{align}

where $\bar{\alpha} \triangleq \frac{\alpha}{n-1}$ when $n \geq 2$ and the second term is omitted when $n = 1$. Here, $\alpha \geq 0$ captures the degree of competitiveness among buyers.

\subsection{The Mechanism Design Problem}

There is a data seller who observes the state of the world, $\omega$, and wishes to sell information about the state to interested buyers. To do so, the seller designs a signaling scheme (also called an experiment) for each buyer. For each buyer, $i \in N$, she defines a set of possible signals, $\mathcal{S}_i$, from which to choose. There is also a message (bid) space $\mathcal{B}$, with bid profile $b = (b_1, \ldots b_n) \in \mathcal{B}$. A bid, for example, may represent a claim about an agent's value and prior belief.

The signaling scheme for buyer $i$ is a mapping $\sigma_i:\Omega \times \mathcal{B} \to \triangle(\mathcal{S}_i)$. Given bid profile $b = (b_1, \ldots b_n) \in \mathcal{B}$, if state $\omega \in \Omega$ is realized, then the seller sends a signal $S_i \sim \sigma_i(\omega, b)$ to buyer $i$. The signaling schemes, one for each buyer, are common knowledge to the buyers. Given this, buyer $i$ updates their belief about $\omega$ and chooses an optimal action.

For notational convenience, we represent $\sigma_i(\cdot, b)$ as an $m \times |\mathcal{S}_i|$ stochastic matrix $\pi_i(b)$, where $m$ is the number of world states. The $j$-th row of $\pi_i(b)$ specifies the likelihood of each signal when state $\omega = \omega_j$ and for bid profile $b$. Thus, we have $\sigma_i(\omega_j, b) = \pi_{i, j}(b)$.

Together with the signaling scheme, the seller designs a payment rule $t_i: \mathcal{B} \to \mathbb{R}_{\geq 0}$ which determines, for each buyer $i$, the amount of payment collected from the buyer as a function of the bid profile. The design goal is to design a set of signaling schemes and payment rules to maximize the expected revenue for the seller. The sequence of interactions between one or more buyers and the seller takes place as follows:

\begin{enumerate}
\item The seller commits to a mechanism, $\mathcal{M} = (\sigma, t)$, where $\sigma = (\sigma_1, \ldots, \sigma_n)$
is the set of signaling schemes and $t = (t_1, \ldots, t_n)$ is the set of payment rules.
\item Each buyer $i$ observes their type, $\rho_i$. The seller observes the state, $\omega$.
\item Each buyer reports a message $b_i$.
\item For every buyer $i$, the seller sends buyer $i$ a signal, $S_i \in \mathcal{S}_i$, generated according to the signaling scheme $\sigma_i(\omega, b)$, and collects payment $t_i(b)$.
\item Each buyer $i$ chooses an action $a_i \in \mathcal{A}_i$, and obtains utility $u_i(a, \omega, \rho_i) - t_i(b)$.
\end{enumerate}

By the revelation principle for dynamic games~\citep[Section~6.3]{myerson1991game},
it is without loss of generality to consider incentive-compatible mechanisms where the message space is the type space of buyers and the size of the signal space is the size of the action space,
i.e., for each $i \in [n]$, $|\mathcal{S}_i| = |\mathcal{A}_i|$. In such mechanisms, the seller designs experiments (via signaling schemes) where each signal corresponds to a different optimal choice of action. Following~\citet{Berg2018}, we can further replace every signal with an action recommended by the seller, so that $\mathcal{S}_i=\mathcal{A}_i$ for each $i$.

\paragraph{Incentive Compatibility} A mechanism $(\sigma, t)$ is {\em Bayesian incentive compatible} (BIC) if every buyer maximizes their expected utility (over other agents' reports) by reporting their true type as well and then following the recommended actions. For notational convenience, let $\mathbb{E}_{-i} := \mathop{\mathbb{E}}_{\rho_{-i} \sim \mathcal{P}_{-i}}$ denote the operator used for computing the interim representations. Here, $\mathcal{P}_{-i}$ represents the distributions on types for all but agent $i$. For each $(\rho_i, \rho_i') \in \mathcal{T}_i\times \mathcal{T}_i$, and for each deviation function $\delta: \mathcal{A}_i \to \mathcal{A}_i$, a BIC mechanism satisfies:
\begin{equation}\label{eqn:bic}
\begin{aligned}
     \mathbb{E}_{-i} \left[\mathop{\mathbb{E}}_{ \substack{a \sim \sigma(\omega, \rho)\\ \omega \sim \theta_i}} \left[u_i( a, \omega, \rho_i) - t_i(\rho) \right]\right] \geq \mathbb{E}_{-i} \left[\mathop{\mathbb{E}}_{ \substack{ a\sim \sigma(\omega, \rho_i', \rho_{-i}) \\\omega \sim \theta_i}} \left[u_i( \delta (a_i), a_{-i}, \omega, \rho_i)- t_i(\rho_i', \rho_{-i}) \right] \right].
\end{aligned}
\end{equation}

In particular,~\eqref{eqn:bic} insists that {\em double deviations} (misreporting the type and disobeying the recommendation) are not profitable, assuming all the other agents report their types truthfully and follow the action recommendations. It is helpful to further define the notions
of obedience and truthfulness, which consider restricted forms of
incentive compatibility.

 A mechanism $(\sigma, t)$ is {\em obedient} if, for any agent, when the agent is restricted to reporting its true type, then it maximizes its expected utility by following the action recommendations, assuming all the other agents report their types truthfully and follow the action recommendations. Obedience requires that a mechanism $(\sigma, t)$ satisfies the following condition, for every agent $i$, every $\rho_i\in \mathcal{T}_i$, and for every deviation function $\delta: \mathcal{A}_i \to \mathcal{A}_i$:
\begin{equation}
    \mathbb{E}_{-i} \left[ \mathop{\mathbb{E}}_{ \substack{a \sim \sigma(\omega, \rho), \\ \omega \sim \theta_i}} \left[u_i( a, \omega, \rho_i) \right] \right] \geq \mathbb{E}_{-i} \left[ \mathop{\mathbb{E}}_{ \substack{a \sim \sigma(\omega, \rho), \\ \omega \sim \theta_i}} \left[u_i( \delta (a_i), a_{-i}, \omega, \rho_i) \right] \right]. \label{eq:interimobed}
\end{equation}
Here, we drop the payment terms on the left and right hand side because they are equivalent
under the action deviations considered under obedience.

A mechanism $(\sigma, t)$ is {\em truthful} if, for any agent, when the agent is restricted to following its signaled action, then it maximizes its expected utility by reporting its true type, assuming all the other agents report their types truthfully and follow the action recommendations. Truthfulness requires that a mechanism $(\sigma, t)$ satisfies the following condition, for every agent $i$, and for every $(\rho_i, \rho_i') \in \mathcal{T}_i\times \mathcal{T}_i$:
\begin{equation}
    \mathbb{E}_{-i} \left[ \mathop{\mathbb{E}}_{ \substack{a \sim \sigma(\omega, \rho), \\ \omega \sim \theta_i}} \left[u_i( a, \omega, \rho_i) - t_i(\rho) \right] \right] \geq \mathbb{E}_{-i} \left[ \mathop{\mathbb{E}}_{ \substack{a \sim \sigma(\omega, \rho_i', \rho_{-i}), \\ \omega \sim \theta_i}\ } \left[u_i( a, \omega, \rho_i)- t_i(\rho_i', \rho_{-i}) \right] \right].
\end{equation}

We also consider the stronger notion of {\em ex post incentive compatible} (IC), which requires, for every agent $i$, and for each $\rho \in \mathcal{T}$, and assuming that every other agent reports its type truthfully and follows the recommended action, then for each misreport $\rho_i' \in \mathcal{T}_i$, and each deviation function, $\delta: \mathcal{A}_i \to \mathcal{A}_i$, the following condition:
\begin{equation}\label{eqn:expost-ic}
\begin{aligned}
    \mathop{\mathbb{E}}_{ \substack{a \sim \sigma(\omega, \rho), \\\omega \sim \theta_i}} \left[u_i( a, \omega, \rho_i) - t_i(\rho) \right] \geq \mathop{\mathbb{E}}_{\substack{a \sim \sigma(\omega, \rho_i', \rho_{-i}), \\ \omega \sim \theta_i}\ } \left[u_i( \delta (a_i), a_{-i}, \omega, \rho_i)- t_i(\rho_i', \rho_{-i}) \right].
\end{aligned}
\end{equation}

 In particular, a mechanism $(\sigma, t)$ is {\em ex post obedient} if it satisfies, for every agent $i$, for each $\rho \in\mathcal{T}$, and each deviation function, $\delta: \mathcal{A}_i \to \mathcal{A}_i$:
\begin{equation}\label{eqn:expost-ob}
   \mathop{\mathbb{E}}_{ \substack{a \sim \sigma(\omega, \rho), \\ \omega \sim \theta_i}} \left[u_i( a, \omega, \rho_i) \right] \geq \mathop{\mathbb{E}}_{ \substack{a \sim \sigma(\omega, \rho), \\ \omega \sim \theta_i}} \left[u_i( \delta (a_i), a_{-i}, \omega, \rho_i) \right].
\end{equation}

We again drop the payment terms on the left and right hand side, because they are equivalent
under the action deviations considered under obedience.
A mechanism is {\em ex post truthful} if it satisfies, for every agent $i$, for each $\rho \in\mathcal{T}$, and each misreport $\rho_i' \in\mathcal{T}_i$:
\begin{equation}\label{eqn:expost-truth}
     \mathop{\mathbb{E}}_{ \substack{a \sim \sigma(\omega, \rho), \\ \omega \sim \theta_i}} \left[u_i( a, \omega, \rho_i) - t_i(\rho) \right] \geq \mathop{\mathbb{E}}_{ \substack{a \sim \sigma(\omega, \rho_i', \rho_{-i}), \\ \omega \sim \theta_i}\ } \left[u_i( a, \omega, \rho_i)- t_i(\rho_i', \rho_{-i}) \right].
\end{equation}

\paragraph{Individual Rationality} \label{sec:ir} A mechanism is {\em interim individually rational} (IIR) if reporting an agent's true type guarantees at least as much expected utility as opting out of the mechanism, assuming all the other agents report their types truthfully and follow the action recommendations. Let $\tilde{\sigma}_{-i}: \Omega \times\mathcal{T}_{-i} \to \Pi_{j \in N, j\not=i}\triangle\mathcal({A}_{j})$ denote the recommendations to the other buyers, which the seller commits to {\em ex ante} when buyer $i$ opts out. We can think about $\tilde{\sigma}_{-i}$ as a
{\em contingency rule}, which
specifies the signals made when only the other buyers participate. The signaling scheme for
each buyer, in this alternative, need not be the same as that when all buyers participate.
For each agent $i$, and each $\rho_i \in\mathcal{T}_i$, IIR requires
\begin{equation}\label{eqn:iir_base}
   \mathbb{E}_{-i} \left[ \mathop{\mathbb{E}}_{ \substack{a \sim \sigma(\omega, \rho), \\\omega \sim \theta_i}} \left[u_i( a, \omega, \rho_i) - t_i(\rho)\right]\right] \geq \max_{\hat{a}_i \in \mathcal{A}_i} \mathbb{E}_{-i} \left[ \mathop{\mathbb{E}}_{ \substack{a_{-i} \sim \tilde{\sigma}_{-i}(\omega, \rho_{-i}),\\\omega \sim \theta_i}} \left[u_i( \hat{a}_i, a_{-i}, \omega, \rho_i) \right] \right].\
\end{equation}

 The right-hand side of this inequality represents buyer $i$'s expected utility from opting out, which we refer to as the {\em utility of the outside option}, denoted by $u_i^{\text{out}}(\rho_i)$. By controlling the recommendations made to participating buyers, this influences non-participating buyers' utility through negative externalities. Hence, the seller can select the contingency rule $\tilde{\sigma}_{-i}$ to minimize the opting-out buyer’s utility (considering that the buyer will pick a best $\hat{a}_i$ in response), making non-participation relatively less attractive.
Minimizing $u_i^{\text{out}}(\rho_i)$ allows the seller to extract higher payments. Rearranging the terms in Equation~\ref{eqn:iir_base}, we get, for each $\rho_i \in\mathcal{T}_i$,:
\begin{equation}
   \mathbb{E}_{-i} [t_i(\rho)] \leq \mathbb{E}_{-i}\left[ \mathop{\mathbb{E}}_{ \substack{a \sim \sigma(\omega, \rho), \\\omega \sim \theta_i}} \left[u_i( a, \omega, \rho_i)\right]\right] - u_i^{\text{out}}.
\end{equation}

Thus, a smaller $u_i^{\text{out}}$ directly increases the maximum payment the seller can charge while still satisfying the buyer’s participation constraint. The optimal choice for the contingency scheme, optimally strengthening the IIR constraint, is, for each $\rho_i \in\mathcal{T}_i$:
\begin{equation}\label{eqn:iir}
   \mathbb{E}_{-i} \left[ \mathop{\mathbb{E}}_{ \substack{a \sim \sigma(\omega, \rho), \\\omega \sim \theta_i}} \left[u_i( a, \omega, \rho_i) - t_i(\rho)\right]\right] \geq \min_{\tilde{\sigma}_{-i}}\; \max_{\hat{a}_i \in \mathcal{A}_i} \;\left( \mathbb{E}_{-i} \left[ \mathop{\mathbb{E}}_{ \substack{a_{-i} \sim \tilde{\sigma}_{-i}(\omega, \rho_{-i}), \\\omega \sim \theta_i} } \left[u_i( \hat{a}_i, a_{-i}, \omega, \rho_i) \right] \right] \right).
\end{equation}

In particular, recognizing that a buyer's utility decreases the more informed other buyers are, the seller can adopt a contingency scheme in which the recommendations made to others reveal full information.

We also consider a stronger version of IR, namely {\em ex post individual rationality} (or simply IR). In this case, for each agent $i$, and each $\rho \in\mathcal{T}$, we require:
\begin{equation}\label{eqn:ir}
 \mathop{\mathbb{E}}_{ \substack{a \sim \sigma(\omega, \rho), \\\omega \sim \theta_i}} \left[u_i( a, \omega, \rho_i) - t_i(\rho)\right] \geq \min_{\tilde{\sigma}_{-i}}\;\max_{\hat{a}_i \in \mathcal{A}_i}\; \left( \mathop{\mathbb{E}}_{ \substack{a_{-i} \sim \tilde{\sigma}_{-i}(\omega, \rho_{-i}), \\\omega \sim \theta_i} } \left[u_i( \hat{a}_i, a_{-i}, \omega, \rho_i) \right] \right).
\end{equation}

We give an illustrative example.
\begin{example}
Consider a setting with $n = 2$ buyers, two states of the world $\Omega = \{0, 1\}$ and two actions for each buyer $\mathcal{A}_1 = \mathcal{A}_2 = \{0, 1\}$. For two states, the prior of an agent $i$ is represented by two numbers, $(\theta_{i0},\theta_{i1})$, with $\theta_{i0}+\theta_{i1}=1$,
giving the prior on state $0$ and the
prior on state $1$.
For this example, each agent has the same, fixed prior, and $\Theta_1 = \Theta_2 = \{ (0.3, 0.7) \}$. Let $\mathcal{V}_1 = \mathcal{V}_2 = U[0, 1]$. Since the priors are fixed, the type of agent $i$ is represented by $v_i$; i.e., we have $\rho_1 = v_1$ and $\rho_2=v_2$. The payoff function for agent $i$ is
\begin{equation}
    u_i( a_i, a_{-i}, \omega, v_i) = v_i \left(\mathbf{1} \{ a_i = \omega\} - \alpha \mathbf{1} \{ a_{-i} = \omega\}\right),
\end{equation}
for some $\omega \in \Omega$. In each state of the world, the dominant strategy, for each of buyers 1 and 2, is to play the action matching the state. Buyer $i$ also incurs a negative externality if the other buyer (denoted by $-i$) chooses the correct action. The intensity of the negative externality is captured by parameter $\alpha \in \mathbb{R}_{\geq 0}$.

Consider a simple mechanism
that just offers the same fixed experiment, or signaling scheme, to each buyer,
and recommends action $1$ with probability $0.9$ if $\omega = 0$, and with probability $0.2$ if $\omega = 1$. This experiment is the same for all bid profiles, with conditional probability matrix:
\begin{align}
    \pi_i(\cdot) = \begin{pmatrix} 0.1 & 0.9 \\ 0.8 & 0.2 \end{pmatrix}.
\end{align}

\paragraph{Obedience:} Since the signaling scheme does not depend on reports, we write $a\sim \sigma$ to denote the recommended action profile $a=(a_1,a_2)$. The expected utility to agent $i$, with value $v_i$ and belief $\theta_i$, if obedient, is given by:
\begin{equation}
\begin{aligned}
    \mathop{\mathbb{E}}_{\substack{a\sim\sigma,\\\omega\sim\theta_i}}[u_i(a_i, a_{-i}, \omega, v_i)]
    &= \mathop{\mathbb{E}}_{\substack{a\sim\sigma,\\\omega\sim\theta_i}}[ v_i \left(\mathbf{1} \{ a_i = \omega\} - \alpha \mathbf{1} \{ a_{-i} = \omega\}\right)]\\
    &= v_i \left( \sum_{j = 0}^{1} \text{Pr}[\omega = j] \text{Pr}[a_i = j \mid \omega = j] \right)- v_i\alpha\mathop{\mathbb{E}}_{\substack{a\sim\sigma,\\\omega\sim\theta_i}}[ \mathbf{1} \{ a_{-i} = \omega\}]\\
    &= 0.17v_i - v_i\alpha\mathop{\mathbb{E}}_{\substack{a\sim\sigma,\\\omega\sim\theta_i}}[ \mathbf{1} \{ a_{-i} = \omega\}].
\end{aligned}
\end{equation}

Since the signaling scheme is fixed, the expression for obedience~\eqref{eq:interimobed} does not depend on the type of the other player. Consider the following deviation function $\delta$ such that agent $i$ takes the exact opposite action to that recommended i.e., $\delta(j) = 1 - j$. The utility to agent $i$ under this deviation is
\begin{equation}
\begin{aligned}
    \mathop{\mathbb{E}}_{\substack{a\sim\sigma,\\\omega\sim\theta_i}}[u_i(\delta(a_i), a_{-i}, \omega, v_i)]
    &= \mathop{\mathbb{E}}_{\substack{a\sim\sigma,\\\omega\sim\theta_i}}[ v_i \left(\mathbf{1} \{ \delta(a_i) = \omega\} - \alpha \mathbf{1} \{ a_{-i} = \omega\}\right)]\\
    &= v_i \left( \sum_{j = 0}^{1} \text{Pr}[\omega = j] \text{Pr}[a_i = 1 - j \mid \omega = j] \right)- v_i\alpha\mathop{\mathbb{E}}_{\substack{a\sim\sigma,\\\omega\sim\theta_i}}[ \mathbf{1} \{ a_{-i} = \omega\}]\\
    &= 0.83v_i - v_i\alpha\mathop{\mathbb{E}}_{\substack{a\sim\sigma,\\\omega\sim\theta_i}}[ \mathbf{1} \{ a_{-i} = \omega\}]\\
    &> \mathop{\mathbb{E}}_{\substack{a\sim\sigma,\\\omega\sim\theta_i}}[u_i(a_i, a_{-i}, \omega, v_i, \theta_i)].
\end{aligned}
\end{equation}
Thus, this simple mechanism is not obedient.

\paragraph{Utility of opting out: }In order to check whether a mechanism is IIR, we first need to compute the utility of the outside option. If the agent opts out, and thus receives no additional information, the best action $\hat{a}_i$ corresponds to the state with the highest prior probability, and thus $\hat{a}_i=1$ since the prior is $\theta_i=(0.3,0.7)$. To minimize buyer $i$'s utility if he opts out, the seller can then recommend the other buyer the correct state with probability $1$. Thus the utility of opting out is given by $v_i ( \max \{ \theta_i \} - \alpha (1)) = v_i (0.7 - \alpha)$.
\end{example}

\section{Optimal Data Market Design in the Single-Buyer Setting}
\label{subsec:rochet}

In this section, we formulate the problem of optimal data market design for a single buyer as an unsupervised learning problem. We study a parametric class of mechanisms, $(\sigma^w, t^w) \in \mathcal{M}$, for parameters, $w \in \mathbb{R}^d$, where $d > 0$. For a single buyer, the goal is to learn parameters $w$ that maximize $\mathbb{E}_{\rho \sim \mathcal{P}}[t^w (\rho)]$, for distribution on type profiles $\mathcal{P}$, such that $(\sigma^w, t^w)$ satisfy {\em ex post }IC and IR.\footnote{For the single bidder settings, {\em ex post } IC and IR are the same as BIC and IIR.} By adopting differentiable loss functions, we can make use of tools from automatic differentiation and stochastic gradient descent (SGD). For notational convenience, we drop the subscript in this case, as there is only one buyer.

\subsection{Neural Network Architecture}

For the single buyer setting, any {\em ex post} IC mechanism can be represented as a menu of experiments. For this, we extend the RochetNet architecture to represent a menu of priced statistical experiments. Specifically, the parameters correspond to a menu of $P$ choices, where each choice $p \in [P]$ is associated with an experiment, with parameters $\gamma^p \in \mathbb{R}^{|\Omega| \times |A| }$, and a price, $\beta^p \in \mathbb{R}$. Given this, we define a {\em menu entry} as,
\begin{align}
    \pi^{p}_{\omega, j} &:= \frac{\exp(\frac{\gamma^p_{\omega, j}}{\tau_\pi})}{\sum_{k \in [m]} \exp(\frac{\gamma^p_{\omega, k}}{\tau_\pi})}, \quad
    t^p := \beta^p
\end{align}
where $\tau_\pi$ is a fixed hyperparameter that helps with optimization stability.

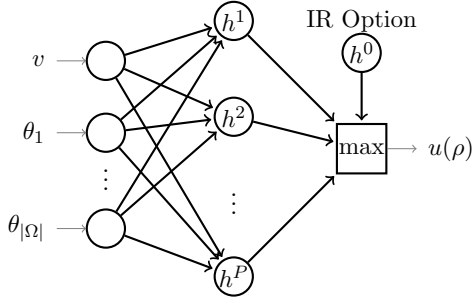
\begin{figure}[t]
\centering
\scalebox{1}{
\centering
\begin{tikzpicture}[scale=0.85, transform shape, shorten >=1pt,->,draw=black!100, node distance=\layersep, thick]
    \tikzstyle{input text}=[circle,minimum size=17pt,inner sep=0pt]
    \tikzstyle{input neuron}=[circle,draw=black!100,minimum size=17pt,inner sep=0pt,thick]
    \tikzstyle{hidden neuron}=[circle,draw=black!100,minimum size=17pt,inner sep=0pt,thick]
    \tikzstyle{hidden text}=[minimum size=22pt,inner sep=0pt,thick]
    \tikzstyle{unit}=[draw=black!100,minimum size=22pt,inner sep=0pt,thick]

    \node[input neuron, pin={[pin edge={<-}]left:$v$}] (I-1) at (0,-1.125) {};
    
    \node[input neuron, pin={[pin edge={<-}]left:$\theta_1$}] (I-2) at (0,-2.25) {};
    \node[input text] (I-34) at (0,-2.875) {$\vdots$};
    \node[input neuron, pin={[pin edge={<-}]left:$\theta_{|\Omega|}$}] (I-3) at (0,-3.75) {};

    \path node[hidden neuron] (H-1) at (2,-0.5) {$h^1$};
    \path node[hidden neuron] (H-2) at (2,-2) {$h^2$};
    \path node[hidden text] (H-0) at (2,-3.25) {\vdots};
    \path node[hidden neuron] (H-3) at (2,-4.5) {$h^P$};
    \path node[hidden neuron] (H-4) at (4,-1) {$h^0$};
    \node[input text] (IR) at (4,-0.5) {$\text{IR Option}$};

    \foreach \src in {1,...,3}
    	\foreach \dest in {1,...,3}
			\path (I-\src) edge (H-\dest); 

	\node[unit, pin={[pin edge={->}]right:$u(\rho)$}] (M) at (4,-2.5) {max};
	
	\foreach \src in {1,...,4}
		\path (H-\src) edge (M); 
\end{tikzpicture}}
\caption{RochetNet: Neural network architecture for learning the optimal menu for a single bidder setting. The neural network takes in as inputs the buyer's payoff $v$ and interim belief $\theta$ (a distribution on world state)
and computes the utility to the buyer for each of $P + 1$ menu options, where option $0$ corresponds to the uninformative experiment.
\label{fig:rochetnet_architecture}}
\end{figure}

The neural network architecture for the single buyer settings is depicted in Figure~\ref{fig:rochetnet_architecture}. For the single buyer setting we drop index $i$ on types and write $\rho=(v,\theta)$ as the type of the buyer.
The input layer takes $\rho = (v, \theta)$ as input and the network computes the $P$ utility values corresponding to each menu entry. We do not impose the obedience constraints explicitly. However, while computing the utility of a menu, we take into consideration the best possible deviating action an agent can take. From the perspective of a buyer, if an action $k$ is recommended by an experiment choice $p$, then $Pr[\omega = j | a_k, \theta] = \frac{\theta_{j} \pi^p_{j, k}}{Pr[a_k]}$. The best action deviation is thus $\delta(k) = \arg \max_{j \in [m]} \frac{ \theta_{j} \pi^p_{j, k}}{Pr[a_k]} = \arg \max_{j \in [m]} \theta_{j} \pi^p_{j, k}$, yielding a utility of $v \cdot \max_{j \in [m]} \frac{ \theta_{j} \pi^p_{j, k}}{Pr[a_k]}$. Taking an expectation over all signals, buyer $i$ receives the following utility for the choice $p \in [P]$:
\begin{equation}\label{eqn:rochet-ob}
h^p(\rho) = v \cdot \left(\sum_{k \in [m]} max_{j \in [m]} \{ \theta_j \pi^p_{j, k}\}\right) - \beta^p
\end{equation}
We also include an additional experiment, $\pi^0$, which corresponds to a {\em null experiment} that generates the same signal regardless of state, and thus provides no useful information. This menu entry has a price of $t^0 = 0$, and thus $h^0(\rho) = v \cdot (\max_{j\in[m]} \theta_j)$. In particular, this menu entry has the same utility as that of opting out.

\begin{lemma}
    Let $p^*(\rho) = \arg \max_{p \in \{0, \ldots, P\}} h^p(\rho)$ denote the optimal menu choice. For each menu option $p$, interpret $\pi^p_\omega$ as the recommendation rule obtained by relabeling each signal by the buyer’s optimal action conditional on that signal. Then the mechanism $\mathcal{M} = (\sigma^w, t^w)$ where $\sigma^w(\omega, \rho) = \pi^{p^*(\rho)}_{\omega}$ and $t^w(\rho) = t^{p^*(\rho)}$ satisfies {\em ex post} IC and IR.
\end{lemma}

This mechanism is {\em ex post} IC as it is agent optimizing, and is IR as it guarantees a buyer at least the utility of opting out. The expected revenue of an {\em ex post} IC and IR mechanism that is parameterized by $\gamma, \beta$ is given by $\mathbb{E}_{\rho \sim \mathcal{P}} [\beta^{p^*(\rho)}]$. Here, $\mathcal{P}$ corresponds to the type distribution of the single buyer.

In our experiments, we initialize RochetNet with a menu of $P = 1000$ choices. This design choice of a large over-parameterized network follows from the recommendation of \citet{dutting2023jacm} (Appendix B.4), who show that a large menu helps recover an optimal mechanism even though only a few options end up being used in the learned mechanism.

After training, a menu option $p\in[P]$ is declared active if there exists at least one type profile $\rho$ in the training set for which $p \;=\; \arg\max_{p'\in\{0, \ldots, P\}} h^{p'}(\rho)$. All inactive options are discarded, and the remaining active set becomes the final, optimized menu.

\subsection{Training Problem}

Rather than minimizing a loss function that measures errors against ground truth labels (as in a supervised learning setting), our goal is to minimize expected negated revenue. To ensure that the objective is differentiable, we replace the argmax operation with a softmax during training. The {\em loss function}, for parameters $\gamma$ (signaling scheme) and $\beta$ (prices), is thus given by $\mathcal{L}(\gamma, \beta) = \mathbb{E}_{\rho \sim \mathcal{P}} \left[ -\sum_{p=1}^{P}\beta^p \tilde{\nabla}^p(\rho)\right],$
where $\tilde{\nabla}^p(\rho) = \text{softmax} (\frac{h^0(\rho)}{\tau_u}, \frac{h^1(\rho)}{\tau_u}, \ldots, \frac{h^P(\rho)}{\tau_u})$, and $\tau_u > 0$ controls the quality of approximation. Note that, at test time, we revert to using the hard max, so as to guarantee {\em ex post} IC and IR from a trained network.

During training, the null experiment $(\pi^{0}, t^0)$ remains fixed, while the parameters $\gamma \in \mathbb{R}^{P \times |\Omega| \times m}$ and $\beta \in \mathbb{R}^P$ are optimized through SGD on the empirical version of the loss calculated over samples $\hat{\mathcal{P}} = \{ \rho^{(1)}, \ldots \rho^{(L)}\}$ drawn i.i.d.~from $\mathcal{P}$.
This assumption of having access to i.i.d.~samples from the buyers' type distribution $\mathcal{P}$ is standard in the automated mechanism design literature. Such an assumption is justified by the increasing availability of historical market data or past interactions from similar contexts, which allows sellers to approximate underlying buyer type distributions and thus facilitate effective training of neural mechanisms (using econometric approaches as needed). The assumption that the seller knows the type distribution of buyers is also standard to the economic theory of optimal mechanism design.

We report our results on a separate test set sampled from $\mathcal{P}$. All reported results are averaged across five independent training runs with different seeds. Each run took 11--12 minutes on a single NVIDIA Tesla V100 GPU.
We refer to Appendix~\ref{app:setup} for more details regarding the choice of hyperparameters.

\section{Optimal Data Market Design in the Multi-Buyer Setting}
\label{sec:multi}

In the multi-buyer setting, the goal is to learn parameters $w \in \mathbb{R}^d$ that maximize $\mathbb{E}_{\rho \sim \mathcal{P}}[\sum_{i=1}^n t^w_i (\rho)]$, where $\mathcal{P}$ is used to represent the joint distribution on types of the $n$ buyers. This optimization problem is solved for a parametric class of mechanisms, $(\sigma^w, t^w) \in \mathcal{M}$, such that $(\sigma^w, t^w)$ satisfy IC (or BIC) and IR (or IIR). By restricting our loss computations to differentiable functions, we can again use tools from automatic differentiation and SGD. For the multi-buyer setting, we use differentiable approximations to represent the rules of the mechanism, and control the degree to which IC constraints are violated during training through the use of an augmented Lagrangian method~\citep{dutting2019}.

\subsection{Neural Network Architecture}

The neural network architecture has two components: one that encodes the experiments and another that encodes payments. We model these components in a straightforward way as feed-forward, fully connected neural networks with Leaky ReLU activation functions.\footnote{Unlike ReLU, which outputs zero for all negative inputs, Leaky ReLU allows a small, nonzero gradient for negative inputs. This prevents ``dead neurons" during training, leading to better gradient flow and more stable training.} The input layer consists of the reported type profile, $\rho$, encoded as a vector. For both IC and BIC settings, the component that encodes experiments outputs a matrix $\pi$ of dimensions $n \times |\Omega| \times |\Omega| $, which represents an experiment that corresponds to each of the $n$ agents. In order to ensure feasibility, i.e., the probability values of sending signals for each agent under each state realization is non-negative and sums up to 1, the neural network first computes an output matrix, $\gamma$, of the same dimension. For all $i, j, \omega$, we then obtain $\pi_{i, \omega, j}$ by computing $\exp(\frac{\gamma_{i,\omega, j}}{\tau_\pi})/\sum_k \exp(\frac{\gamma_{i,\omega, k}}{\tau_\pi})$ where $\tau_\pi$ is a fixed hyperparameter.

For the setting of IC (vs. BIC), we define payments by first computing a {\em normalized payment},
 $\tilde{t}_i^w(\rho) \in [0, 1]$, for each buyer $i$,
 using a sigmoid activation unit. The payment $t_i^w(\rho)$ is computed as follows:
 \begin{equation}\label{eqn:ir-net}
     t_i^w(\rho) = \tilde{t}_i^w(\rho) \cdot v_i \cdot \left( \sum_{k \in [m]}\pi_{i,k,k}^w(\rho)\theta_{i,k} - \bar{\alpha} \sum_{k \in [m]}\sum_{j \in [n]\setminus i}\pi_{j,k,k}^w(\rho)\theta_{i,k} - \left(\max_k \theta_{i,k} - \alpha\right)\right).
 \end{equation}

 \begin{restatable}{lemma}{irnet}
 \label{lem:ir-net}
     Any mechanism $\mathcal{M} = (\sigma^w, t^w)$ where $\sigma^w_i(\omega, \rho) = \left(\pi_{i, \omega}^w(\rho)\right)$ and $t^w$ satisfies Eqn~\ref{eqn:ir-net} is ex post IR for any $w \in \mathbb{R}^d$, provided that the induced experiment $\pi_i^w(\rho)$ is obedient with respect to buyer $i$'s posterior beliefs.
 \end{restatable}

For the BIC setting since only {\em interim} payment is required for computing all the constraints and revenue, we can replace $\mathbb{E}_{-i} [t_i(\rho_i, \rho_{-i})]$ by $t_i(\rho_i)$. For this, we compute an interim normalized payment, $\tilde{t}_i \in [0, 1]$, for each buyer $i$ by using a sigmoid unit. We compute the {\em interim} payment as:
 \begin{equation}\label{eqn:iir-net}
     t_i^w(\rho_i) = \tilde{t}_i^w(\rho_i) \cdot v_i \cdot \mathbb{E}_{-i}\left[ \sum_{k \in [m]}\pi_{i,k,k}^w(\rho)\theta_{i,k} - \bar{\alpha} \sum_{k \in [m]}\sum_{j \in [n]\setminus i}\pi_{j,k,k}^w(\rho)\theta_{i,k} - \left(\max_k \theta_{i,k} - \alpha\right)\right].
 \end{equation}

  \begin{restatable}{lemma}{iirnet}
  \label{lem:iir-net}
     Any mechanism $\mathcal{M} = (\sigma^w, t^w)$ where $\sigma^w_i(\omega, \rho) = \left(\pi_{i, \omega}^w(\rho)\right)$ and $t^w$ satisfies Eqn~\ref{eqn:iir-net} is interim IR for any $w \in \mathbb{R}^d$, provided that the induced experiment $\pi_i^w(\rho)$ is obedient with respect to buyer $i$'s posterior beliefs.
  \end{restatable}

\begin{figure}[t]
\centering
\scalebox{1}{
\centering

\begin{tikzpicture}[scale=0.6,transform shape, shorten >=1pt,->,draw=black!100, node distance=\layersep, thick]
   
   \def\x{0}     
   \def\y{9}   
   
    \tikzstyle{input text}=[draw=white,minimum size=22pt,inner sep=0pt]
    \tikzstyle{input neuron}=[circle,draw=black!100,minimum size=17pt,inner sep=0pt,thick]
    \tikzstyle{hidden neuron1}=[draw=black!100,minimum size=25pt,inner sep=0pt,thick]
    \tikzstyle{hidden neuron}=[circle,draw=black!100,minimum size=25pt,inner sep=0pt,thick]
    \tikzstyle{hidden text}=[draw=white,minimum size=22pt,inner sep=0pt,thick]
    \tikzstyle{unit}=[circle,draw=black!100,minimum size=25pt,inner sep=0pt,thick]
    
    \node[input text] (I-0) at (0,-0.65) {$\vdots$};
    \node[input text] (I-0) at (0,-2.45) {$\vdots$};
    \node[input text] (I-0) at (0,-4.2) {$\vdots$};
    \node[input neuron] (I-1) at (\x, 0.) {};
    \node[input neuron] (I-2) at (\x,-1.5) {};
    \node[input neuron] (I-3) at (\x,-3.5) {};
    \node[input neuron] (I-4) at (\x,-5) {};

    \draw [-, decorate, decoration={brace, mirror},xshift=-0.5cm,yshift=0pt] (\x,0.15) -- (\x,-1.65);
    \draw [-, decorate, decoration={brace, mirror},xshift=-0.5cm,yshift=0pt] (\x,-3.35) -- (\x,-5.15);

    \draw [-, decorate, decoration={brace, mirror}, xshift=-0.5cm, yshift=0pt] (\x, 0.15) -- (\x, -1.65) node[midway, left,xshift=-2pt] {$\rho_1$};
    \draw [-, decorate, decoration={brace, mirror}, xshift=-0.5cm, yshift=0pt] (\x, -3.35) -- (\x, -5.15) node[midway, left,xshift=-2pt] {$\rho_n$};

    \path node[hidden neuron] (H-1) at (\x+1.5,-0.5) {$h^{(1)}_1$};
    \path node[hidden neuron] (H-2) at (\x+1.5,-2) {$h^{(1)}_2$};
    \path node[hidden text] (H-0) at (\x+1.5,-3.25) {\vdots};
    \path node[hidden neuron] (H-3) at (\x+1.5,-4.5) {$h^{(1)}_{J_1}$};
    
    \path node[hidden neuron] (J-1) at (\x+3,-0.5) {$h^{(R)}_1$};
    \path node[hidden neuron] (J-2) at (\x+3,-2) {$h^{(R)}_2$};
    \path node[hidden text] (J-0) at (\x+3,-3.25) {\vdots};
    \path node[hidden neuron] (J-3) at (\x+3,-4.5) {$h^{(R)}_{J_{R}}$};
    
    \node[input text] (S-0) at (\x+4.5,-0.65) {$\vdots$};
    \node[input text] (S-0) at (\x+4.5,-2.45) {$\vdots$};
    \node[input text] (S-0) at (\x+4.5,-4.2) {$\vdots$};

    \path node[unit] (S-1) at (\x + 4.5,-0) {$\pi^1_{11}$};
    \path node[unit] (S-2) at (\x + 4.5,-1.5) {$\pi^1_{|\Omega|1}$};
    \path node[unit] (S-3) at (\x +4.5,-3.5) {$\pi^n_{11}$};
    \path node[unit] (S-4) at (\x + 4.5, -5) {$\pi^n_{|\Omega|1}$};

    \node[input text] (S-0) at (\x+6.0,-0.65) {$\vdots$};
    \node[input text] (S-0) at (\x+6.0,-2.45) {$\vdots$};
    \node[input text] (S-0) at (\x+6.0,-4.2) {$\vdots$};

    \path node[unit] (T-1) at (\x + 6.0,-0) {$\pi^1_{1m}$};
    \path node[unit] (T-2) at (\x + 6.0,-1.5) {$\pi^1_{|\Omega|m}$};
    \path node[unit] (T-3) at (\x + 6.0,-3.5) {$\pi^n_{1m}$};
    \path node[unit] (T-4) at (\x + 6.0, -5) {$\pi^n_{|\Omega|m}$};

    \foreach \src in {1,...,4}
    	\foreach \dest in {1,...,3}
			\path (I-\src) edge (H-\dest); 
			
	\foreach \src in {1,...,3}
		\foreach \dest in {1,...,4}
			\path (J-\src) edge (S-\dest); 
	
	\node at  ($(H-2.south east) + (0.45,0)$) {\ldots};
	
     \draw[thin] ($(S-1.north west)+(-0.3,0.4)$)  rectangle ($(T-2.south east)+(0.3,-0.4)$) ;
     \node (N-1) at  ($(S-1.north west) + (1.0,0.8)$) {$\text{row-wise}$};
     \node (N-1) at  ($(S-1.north west) + (1.0,0.6)$) {$\text{softmax}$};

     \draw[thin] ($(S-3.north west)+(-0.3,0.4)$)  rectangle ($(T-4.south east)+(0.3,-0.4)$) ;
    \node at  ($(S-4.south west) + (1.0,-0.55)$) {$\text{row-wise}$};
    \node at  ($(S-4.south west) + (1.0,-0.75)$) {$\text{softmax}$};
	
    \draw[thick,dashed] ($(S-1.north west)+(-0.2,0.3)$)  rectangle ($(T-1.south east)+(0.2,-0.3)$) ;
    \draw[thick,dashed] ($(S-2.north west)+(-0.2,0.3)$)  rectangle ($(T-2.south east)+(0.2,-0.3)$) ;
    \draw[thick,dashed] ($(S-3.north west)+(-0.2,0.3)$)  rectangle ($(T-3.south east)+(0.2,-0.3)$) ;
    \draw[thick,dashed] ($(S-4.north west)+(-0.2,0.3)$)  rectangle ($(T-4.south east)+(0.2,-0.3)$) ;

    \tikzstyle{input text}=[draw=white,minimum size=22pt,inner sep=0pt]
    \tikzstyle{input neuron}=[circle,draw=black!100,minimum size=17pt,inner sep=0pt,thick]
    \tikzstyle{hidden neuron1}=[draw=black!100,minimum size=25pt,inner sep=0pt,thick]
    \tikzstyle{hidden neuron}=[circle,draw=black!100,minimum size=25pt,inner sep=0pt,thick]
    \tikzstyle{hidden text}=[draw=white,minimum size=22pt,inner sep=0pt,thick]
    \tikzstyle{unit}=[draw=black!100,minimum size=20pt,inner sep=0pt,thick]
     
    \node[input text] (I-0) at (\y,-0.65) {$\vdots$};
    \node[input text] (I-0) at (\y,-2.45) {$\vdots$};
    \node[input text] (I-0) at (\y,-4.2) {$\vdots$};
    \node[input neuron] (I-1) at (\y,0) {};
    \node[input neuron] (I-2) at (\y,-1.5) {};
    \node[input neuron] (I-3) at (\y,-3.5) {};
    \node[input neuron] (I-4) at (\y,-5) {};
    
    \draw [-, decorate, decoration={brace, mirror}, xshift=-0.6cm, yshift=0pt] ([xshift=-5pt]I-1.west) -- ([xshift=-5pt]I-2.west) node[midway, left, xshift=-2pt] {$\rho_1$};
    \draw [-, decorate, decoration={brace, mirror}, xshift=-0.6cm, yshift=0pt] ([xshift=-5pt]I-3.west) -- ([xshift=-5pt]I-4.west) node[midway, left, xshift=-2pt] {$\rho_n$};

    \path node[hidden neuron] (H-1) at (\y+1.5,-0.5) {$c^{(1)}_1$};
    \path node[hidden neuron] (H-2) at (\y+1.5,-2) {$c^{(1)}_2$};
    \path node[hidden text] (H-0) at (\y+1.5,-3.25) {\vdots};
    \path node[hidden neuron] (H-3) at (\y+1.5,-4.5) {$c^{(1)}_{J'_1}$};
    
    \path node[hidden neuron] (J-1) at (\y+3,-0.5) {$c^{(T)}_1$};
    \path node[hidden neuron] (J-2) at (\y+3,-2) {$c^{(T)}_2$};
    \path node[hidden text] (J-0) at (\y+3,-3.25) {\vdots};
    \path node[hidden neuron] (J-3) at (\y+3,-4.5) {$c^{(T)}_{J'_{T}}$};
    
    \path node[unit] (S-1) at (\y+4.5,-0.5) {$\sigma$};
    \path node[unit] (S-2) at (\y+4.5,-2) {$\sigma$};
    \path node[hidden text] (S-0) at (\y+4.5,-3.25) {\vdots};
    \path node[unit] (S-3) at (\y+4.5,-4.5) {$\sigma$};
    
    \path node[hidden text] (T-1) at (\y+6,-0.5) {$\tilde{t}_1$};
    \path node[hidden text] (T-2) at (\y+6,-2) {$\tilde{t}_2$};
    \path node[hidden text] (T-0) at (\y+6,-3.25) {\vdots};
    \path node[hidden text] (T-3) at (\y+6,-4.5) {$\tilde{t}_n$};
    
    \path node[hidden neuron, pin={[pin edge={->}]right:$\displaystyle {t}_1$}] (M-1) at (\y+7.5,-0.5) {$\times$};
    \path node[hidden neuron, pin={[pin edge={->}]right:$\displaystyle{t}_2$}] (M-2) at (\y+7.5,-2) {$\times$};
    \path node[hidden text] (M-0) at (\y+7.5,-3.25) {\vdots};
    \path node[hidden neuron, pin={[pin edge={->}]right:$\displaystyle {t}_n $}] (M-3) at (\y+7.5,-4.5) {$\times$};

    \foreach \src in {1,...,4}
    	\foreach \dest in {1,...,3}
			\path (I-\src) edge (H-\dest); 
			
	\foreach \src in {1,...,3}
		\foreach \dest in {1,...,3}
			\path (J-\src) edge (S-\dest); 
			
	\foreach \src in {1,...,3}
		\path (S-\src) edge (T-\src); 
	
	\foreach \src in {1,...,3}
		\path (T-\src) edge (M-\src); 
	
	\node at  ($(H-2.south east) + (0.45,0)$) {\ldots};

   	 \node[input text] (temp-1) at (\x+6.25,-0.0) {}; 
   	 \draw (temp-1) edge[bend right=-25,solid,line width=0.6mm,dotted,color=gray] node[below,color=black]{$\pi^1$} (M-1); 
	
	 \node[hidden text] (temp-2) at (\y+6.6,0.5) {$\mathbf \rho$};
	 \draw[<-,solid,line width=0.5mm,dotted,color=gray] (M-1)  -- (\y+6.7,0.43); 
	 
	 \node[input text] (temp-3) at (\x+6.25,-2.5) {}; 
   	 \path (temp-3) edge[bend right=15,solid,line width=0.6mm,dotted,color=gray]  (M-2);
   	 
   	 \node[hidden text] (temp-4) at (\y+6.6,-1.05) {$\mathbf \rho$};
	 \draw[<-,solid,line width=0.5mm,dotted,color=gray] (M-2)  -- (\y+6.7,-1.13); 
	 
	 \node[input text] (temp-5) at (\x+6.25,-4.2) {}; 
   	 \path (temp-5) edge[bend right=25,solid,line width=0.6mm,dotted,color=gray]  
node[auto,color=black]{$\pi^n$}    	 
   	 (M-3);
   	 
   	 \node[hidden text] (temp-6) at (\y+6.6,-3.60) {$\mathbf \rho$};
	\draw[<-,solid,line width=0.6mm,dotted,color=gray] (M-3)  -- (\y+6.7,-3.68);

    \draw [-,dashed] (7.0,1.5) -- (7.0,-6.5);
    
    \node at  (2.7,1.5) {\textbf{Experiment Network} $\sigma$};	 
    \node at  (13.2,1.5) {\textbf{Payment Network} $t$};

    \node[unit] (metrics) at (6.5,-7.5) {~\textit{Metrics}: $rev$, $rgt_1,\ldots,rgt_n$~~~};
	\node (alloc) at (\x+2,-5.9) {};
	\path (alloc) edge[bend right=25,solid,line width=1.5mm]  node[below]{$w_\sigma$~~~} (metrics); 	 
	\node (pay) at (\x+11,-5.9) {};
	\path (pay) edge[bend left=25,solid,line width=1.5mm] node[below]{~~~$w_t$}  (metrics); 	 
		 
\end{tikzpicture}}
\caption{RegretNet: Neural network architecture for the multi-buyer {\em ex post} IC setting. The inputs are the reported types $\rho$ of each buyer. The experiment network outputs an experiment of size $|\Omega|\times|\Omega|$ for each buyer, where a row-wise softmax operation ensures the experiments are valid. The payment network outputs the normalized payment $\tilde{t}_i^w(\rho)\in [0,1]$ through a sigmoid activation unit that is used to compute payment via Equation~\ref{eqn:ir-net}.
\label{fig:regretnet_architecture}}
\end{figure}
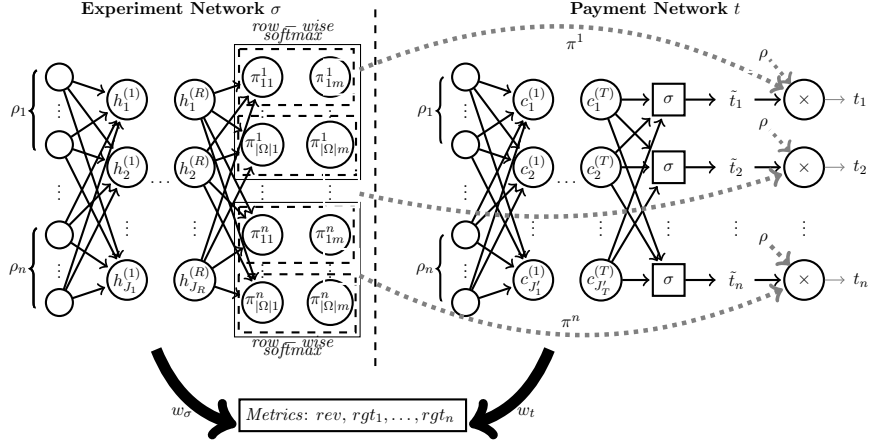

The neural network architecture for the multi-buyer {\em ex post} IC setting is depicted in Figure~\ref{fig:regretnet_architecture}. For the BIC setting, the experiment network $\sigma$ remains the same and the interim experiments are computed through sampling. The payment network only takes the type corresponding to the buyer as input as we can replace $\mathbb{E}_{\rho_{-i} \sim \mathcal{P}_{-i}} [t_i(\rho_i, \rho_{-i})]$ with just $t_i(\rho_i)$. We also have separate payment networks for each buyer if their type distributions are different.

\subsection{Training Problem}
The training problem here amounts to maximizing expected revenue while enforcing incentive constraints. We operationalize this by minimizing the negated expected revenue subject to IC (or BIC) and IR (or IIR). Following~\citet{dutting2019}, we quantify violations of incentive compatibility using \emph{regret}, and optimize through the augmented Lagrangian framework. Intuitively, regret captures the maximum utility gain a buyer could obtain by deviating from truthful reporting and/or recommended actions, assuming others remain truthful and obedient.

We define the expected {\em ex post} regret for a buyer $i$ as $\text{IC-RGT}_i^w = \mathbb{E}_{\rho \sim \mathcal{P}} \left[ \max_{\rho'_i \in \mathcal{T}_i} \mathrm{ic{\text -}rgt}^w_i(\rho'_i, \rho) \right]$ where $\mathrm{ic{\text -}rgt}^w_i(\rho'_i, \rho)$ is defined as:
\begin{equation}\label{eq::rgtep}
\begin{aligned}
    \mathrm{ic{\text -}rgt}^w_i(\rho'_i, \rho)
    &= v_i \cdot \Bigg(
        \sum_{k \in [m]} \max_{k' \in [m]}
            \left\{ \pi_{i,k',k}^w(\rho'_i, \rho_{-i}) \, \theta_{i,k'} \right\}
        - \bar{\alpha} \sum_{k \in [m]} \sum_{j \in [n]\setminus i}
            \pi_{j,k,k}^w(\rho'_i, \rho_{-i}) \, \theta_{i,k}
      \Bigg) \\[6pt]
    &\quad - \, t^w_i(\rho'_i, \rho_{-i}) \\[6pt]
    &\quad - \, v_i \cdot \Bigg(
        \sum_{k \in [m]} \pi_{i,k,k}^w(\rho) \, \theta_{i,k}
        - \bar{\alpha} \sum_{k \in [m]} \sum_{j \in [n]\setminus i}
            \pi_{j,k,k}^w(\rho) \, \theta_{i,k}
      \Bigg) \\[6pt]
    &\quad + \, t^w_i(\rho).
\end{aligned}
\end{equation}

\begin{restatable}{lemma}{icnet}
 \label{lem:ic-net}
     Any mechanism $\mathcal{M} = (\sigma^w, t^w)$ where $\sigma^w_i(\omega, \rho) = \left(\pi_{i, \omega}^w(\rho)\right)$ is ex post IC if and only if $\text{IC-RGT}_i^w = 0~\forall i \in [n]$, except for measure zero events.
 \end{restatable}

This leads to the following Lagrangian objective for the setting with ex post IC constraints:
\begin{equation}
    \mathcal{L}(w, \lambda, \zeta) = \mathbb{E}_{\rho \sim \mathcal{P}}\left[-\sum_{i=1}^n t^i_w (\rho)\right] + \sum_{i=1}^n\lambda^{\mathrm{rgt}}_i \cdot\text{IC-RGT}^i_w + \sum_{i=1}^n \zeta^{\mathrm{rgt}}_i \cdot\left(\text{IC-RGT}^w_i\right)^2
\end{equation}

We define the {\em interim} regret for a buyer $i$ as $\text{BIC-RGT}_i^w = \mathbb{E}_{\rho \sim \mathcal{P}} \left[ \max_{\rho'_i \in \mathcal{T}_i} \mathrm{bic{\text -}rgt}^w_i(\rho'_i, \rho_i) \right]$ where $\mathrm{bic{\text -}rgt}^w_i(\rho'_i, \rho_i)$ is defined as:
\begin{equation}\label{eq:bicreg}
\begin{aligned}
    \mathrm{bic{\text -}rgt}^w_i(\rho'_i, \rho_i)
    &= v_i \cdot \Bigg(
        \sum_{k \in [m]} \max_{k' \in [m]}
            \mathbb{E}_{-i} \Big[
                \pi_{i,k',k}^w(\rho'_i, \rho_{-i}) \, \theta_{i,k'}
                - \bar{\alpha} \sum_{j \in [n]\setminus i}
                    \pi_{j,k,k}^w(\rho'_i, \rho_{-i}) \, \theta_{i,k}
            \Big]
      \Bigg) \\[6pt]
    &\quad - \, t^w_i(\rho'_i) \\[6pt]
    &\quad - \, v_i \cdot \mathbb{E}_{-i} \Bigg[
        \sum_{k \in [m]} \pi_{i,k,k}^w(\rho_i, \rho_{-i}) \, \theta_{i,k}
        - \bar{\alpha} \sum_{k \in [m]} \sum_{j \in [n]\setminus i}
            \pi_{j,k,k}^w(\rho_i, \rho_{-i}) \, \theta_{i,k}
      \Bigg] \\[6pt]
    &\quad + \, t^w_i(\rho_i).
\end{aligned}
\end{equation}

\begin{restatable}{lemma}{bicnet}
 \label{lem:bic-net}
     Any mechanism $\mathcal{M} = (\sigma^w, t^w)$ where $\sigma^w_i(\omega, \rho) = \left(\pi_{i, \omega}^w(\rho)\right)$ is interim IC if and only if $\text{BIC-RGT}_i^w = 0~\forall i \in [n]$, except for measure zero events.
 \end{restatable}

This leads to the following Lagrangian objective for the setting with BIC constraints:
\begin{equation}
    \mathcal{L}(w, \lambda, \zeta) = \mathbb{E}_{\rho \sim \mathcal{P}}\left[-\sum_{i=1}^n t^w_i (\rho_i)\right] + \sum_{i=1}^n\lambda^{\mathrm{rgt}}_i \cdot\text{BIC-RGT}^w_i + \sum_{i=1}^n \zeta^{\mathrm{rgt}}_i \cdot\left(\text{BIC-RGT}^w_i\right)^2
\end{equation}

Although regret is a natural way to enforce BIC, in our setting it can be replaced with a simpler formulation in settings where beliefs are fixed. Specifically,~\citet{Bonatti2022} show that when types are independent and valuations are linear, BIC is equivalent to obedience and truthfulness under fixed beliefs. This allows us to replace regret constraints with simpler obedience and truthfulness violation measures while preserving incentive compatibility in these settings. In this sense, double deviations do not introduce additional constraints.

Because $v_i>0$, we measure normalized obedience violations by setting
$\rho_i'=\rho_i$ in Equation~\ref{eq:bicreg}, canceling the common
payment and externality terms, and dividing by $v_i$:
\begin{equation}\label{eq::bicob}
\mathrm{obv}_i^w(\rho_i)
=
\sum_{k\in[m]}
\left(
\max_{k'\in[m]}
\mathbb E_{-i}
\left[
\pi_{i,k',k}^w(\rho_i,\rho_{-i})\theta_{i,k'}
\right]
-
\mathbb E_{-i}
\left[
\pi_{i,k,k}^w(\rho_i,\rho_{-i})\theta_{i,k}
\right]
\right).
\end{equation}

To measure truthfulness violations, we consider misreports while assuming the buyer remains obedient. For convenience, define the interim downstream payoff of buyer $i$ as
\[
x_i(\rho_i) := \mathbb{E}_{-i}\!\left[
    \sum_{k \in [m]} \pi_{i,k,k}^w(\rho_i, \rho_{-i}) \,\theta_{i,k}
    - \bar{\alpha} \sum_{k \in [m]} \sum_{j \in [n]\setminus i} \pi_{j,k,k}^w(\rho_i, \rho_{-i})\,\theta_{i,k}
\right].
\]
Using this shorthand, the truthfulness violation can be written compactly as
\begin{equation}\label{eq::bictruth}
\mathrm{trv}^w_i(\rho_i)
= \max_{\rho'_i} \Big(v_i \, x_i(\rho'_i) - t^w_i(\rho'_i) \Big)
 - \Big(v_i \, x_i(\rho_i) - t^w_i(\rho_i)\Big).
\end{equation}

The corresponding augmented Lagrangian objective is then:
\begin{equation}
\begin{aligned}
    \mathcal{L}(w, \lambda, \zeta)
    &= \mathbb{E}_{\rho \sim \mathcal{P}} \Bigg[
        -\sum_{i=1}^n
            t^w_i (\rho_i) \Bigg]
            + \mathbb{E}_{\rho \sim \mathcal{P}} \Bigg[ \sum_{i=1}^n \left(\lambda^{\mathrm{obv}}_i \cdot \mathrm{obv}^w_i(\rho_i)
            + \lambda^{\mathrm{trv}}_i \cdot \mathrm{trv}^w_i(\rho_i) \right)\Bigg] \\[6pt]
    &\hspace{3cm}
            + \mathbb{E}_{\rho \sim \mathcal{P}} \Bigg[ \sum_{i=1}^n \left(\zeta^{\mathrm{obv}}_i \cdot \big(\mathrm{obv}^w_i(\rho_i)\big)^2
            + \zeta^{\mathrm{trv}}_i \cdot \big(\mathrm{trv}^w_i(\rho_i)\big)^2 \right)
    \Bigg].
\end{aligned}
\end{equation}

In practice, the expectations in our Lagrangian objective cannot be computed exactly. Instead, we approximate them using a finite set of training samples $\hat{\mathcal{P}} = \{\rho^{(1)}, \ldots, \rho^{(L)}\}$. In the BIC setting, additional expectations are required to compute interim values. For each agent $i$, we therefore draw a separate subset $\hat{\mathcal{P}}_{-i} = \{\rho_{-i}^{(1)}, \ldots, \rho_{-i}^{(K)}\}$. Given these empirical distributions, we evaluate both the loss (negated revenue) and the regret terms, and then optimize the resulting objective using the augmented Lagrangian method.

A key structural feature of the formulation is that the maximization over action deviations is handled analytically inside the regret definitions. With each buyer's action space finite and the utility linear in the induced experiment, the best deviation after receiving a recommendation can be expressed in closed form as a pointwise maximization over actions, as in Equations~\ref{eq::rgtep} and~\ref{eq:bicreg}. Thus, we do not numerically optimize over the full combinatorial space of deviation functions.

The remaining inner maximization is over continuous type
misreports and is generally non-concave. Following~\citet{dutting2023jacm}, we solve this approximately using gradient ascent in the report space. The network parameters are held fixed while we optimize over continuous type misreports, with the maximization over discrete action deviations handled analytically inside the regret expressions.~Although this search
yields an approximation to
regret, similar best-response procedures have been effective in prior
work and again perform well empirically in our setting, including closely
recovering known theoretical mechanisms.

To obtain a better numerical estimate of regret, we consider multiple candidate misreports and warm start gradient ascent from the best candidate. In the BIC setting, however, evaluating each candidate misreport requires computing its interim utility over multiple samples of the other buyers' types, making this search computationally expensive. To overcome this challenge, we exploit a simple but powerful observation: in the BIC setting, violations for buyer $i$ depend only on $\rho_i$, not the joint profile. This means we can reuse minibatch samples as candidate misreports.
If $\rho_i$ is the true report, then any other type $\rho'_i$ appearing in the same batch can be treated as a possible deviation. Since experiments and payments for these samples are already computed, this provides a pool of “free” candidate misreports without incurring additional forward passes. We then select the best-performing candidate as a warm start for the inner maximization problem, cutting the number of forward passes by a factor equal to the number of sampled misreports.

The reuse trick is only valid in the BIC setting. In the {\em ex post} IC case, regret depends on the full joint profile, so reusing minibatch samples would require finding two profiles $(\rho^p_i, \rho^p_{-i})$ and $(\rho^q_i, \rho^q_{-i})$ with identical opponents’ types ($\rho^p_{-i} = \rho^q_{-i}$). Such coincidences are exceedingly rare, giving minibatch reuse little power. At the same time, we do not need to compute {\em interim} values in the {\em ex post} IC setting, which means misreport sampling is already
computationally feasible there.

To rigorously ground the sample-based optimization, we adapt the statistical-learning analysis of \citet[Theorem~2.2 and Appendix~D.2]{dutting2023jacm}. Relative to the auction setting, the only substantive changes are that buyer-level regret now optimizes jointly over a type misreport and an action-deviation rule, and that utilities depend on other buyers' experiments through the negative-externality term. The proof template from \citet{dutting2023jacm} carries over once we establish a uniform Lipschitz bound for these data-market utilities in the mechanism metric.

Throughout this discussion, \(n\) denotes the number of buyers, \(m=|\Omega|\) the number of states (and, in the matching-payoff setting, actions), \(v_{\max}\) an upper bound on buyer values, and \(\alpha\) the negative-externality parameter from Section~\ref{sec:prelim}. We work on a compact reported type domain
\[
\mathcal{T}=\prod_{i=1}^n \mathcal{T}_i,
\qquad
\mathcal{T}_i\subseteq [0,v_{\max}]\times\triangle(\Omega),
\]
and all misreports are taken from the corresponding set \(\mathcal{T}_i\).
Let \(\mathcal{M}\) be a class of data-market mechanisms \((\sigma,t)\) on \(\mathcal{T}\). We measure the distance between two mechanisms using the \(\ell_{\infty,1}\) metric,
\[
d((\sigma,t),(\sigma',t'))
=
\max_{\rho\in\mathcal{T}}
\sum_{i\in[n]}
\left(
\sum_{j,k\in[m]}
|\pi_{i,j,k}(\rho)-\pi'_{i,j,k}(\rho)|
+
|t_i(\rho)-t_i'(\rho)|
\right).
\]
For any \(\epsilon>0\), let \(\mathcal{N}_\infty(\mathcal{M},\epsilon)\) denote the minimum number of balls of radius \(\epsilon\) needed to cover \(\mathcal{M}\) under this metric. In the BIC setting, we identify the interim payment with its lift to \(\mathcal{T}\), namely \(t_i(\rho):=t_i(\rho_i)\).

The following results bound the gaps between expected and
sample revenue and between expected and sample regret, uniformly over
the specified mechanism class. Regret in these results is defined
using exact global best responses and, in the BIC setting, exact
interim expectations. The second theorem bounds the generalization
term for our neural-network class and shows that it vanishes with
sample size.

{\color{black}
\begin{restatable}[Generalization Bounds]{theorem}{genbound}
\label{thm:gen_bounds}
Let \(\mathcal{M}\) be a class of data-market mechanisms \((\sigma,t)\) on \(\mathcal{T}\) such that, for every mechanism \((\sigma,t)\in\mathcal{M}\), every buyer \(i\), and every \(\rho\in\mathcal{T}\),
\[
|t_i(\rho)|\le v_{\max}(1+\alpha).
\]
Fix \(\delta\in(0,1)\), and let
\[
c:=\max\{1,\;4v_{\max}(1+\alpha)\}.
\]

With probability at least \(1-\delta\) over the draw of a sample
\[
\hat{\mathcal{P}}=\{\rho^{(1)},\ldots,\rho^{(L)}\}
\]
of \(L\) type profiles from \(\mathcal{P}\), the following hold simultaneously for every mechanism \((\sigma,t)\in\mathcal{M}\):

\begin{align*}
\mathbb{E}_{\rho\sim\mathcal{P}}
\!\left[\sum_{i=1}^n t_i(\rho)\right]
&\ge
\frac{1}{L}\sum_{\ell=1}^L\sum_{i=1}^n
t_i(\rho^{(\ell)})
-2n\Delta_L
-4nc\sqrt{\frac{2\log(8/\delta)}{L}},
\\
\frac{1}{n}\sum_{i=1}^n\mathrm{RGT}_i
&\le
\frac{1}{n}\sum_{i=1}^n\widehat{\mathrm{rgt}}_i
+2\Delta_L
+4c\sqrt{\frac{2\log(8/\delta)}{L}}.
\end{align*}

where $\mathrm{RGT}_i$ denotes
$\mathrm{IC\text{-}RGT}_i$ (expected {\em ex post} regret) in the
{\em ex post} IC setting and $\mathrm{BIC\text{-}RGT}_i$ in the BIC
setting, while $\widehat{\mathrm{rgt}}_i$ denotes the corresponding
sample regret. Moreover,

\[
\Delta_L
=
\inf_{\epsilon>0}
\left\{
\frac{\epsilon}{n}
+
2c\sqrt{
\frac{
2\log\!\left(
\mathcal{N}_\infty\!\left(\mathcal{M},\frac{\epsilon}{2nc}\right)
\right)
}{L}
}
\right\}.
\]
\end{restatable}

The proof, deferred to Appendix~\ref{app:generalization_bounds}, follows \citet[Appendix~D.2]{dutting2023jacm}. Step~1 and Step~3 of their regret argument are unchanged after enlarging the maximizing domain from misreports alone to misreport/deviation pairs; the only model-specific ingredient is a buyer-utility Lipschitz bound of order \(c\) in the mechanism metric.

\begin{restatable}[Network Capacity Bound]{theorem}{coverbound}
\label{thm:cover_bound}
Let \(\mathcal{M}_{\mathrm{NN}}\) be the class of data-market mechanisms induced by the bounded-activation counterpart of our neural-network architecture: the experiment network and normalized-payment network have \(R\) hidden layers, width at most \(K\), parameter counts \(d_\pi\) and \(d_t\), respectively, and the full parameter vector satisfies \(\|w\|_1\le W\) for some \(W\ge1\). Assume the hidden-layer activations are bounded and \(1\)-Lipschitz. If, in the BIC setting, separate normalized-payment networks are used for different buyers, let \(d_t\) denote the total number of trainable parameters across all such networks.

There exists a constant \(C>0\), depending only on universal constants, fixed architecture hyperparameters (including \(\tau_\pi\)), and on \((v_{\max},\alpha)\), such that for every \(\epsilon\in(0,1]\),
\[
\log \mathcal{N}_\infty(\mathcal{M}_{\mathrm{NN}},\epsilon)
\le
C(d_\pi+d_t)\Big(
(R+1)\log(2W)+\log\max\{K,nm^2\}+\log(1/\epsilon)
\Big).
\]

Then, for any sample size \(L\ge 1\), let \(\Delta_L\) denote the complexity term from Theorem~\ref{thm:gen_bounds} corresponding to \(\mathcal{M}=\mathcal{M}_{\mathrm{NN}}\) (or any subclass thereof). There exists a constant \(C_2>0\), depending only on universal constants, fixed architecture hyperparameters (including \(\tau_\pi\)), and on \((v_{\max},\alpha)\), such that
\[
\Delta_L
\le
\frac{1}{n\sqrt L}
+
C_2\sqrt{
\frac{
(d_\pi+d_t)\big((R+1)\log(2W)+\log(L\max\{K,nm^2\})\big)
}{L}
}.
\]
In particular, \(\Delta_L\to 0\) as \(L\to\infty\).
\end{restatable}
}
The proof adapts \citet[Lemma~D.2, Lemma~D.3, and Theorem~D.1]{dutting2023jacm}. We first cover the experiment network and the normalized-payment map separately, and then combine these covers using the fact that the actual payment is a normalized payment multiplied by an experiment-dependent scaling term whose absolute value is at most \(v_{\max}(1+\alpha)\). For the experiment network, the final normalization is the temperature-\(\tau_\pi\) softmax \(x\mapsto \mathrm{softmax}(x/\tau_\pi)\), whose \(\ell_1\)-Lipschitz constant is \(1/\tau_\pi\); because \(\tau_\pi\) is a fixed hyperparameter, this factor is absorbed into the constant in the covering-number bound. The same blockwise softmax calculation as in \citet{dutting2023jacm} then applies.

As in \citet{dutting2023jacm}, the resulting bound implies \(\Delta_L\to 0\) as \(L\to\infty\). We defer the full proofs to Appendix~\ref{app:generalization_bounds}.

These results address statistical generalization within the
specified mechanism class. They do not quantify the loss from
restricting attention to that class, optimization error from non-convex
training, or error in the numerical best-response search. In the BIC
implementation, Monte Carlo approximation of interim expectations is
an additional source of sampling error. The close
recovery of known theoretical benchmarks and the new
optimality result obtained from a learned design provide complementary validation of the framework beyond this statistical guarantee.

Theorem~\ref{thm:gen_bounds} gives a uniform bound over the mechanism
class with Theorem~\ref{thm:cover_bound} controlling the associated
complexity term.
For a single, fixed mechanism, selected independently of samples,
the expected {\em ex post} regret admits a simpler
bound without making use of the covering-number term. This is stated as Proposition~\ref{prop:heldout-regret}, again
under the assumption of exact global best responses.
\begin{proposition}[Hold-out Regret Evaluation]
\label{prop:heldout-regret}
Under the assumptions of Theorem~\ref{thm:gen_bounds}, let
$M\in\mathcal M$ be a possibly random mechanism selected independently
of an evaluation sample
$S=\{\rho^{(1)},\ldots,\rho^{(L)}\}$ drawn i.i.d.\ from
$\mathcal P$. Let $r_i^M(\rho)$ denote buyer $i$'s exact pointwise
ex post regret, and define
\[
\overline{\mathrm{RGT}}(M)
:=
\mathbb E_{\rho\sim\mathcal P}
\left[
\frac{1}{n}\sum_{i=1}^n r_i^M(\rho)
\right],
\qquad
\widehat{\overline{\mathrm{rgt}}}_S(M)
:=
\frac{1}{L}\sum_{\ell=1}^L
\frac{1}{n}\sum_{i=1}^n r_i^M(\rho^{(\ell)}).
\]
Then, for every $\delta\in(0,1)$, with probability at least
$1-\delta$ over $S$,
\[
\overline{\mathrm{RGT}}(M)
\le
\widehat{\overline{\mathrm{rgt}}}_S(M)
+
c\sqrt{\frac{\log(1/\delta)}{2L}},
\]
where $c=\max\{1,4v_{\max}(1+\alpha)\}$ as in
Theorem~\ref{thm:gen_bounds}.
\end{proposition}
\begin{proof}
{\bf Proof.} Conditional on $M$, let
\[
X_\ell
:=
\frac{1}{n}\sum_{i=1}^n r_i^M(\rho^{(\ell)}).
\]
The range calculation in the proof of
Theorem~\ref{thm:gen_bounds} gives $0\le X_\ell\le c$.
Conditional on $M$, the variables $X_1,\ldots,X_L$ are i.i.d.,
with mean $\overline{\mathrm{RGT}}(M)$ and sample mean
$\widehat{\overline{\mathrm{rgt}}}_S(M)$. Therefore, the one-sided
Hoeffding inequality gives
\[
\Pr\!\left(
\overline{\mathrm{RGT}}(M)
-
\widehat{\overline{\mathrm{rgt}}}_S(M)
>
\varepsilon
\,\middle|\, M
\right)
\le
\exp\!\left(-\frac{2L\varepsilon^2}{c^2}\right).
\]
Taking
$\varepsilon=c\sqrt{\log(1/\delta)/(2L)}$ proves the conditional
bound. Since $M$ is selected independently of $S$, integrating the
conditional probability over $M$ proves the stated unconditional
bound.
\end{proof}
\medskip

Later, in~Section~\ref{sec:expost62}, we give a numerical
illustration of Proposition~\ref{prop:heldout-regret}.

\section{Experimental Results for the Single Buyer Setting}

In this section, we demonstrate the use of RochetNet to recover known existing results from the economic theory literature. Further, we give representative examples of how we can characterize the informativeness of the menu options as we change different properties of the prior distribution. Additionally, we also show how we can use this approach to conjecture the optimal menu structure for settings outside the reach of the current theory. Further, this builds economic intuition in regard to the structure of optimal market designs.

For both the single buyer setting in this section, and the multi-buyer setting in Section~\ref{sec:multi-exp}, we report our results on a separate test set sampled from $\mathcal{P}$. All reported results are averaged across five independent training runs with different seeds. Each run took 11--12 minutes on a single NVIDIA Tesla V100 GPU.
We refer to Appendix~\ref{app:setup} for more details regarding the choice of hyperparameters.

\subsection{Buyers with world prior heterogeneity}

We first show that RochetNet can recover the optimal menu for all the continuous distribution settings considered in \citet{Berg2018} where the input distributions are a continuum of types. We specifically consider the following settings:
\begin{enumerate}[label=\Alph*.,ref=\Alph*]
    \item Single buyer, Binary State, and Binary Actions, where the payoffs $v = 1$ and the interim beliefs $\theta \sim U[0, 1]$.
    \label{SI}
    \item Single buyer, Binary State, and Binary Actions, where the payoffs $v = 1$ and the interim beliefs are drawn from an equal weight mixture of $\text{Beta}(8, 30)$ and $\text{Beta}(60, 30)$.
    \label{SII}
\end{enumerate}

\begin{table}[t]
\centering
\begin{tabular}{ccccc}
\midrule[\heavyrulewidth]
\multirow{2}{*}{Distribution} & \multicolumn{2}{c}{Rochet Menu} & Rochet & \multirow{2}{*}{Opt} \\ \cline{2-3}
                            & Experiment & Price & $\text{rev}$ & $\text{rev}$ \\ \midrule[\heavyrulewidth] \addlinespace[1.ex]
Setting \ref{SI} & $\begin{bmatrix} 1.00 & 0.00 \\ 0.00 & 1.00 \end{bmatrix}$ & 0.25 & 0.125 & 0.125 \\ \addlinespace[1.ex]\hline \addlinespace[1.ex]
\multirow{3}{*}{Setting \ref{SII}} & $\begin{bmatrix} 0.78 & 0.22 \\ 0 & 1.00 \end{bmatrix}$ & 0.14 & \multirow{3}{*}{0.167} & \multirow{3}{*}{0.167} \\ \addlinespace[1.ex]
                           & $\begin{bmatrix} 1.00 & 0.00 \\ 0.00 & 1.00 \end{bmatrix}$ & 0.26 & & \\\addlinespace[1.ex]
\midrule[\heavyrulewidth]
\end{tabular}\\
\caption{Menu(s) and associated prices learned by RochetNet, RochetNet revenue, and optimal revenue on test data in Settings~\ref{SI} and~\ref{SII}.\label{tab:rochet-exp} RochetNet recovers the optimal menus for both settings. The standard deviation was consistently below the rounding threshold (four significant digits) and is therefore not reported.}\label{tab:menus}
\end{table}

The optimal menus for each of these settings are given by \citet{Berg2018}. For the first setting, the optimal menu consists of a fully informative experiment with a price of $0.25$. For the second setting, the seller offers two menu options: a partially informative experiment and a fully informative experiment. In each case, RochetNet recovers the exact optimal menu. We describe the optimal menus, associated prices, revenue, and RochetNet revenue in Table~\ref{tab:menus}.

We also run experiments where we use RochetNet to study how the informativeness of learned experiments vary when we change different properties of the prior distributions. In the binary-state, binary-action setting we define the {\em informativeness} of an experiment $\pi$ as $I \triangleq \bigl|\,\pi_{1,1}-\pi_{2,1}\bigr|$. With this, $I=0$ corresponds to a completely uninformative experiment and $I=1$ to a fully informative one.

This differs from the informativeness measure Q in \citet{Berg2018}, where $Q = \pi_{1,1}$. This characterization applies only within the restricted class of concentrated, non-dispersed experiments. Since we do not impose such restrictions on the menu options learned by RochetNet, we adopt the more general metric $I$, which applies to any experiment. In the discrete version of these settings studied by \citet{Berg2018}, the low-type experiment is concentrated ($\pi_{2,1}=0$ and hence $\pi_{2,2}=1$) and the two measures of informativeness coincide.

We consider the following settings:

\begin{enumerate}[label=\Alph*.,start=3,ref=\Alph*]
    \item Single buyer, Binary State, and Binary Actions, where the payoffs $v_1= 1$ and the interim beliefs are drawn from a mixture of $\text{Beta}(8, 30)$ and $\text{Beta}(60, 30)$ weighted by $(1 - \gamma, \gamma)$ for $\gamma \in [0, 1]$.
    \label{SIII}
    \item Single buyer, Binary State, and Binary Actions, where the payoffs $v_1 = 1$ and the interim beliefs are drawn from an equal weight mixture of $\text{Beta}(a_1, b_1)$ and $\text{Beta}(60, 30)$. We vary $a_1, b_1$ so that $\text{Beta}(a_1, b_1)$ remains the low type and $\text{Beta}(60, 30)$ the high type, while the mode of $\text{Beta}(a_1, b_1)$ decreases (therefore the precision of the low type's prior belief $\lvert \theta^L_1 -0.5 \rvert$ increases). Simultaneously, we ensure that the probability density of values at the modes of these distributions stays constant while we alter $a_1, b_1$.
    \label{SIV}
\end{enumerate}

In particular, the modes of the two distributions are $\theta^L = \left(\frac{a_1-1}{a_1+b_1-2}, \frac{b_1-1}{a_1+b_1-2}\right)$ and $\theta^H = \left(\frac{59}{88}, \frac{29}{88}\right)$ in the case of Experiment~\ref{SIV}.\footnote{This also requires $|\frac{a_1-1}{a_1+b_1-2}-0.5|\geq |\frac{59}{88}-0.5|$, otherwise $\text{Beta}(a_1,b_1)$ becomes the high type instead. For our experiments, we only consider data points at which this condition holds. In other words, we only consider combinations of $a_1,b_1$ such that $\text{Beta}(a_1,b_1)$ is the low type with the mode at $\theta^L$ and $\text{Beta}(60,30)$ is the high type with the mode at $\theta^H$.}
$\theta^L$ and $\theta^H$ are non-congruent types, i.e., without the supplemental information, a buyer with type $\theta^L$ takes action $1$ while $\theta^H$ takes action $0$. Moreover, $\theta^H$ also values a fully informative experiment more than $\theta^L$. In the first experiment, we change the likelihood of type $\theta^H$ with respect to type $\theta^L$. In the second experiment, we change the precision of the belief of one type while keeping the other fixed (we vary values of $a_1, b_1$ while numerically ensuring that the probability distribution function of all the plotted points have the same height at the mode, so that the type distributions have comparable shape as we vary the low type's precision).
\begin{figure*}[t]
  \centering
  \includegraphics[width=0.4\textwidth]{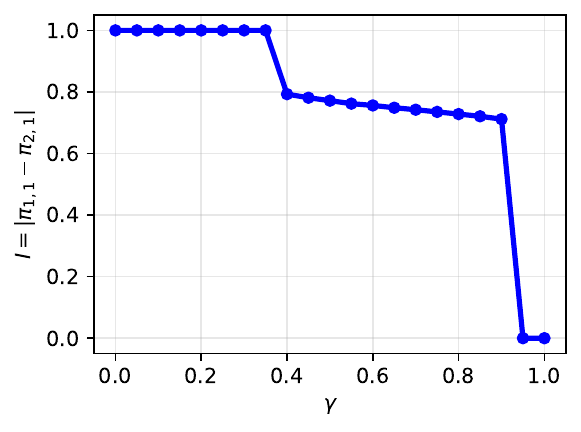}
  \includegraphics[width=0.4\textwidth]{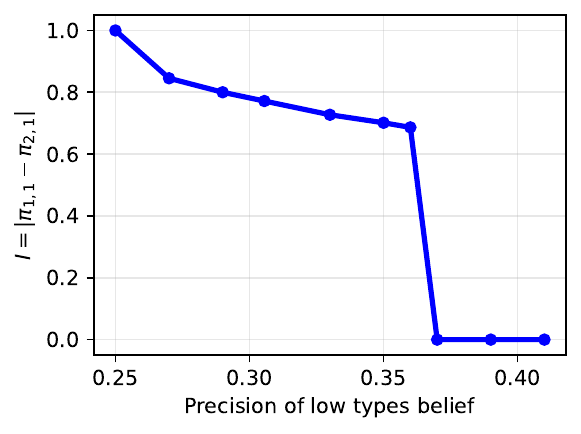}
  \caption{Results for Experiment~\ref{SIII} (left) and Experiment~\ref{SIV} (right). For both experiments, the vertical axis is the informativeness $I$ of the experiment used by $\theta^L$. For Experiment~\ref{SIII}, the horizontal axis is the frequency of the high type, and for Experiment~\ref{SIV}, the horizontal axis is the precision of the low type's prior belief $\lvert \theta^L_1 -0.5 \rvert$.}
  \label{fig:addrochetnet}
\end{figure*}

Figure~\ref{fig:addrochetnet} reports $I$ for the two settings. For Setting~\ref{SIII}, increasing the high type's frequency decreases the low type experiment's informativeness. For Setting~\ref{SIV}, increasing the precision of the low type's prior belief decreases the low type experiment's informativeness.

Proposition 3 in \citet{Berg2018} states that informativeness of the low type decreases with the frequency of the high type or with the precision of the low type's prior belief $|\theta^L_1 - 0.5|$. It is interesting, then, that we observe the same behavior in the RochetNet results for a continuous analog of their discrete distribution set-up. This suggests a new target for economic theory.\footnote{For additional validation, we also evaluate RochetNet on the corresponding discrete two-type benchmarks that underlie these comparative statics. The learned mechanisms exhibit behavior that is qualitatively consistent with existing theory~\citep[Proposition 3]{Berg2018}: informativeness of the low type experiment decreases with (i) increasing the frequency of the high type and (ii) increasing the precision of the low type’s prior. We defer these results to Appendix~\ref{app:discrete-two-type}.}

\subsection{Buyers with payoff and world prior heterogeneity}

We are aware of
no theoretical characterization of optimal data market designs
 when both $v$ and $\theta$ vary.
 In such cases, we can use RochetNet to conjecture the structure of an
optimal solution. For this, we consider the following settings with enlarged buyer types (in Setting~\ref{SVI}, the distribution of $v$ is parameterized by $c$):
\begin{enumerate}[label=\Alph*.,start=5,ref=\Alph*]
    \item Single buyer, Binary State, and Binary Actions, where $v \sim U[0,1 ]$ and the interim beliefs are drawn from $U[0, 1]$.
    \label{SV}
    \item Single buyer, Binary State, and Binary Actions, where the $v \sim U[c, 1]$ and the interim beliefs are drawn from an equal weight mixture of $\text{Beta}(8, 30)$ and $\text{Beta}(60, 30)$ for $c \in [0, 1]$.
    \label{SVI}
\end{enumerate}

\begin{figure}[t]
    \centering
    \includegraphics[scale=0.50]{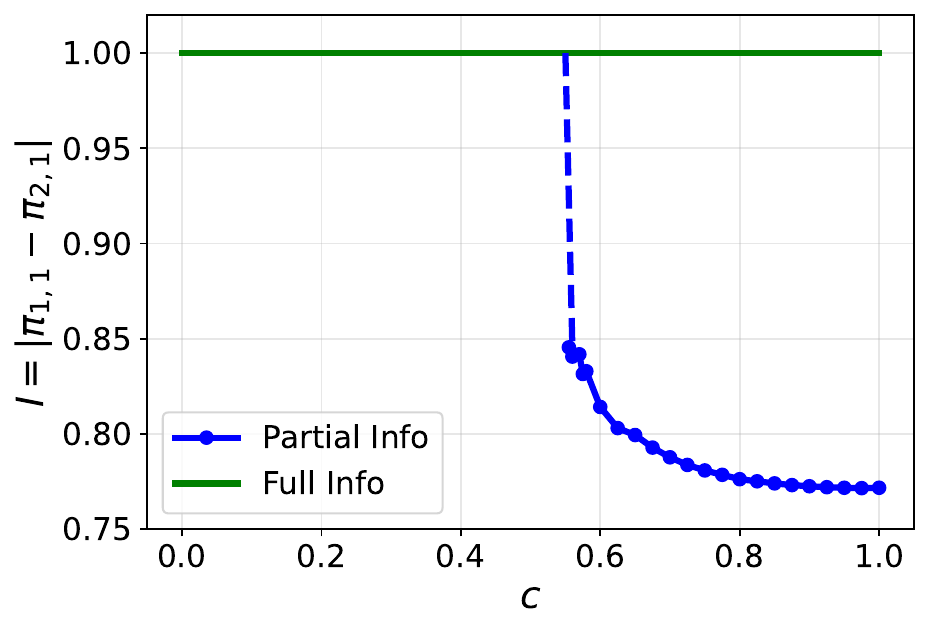}
    \caption{The informativeness $I$ of the menu options learned by RochetNet for Setting~\ref{SVI}, varying the parameter $c$ that instantiates the economic environment.}\label{fig:rochet-en}
  \end{figure}

For Setting~\ref{SV}, RochetNet learns a menu consisting of a
single fully informative experiment with a price of $0.14$. For Setting~\ref{SVI}, RochetNet learns a menu consisting of a single fully informative experiment when $c< 0.55$. For higher values of $c$, RochetNet learns an additional partially informative menu option. Further, in Figure~\ref{fig:rochet-en}, we show how the informativeness $I$ changes with $c$.

\section{Experimental Results for the Multi-Buyer Setting}
\label{sec:multi-exp}

In this section, we study three multi-buyer settings that illustrate different uses of RegretNet. We begin with the BIC setting, where we show that RegretNet recovers known theoretically optimal mechanisms. We then turn to an {\em ex post} IC setting for which no theoretical characterization is available. Here, we use the learned mechanisms to conjecture the structure of an optimal design, and then prove that the conjectured design is indeed optimal.
Lastly, we study a setting in which analytically characterizing the optimal mechanism is difficult, and use RegretNet to examine how revenue varies with the intensity of negative externalities. Together, these results illustrate how the framework can be used not only to recover existing theory, but also to suggest new theoretical results and to develop economic intuition in analytically challenging settings.

Throughout this section, reported regret values are numerical estimates obtained using the approximate best-response procedure described in Section~\ref{sec:multi}. These estimates should be interpreted as diagnostics of incentive alignment rather than certified upper bounds on exact regret.

\subsection{BIC settings}

\citet{Bonatti2022} study the multi-buyer market design problem with two buyers, binary actions, and fixed interim beliefs, the same for each agent. They show that a deterministic signaling scheme is optimal and thus characterize their solution in terms of recommending a correct action (selling a fully informative menu). Since the obedience constraints are defined on the interim representation, the optimal mechanism is also sometimes able to recommend incorrect actions to a buyer (in effect, sending recommendations that are opposed to the realized state). We consider the settings studied by~\citet{Bonatti2022}:
\begin{enumerate}[label=\Alph*.,start=7,ref=\Alph*]
    \item The payoffs $v_1, v_2$ are sampled from the exponential distribution with $\lambda = 1$. The common and fixed interim beliefs are given by either $\theta = (0.5, 0.5)$ or $\theta =(0.75, 0.25)$. $\alpha$ is either 0.5 or 2.0.
    \label{SVII}
\end{enumerate}
We also consider the following additional settings with common and fixed interim beliefs given by $ \theta = (0.5, 0.5)$ or $\theta = (0.75, 0.25)$ and when $\alpha$ is either 0.5 or 2.0.
\begin{enumerate}[label=\Alph*.,start=8,ref=\Alph*]
    \item The payoffs $v_1, v_2$ are sampled from a uniform distribution over the unit interval i.e. $U[0, 1]$.
    \label{SVIII}
    \item The payoffs $v_1, v_2$ are sampled uniformly from the intervals $U[0, 1]$ and $U[0, 2]$ respectively.
    \label{SIX}
\end{enumerate}

\begin{table}[t]
\centering
\small
\begin{tabular}{ccccccc}
\toprule
\multirow{2.5}{*}{Setting} & \multirow{2.5}{*}{$\alpha$} & \multirow{2.5}{*}{$\theta$} & \multicolumn{2}{c}{RegretNet} & Opt \\
\cmidrule(lr){4-5}
& & & {rev} & {rgt} & {rev} \\
\midrule
\multirow{4}{*}{~\ref{SVII}}
    & \multirow{2}{*}{0.5} & (0.50, 0.50) & 0.64 $\pm$ 0.006 & \multirow{4}{*}{$<0.001$} & 0.632 \\
    & & (0.75, 0.25) & 0.453 $\pm$ 0.004 & & 0.448 \\
    & \multirow{2}{*}{2.0} & (0.50, 0.50) & 1.624 $\pm$ 0.014 & & 1.613 \\
    & & (0.75, 0.25) & 1.417 $\pm$ 0.024 & & 1.415 \\
\midrule
\multirow{4}{*}{~\ref{SVIII}}
    & \multirow{2}{*}{0.5} & (0.50, 0.50) & 0.402 $\pm$ 0.004 & \multirow{4}{*}{$<0.001$} & 0.396 \\
    & & (0.75, 0.25) & 0.240 $\pm$ 0.007 & & 0.237 \\
    & \multirow{2}{*}{2.0} & (0.50, 0.50) & 1.052 $\pm$ 0.007 & & 1.042 \\
    & & (0.75, 0.25) & 0.747 $\pm$ 0.026 & & 0.750 \\
\midrule
\multirow{4}{*}{~\ref{SIX}}
    & \multirow{2}{*}{0.5} & (0.50, 0.50) & 0.617 $\pm$ 0.007 & \multirow{4}{*}{$<0.001$} & 0.609 \\
    & & (0.75, 0.25) & 0.373 $\pm$ 0.010 & & 0.369 \\
    & \multirow{2}{*}{2.0} & (0.50, 0.50) &1.607 $\pm$ 0.009 & & 1.594 \\
    & & (0.75, 0.25) & 1.149 $\pm$ 0.039& & 1.135 \\
\bottomrule
\end{tabular}
\caption{Test revenue and test regret obtained by RegretNet for the BIC settings. We report the mean and standard deviation across five independent runs.}
\label{tab:bic-rev-rgt}
\end{table}

\begin{figure}[htpb]
\centering

\begin{subfigure}{\linewidth}
  \centering
  \begin{subfigure}{.44\linewidth}
    \includegraphics[width=\linewidth, clip,trim=0 20 0 0]{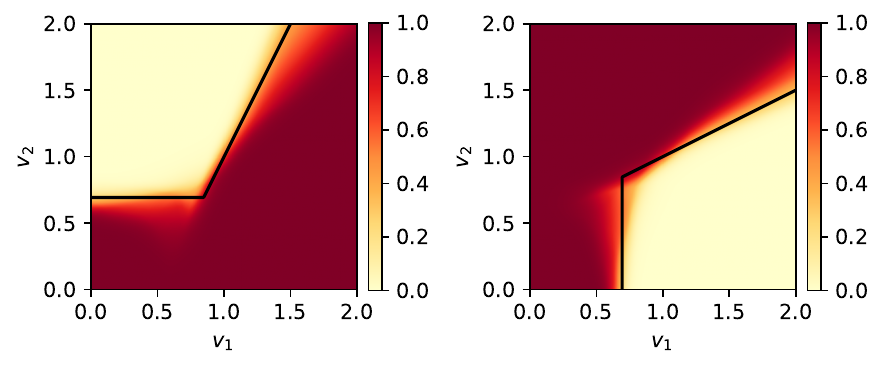}
    \centering \par \scriptsize {$\theta=0.50, \alpha=0.50$, MAE: 0.047}
  \end{subfigure}%
  \begin{subfigure}{.44\linewidth}
    \includegraphics[width=\linewidth, clip,trim=0 20 0 0]{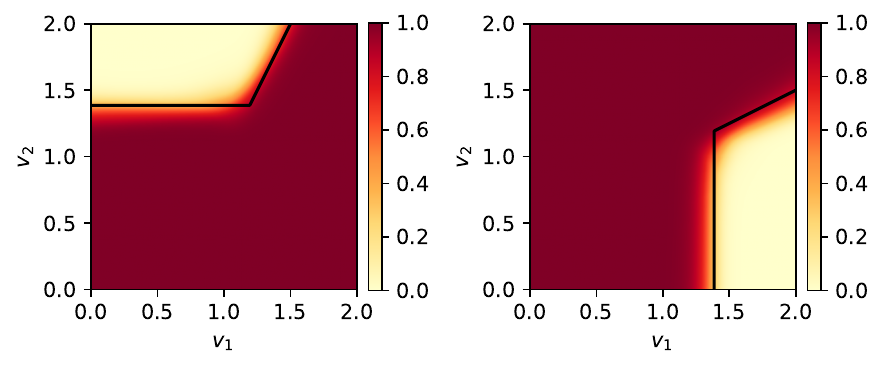}
    \centering \par \scriptsize {$\theta=0.75, \alpha=0.50$, MAE: 0.036}
  \end{subfigure}

  \begin{subfigure}{.44\linewidth}
    \includegraphics[width=\linewidth, clip,trim=0 20 0 0]{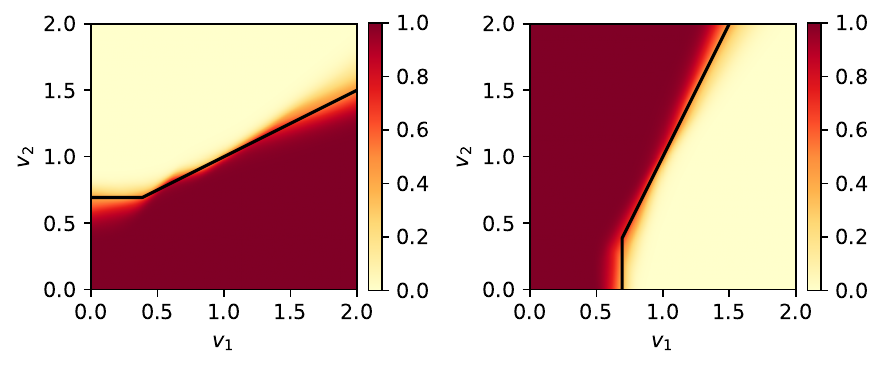}
    \centering \par \scriptsize {$\theta=0.50, \alpha=2.00$, MAE: 0.033}
  \end{subfigure}%
  \begin{subfigure}{.44\linewidth}
    \includegraphics[width=\linewidth, clip,trim=0 20 0 0]{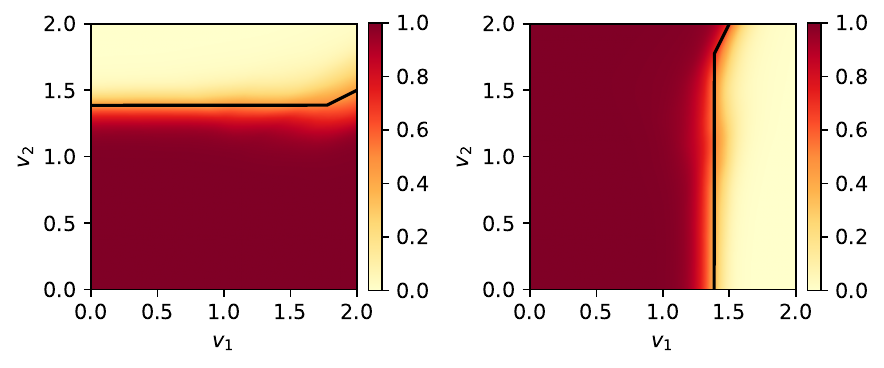}
    \centering \par \scriptsize {$\theta=0.75, \alpha=2.00$, MAE: 0.055}
  \end{subfigure}
  \caption[size=small]{Setting~\ref{SVII}: EXP}
  \label{fig:bic-exp}
\end{subfigure}

\begin{subfigure}{\linewidth}
  \centering
  \begin{subfigure}{.44\linewidth}
    \includegraphics[width=\linewidth, clip,trim=0 20 0 0]{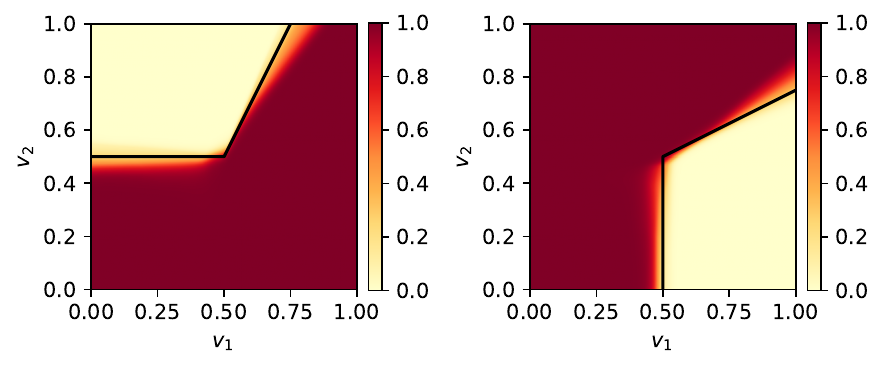}
    \centering \par \scriptsize {$\theta=0.50, \alpha=0.50$, MAE: 0.027}
  \end{subfigure}%
  \begin{subfigure}{.44\linewidth}
    \includegraphics[width=\linewidth, clip,trim=0 20 0 0]{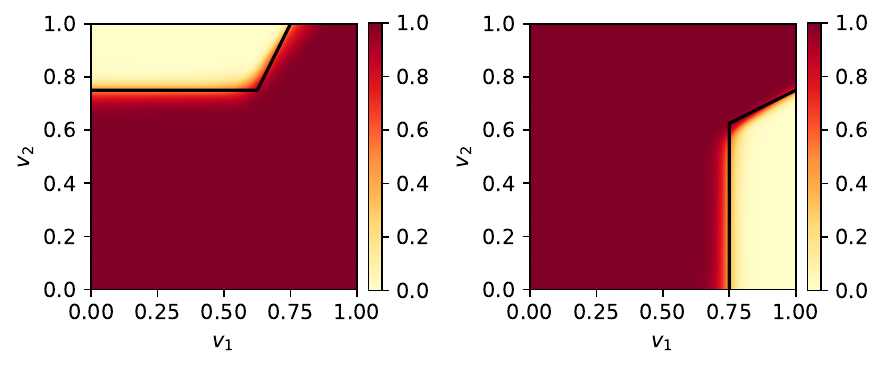}
    \centering \par \scriptsize {$\theta=0.75, \alpha=0.50$, MAE: 0.025}
  \end{subfigure}

  \begin{subfigure}{.44\linewidth}
    \includegraphics[width=\linewidth, clip,trim=0 20 0 0]{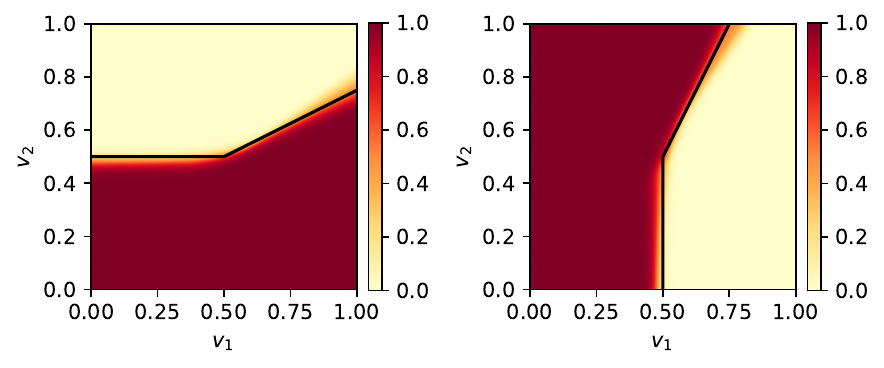}
    \centering \par \scriptsize {$\theta=0.50, \alpha=2.00$, MAE: 0.023}
  \end{subfigure}%
  \begin{subfigure}{.44\linewidth}
    \includegraphics[width=\linewidth, clip,trim=0 20 0 0]{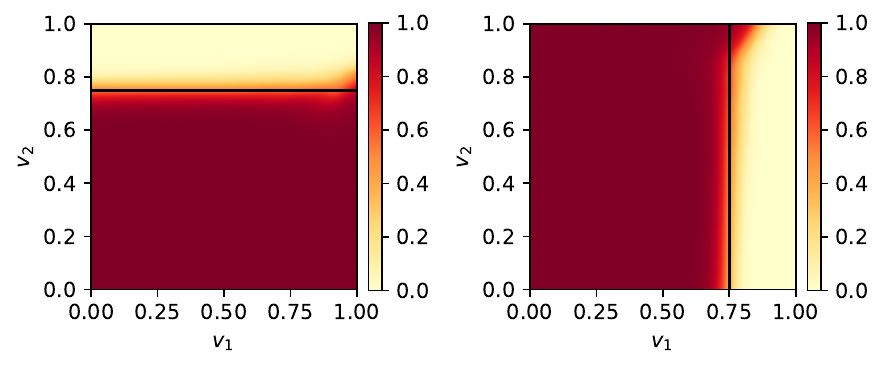}
    \centering \par \scriptsize {$\theta=0.75, \alpha=2.00$, MAE: 0.04}
  \end{subfigure}
  \caption{Setting~\ref{SVIII}: UNF}
  \label{fig:bic-unf}
\end{subfigure}

\begin{subfigure}{\linewidth}
  \centering
  \begin{subfigure}{.44\linewidth}
    \includegraphics[width=\linewidth, clip,trim=0 20 0 0]{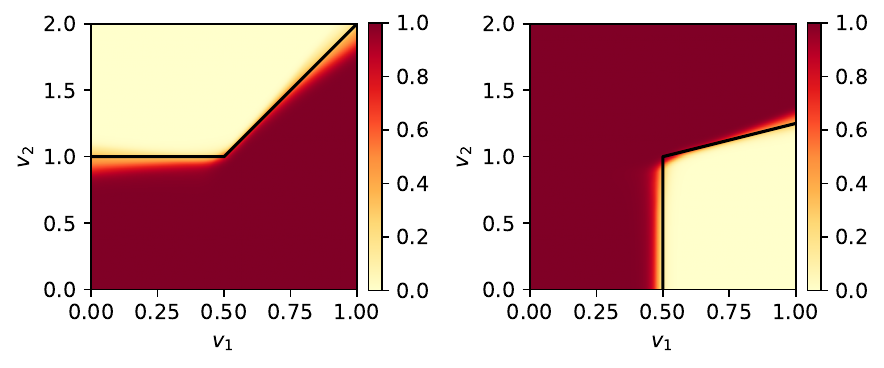}
    \centering \par \scriptsize $\theta=0.50, \alpha=0.50$ MAE: 0.024
  \end{subfigure}%
  \begin{subfigure}{.44\linewidth}
    \includegraphics[width=\linewidth, clip,trim=0 20 0 0]{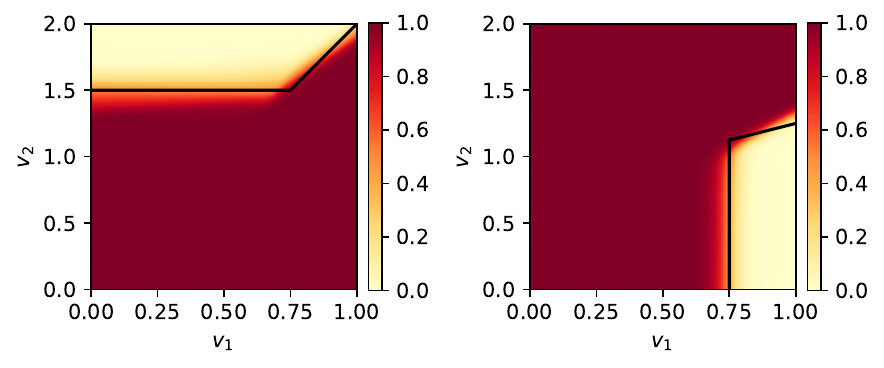}
    \centering \par \scriptsize{$\theta=0.75, \alpha=0.50$, MAE: 0.036}
  \end{subfigure}

  \begin{subfigure}{.44\linewidth}
    \includegraphics[width=\linewidth, clip,trim=0 20 0 0]{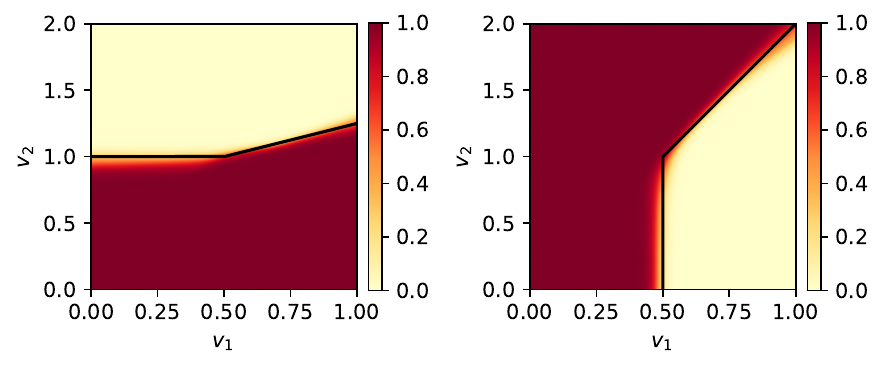}
    \centering \par \scriptsize {$\theta=0.50, \alpha=2.00$, MAE: 0.033}
  \end{subfigure}%
  \begin{subfigure}{.44\linewidth}
    \includegraphics[width=\linewidth, clip,trim=0 20 0 0]{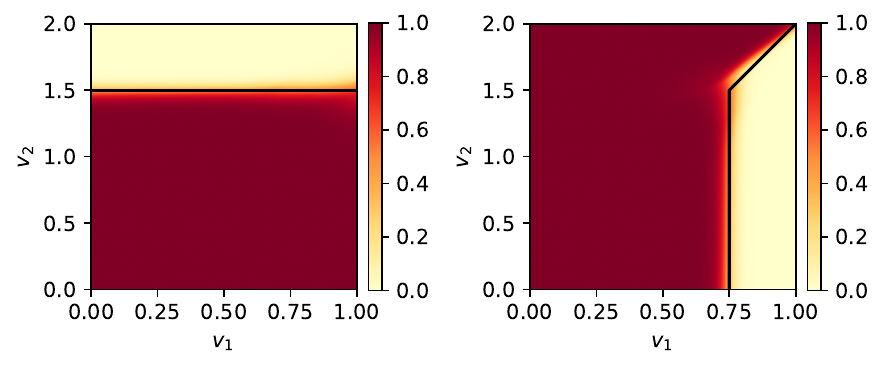}
    \centering \par \scriptsize {$\theta=0.75, \alpha=2.00$, MAE: 0.055}
  \end{subfigure}
  \caption{Setting~\ref{SIX}: Asym UNF}
  \label{fig:bic-asym_unf}
\end{subfigure}

\caption{\label{fig:multi-bic} Experiments learned for Settings~\ref{SVII} (\textbf{top}), ~\ref{SVIII} (\textbf{middle}), and \ref{SIX} (\textbf{bottom}) for BIC constraints. The plot shows the probability of recommending the correct action to buyer $1$ (left subplot) and buyer $2$ (right subplot) for varying values of $v_1$ (x-axis) and $v_2$ (y-axis). For each setting, the theoretically optimal experiments are represented by solid lines separating the regions. We include in the captions the Mean Absolute Error (MAE) between the optimal recommendation and learned recommendation.
 \label{fig:multi-plot}}
\end{figure}

The test revenue and regret for these settings are given in Table~\ref{tab:bic-rev-rgt} with the designs are shown in Figure~\ref{fig:multi-plot}. The theoretically optimal design contains two kinds of recommendations, those recommending the correct action (the optimal action to take given the realized state) and those recommending incorrect actions, separated by the black solid lines. The heatmaps depict our computational results, which show the probability of recommending the correct action across the $(v_1, v_2)$ grid. To quantify the closeness of the learned design to the theoretical optimum, we compute the {\em mean absolute error (MAE)} between the optimal rule and the learned probabilities at each grid point, and report these values in Figure~\ref{fig:multi-plot}. Both visually and in terms of expected revenue, the results confirm that our approach successfully recovers the optimal revenue and closely matches the structure of the optimal experiment design.

\paragraph{Studying the effect of varying the intensity of externalities.}

Following~\citet{Bonatti2022}, we
can study the effect of the negative externality parameter $\alpha$ on the revenue for the BIC, multi-buyer setting with a common and known prior for each buyer and payoffs sampled from $U[0, 1]$.
 We also consider the case where the payoffs are constant, but the interim beliefs are independently sampled from $U[0, 1]$. While there is no known analytical characterization for the latter, we show how easy it is to study the models learned by RegretNet to analyze economic properties of interest.
\begin{figure}[t]
    \centering
        \includegraphics[width=0.35\textwidth]{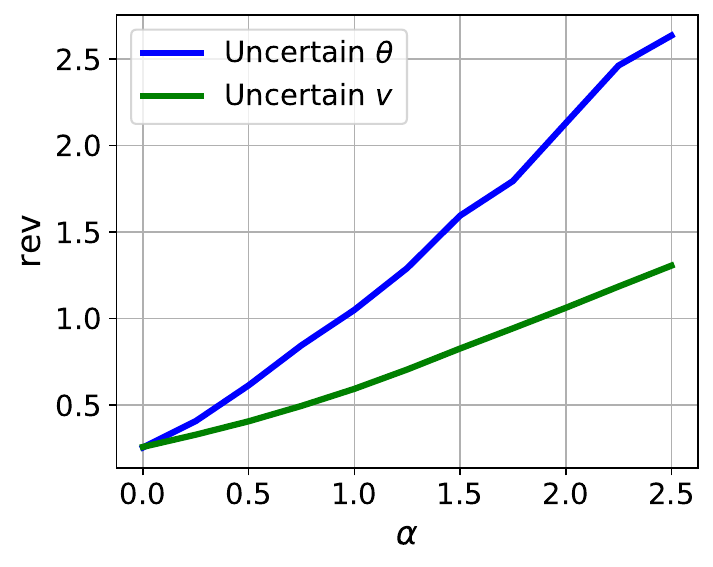}
    \caption{Effect of $\alpha$ on revenue.}
    \label{fig:rev_v_alpha}
\end{figure}
Figure~\ref{fig:rev_v_alpha} shows the effect of $\alpha$ on the revenue.
In the context of uncertain priors, we note that the effect on revenue is similar to the setting where payoffs are uncertain. As we enhance competition intensity via $\alpha$, the revenue grows. Even though the buyers are not recommended the correct action, the seller manages to generate revenue by threatening to share exclusive information with a competitor. This effect becomes stronger in the settings with uncertain priors.

\subsection{\emph{Ex post} IC settings}

\begin{figure}[htpb]
\centering

\begin{subfigure}{\linewidth}
  \centering
  \begin{subfigure}{.44\linewidth}
    \includegraphics[width=\linewidth, clip,trim=0 20 0 0]{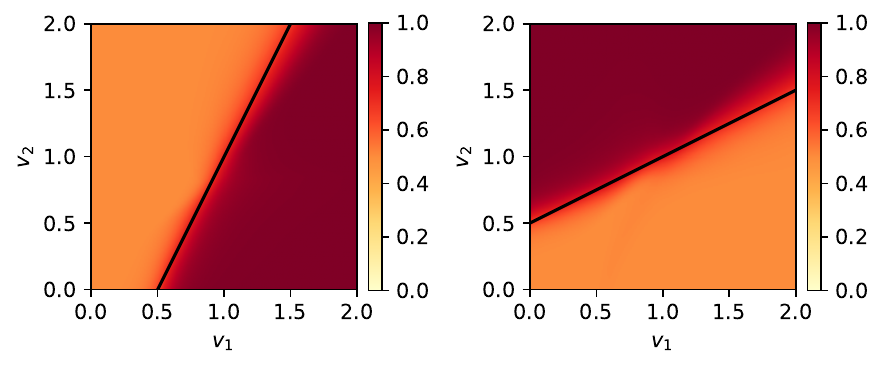}
    \centering \par \scriptsize {$\theta=0.50, \alpha=0.50$, MAE: 0.027}
  \end{subfigure}%
  \begin{subfigure}{.44\linewidth}
    \includegraphics[width=\linewidth, clip,trim=0 20 0 0]{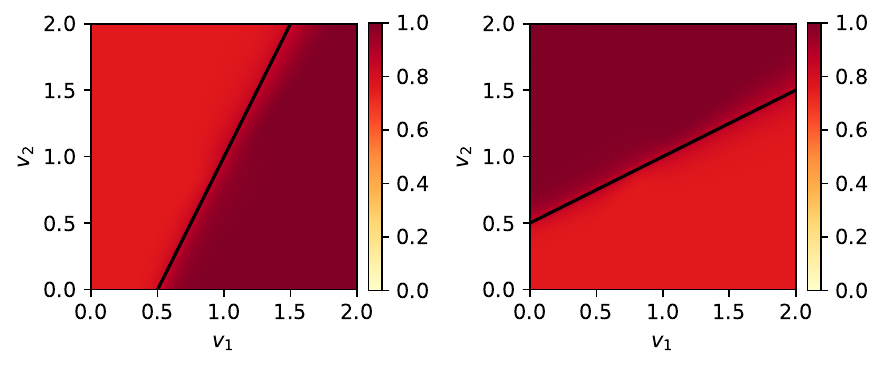}
    \centering \par \scriptsize {$\theta=0.75, \alpha=0.50$, MAE: 0.013}
  \end{subfigure}

  \begin{subfigure}{.44\linewidth}
    \includegraphics[width=\linewidth, clip,trim=0 20 0 0]{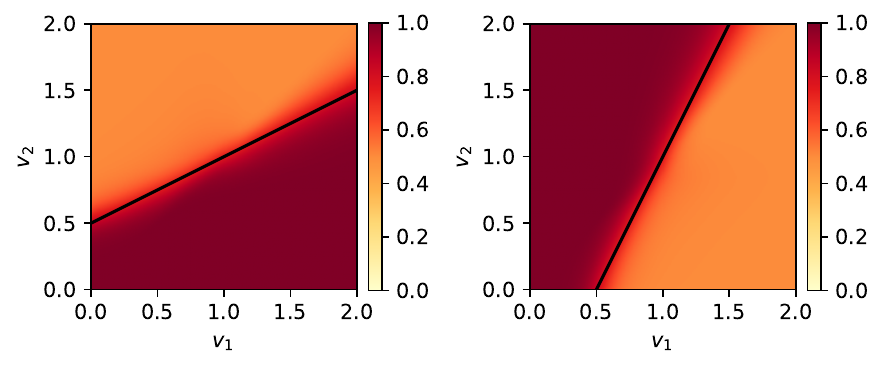}
    \centering \par \scriptsize {$\theta=0.50, \alpha=2.00$, MAE: 0.028}
  \end{subfigure}%
  \begin{subfigure}{.44\linewidth}
    \includegraphics[width=\linewidth, clip,trim=0 20 0 0]{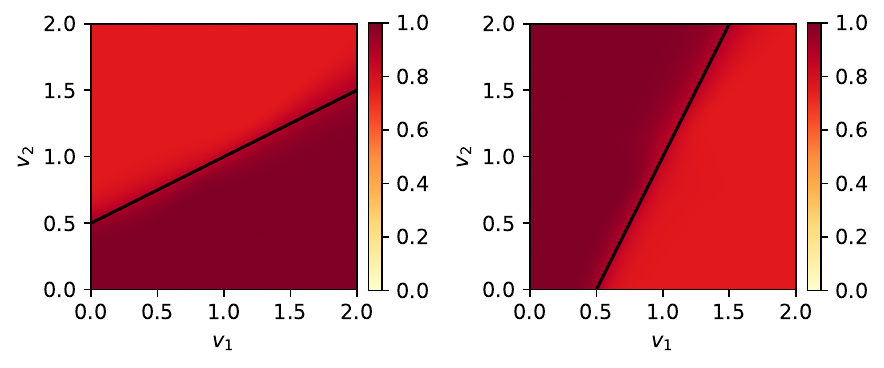}
    \centering \par \scriptsize {$\theta=0.75, \alpha=2.00$, MAE: 0.014}
  \end{subfigure}
  \caption{Setting~\ref{SVII}: EXP}
\end{subfigure}

\begin{subfigure}{\linewidth}
  \centering
  \begin{subfigure}{.44\linewidth}
    \includegraphics[width=\linewidth, clip,trim=0 20 0 0]{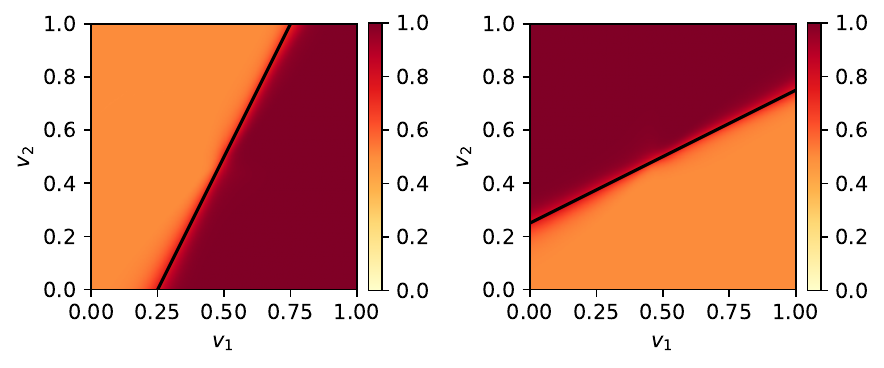}
    \centering \par \scriptsize {$\theta=0.50, \alpha=0.50$, MAE: 0.015}
  \end{subfigure}%
  \begin{subfigure}{.44\linewidth}
    \includegraphics[width=\linewidth, clip,trim=0 20 0 0]{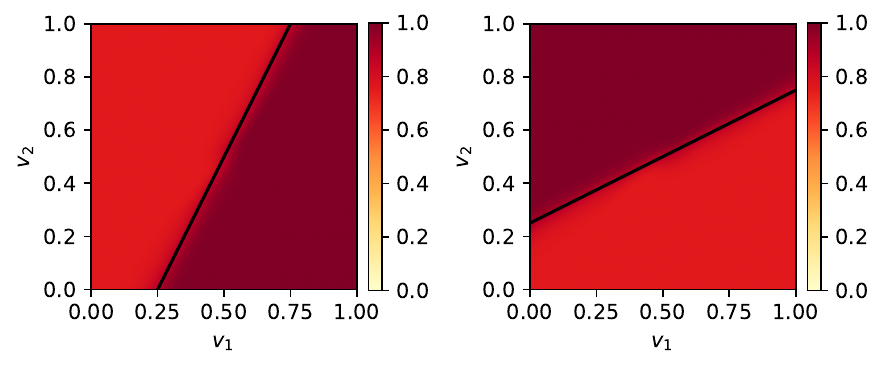}
    \centering \par \scriptsize {$\theta=0.75, \alpha=0.50$, MAE: 0.009}
  \end{subfigure}

  \begin{subfigure}{.44\linewidth}
    \includegraphics[width=\linewidth, clip,trim=0 20 0 0]{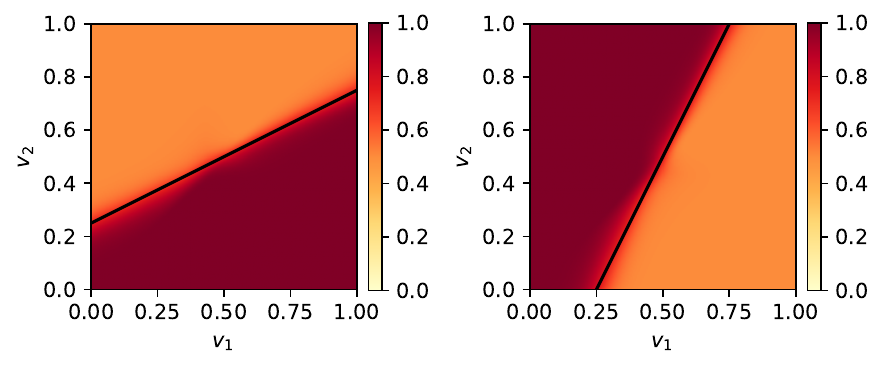}
    \centering \par \scriptsize {$\theta=0.50, \alpha=2.00$, MAE: 0.017}
  \end{subfigure}%
  \begin{subfigure}{.44\linewidth}
    \includegraphics[width=\linewidth, clip,trim=0 20 0 0]{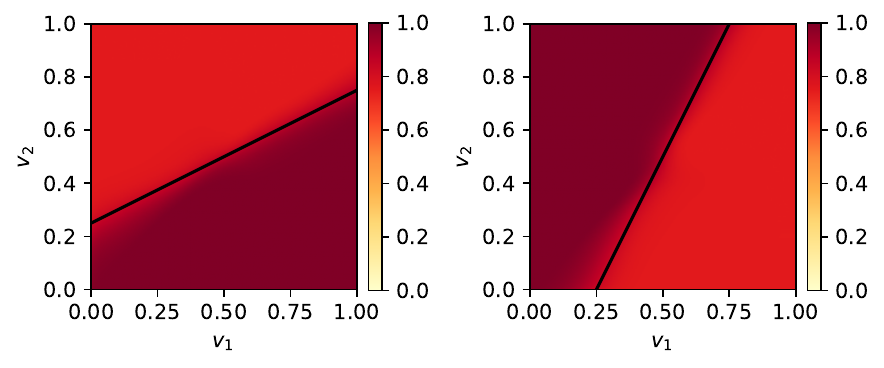}
    \centering \par \scriptsize {$\theta=0.75, \alpha=2.00$, MAE: 0.011}
  \end{subfigure}
  \caption{Setting~\ref{SVIII}: UNF}
\end{subfigure}

\begin{subfigure}{\linewidth}
  \centering
  \begin{subfigure}{.44\linewidth}
    \includegraphics[width=\linewidth, clip,trim=0 20 0 0]{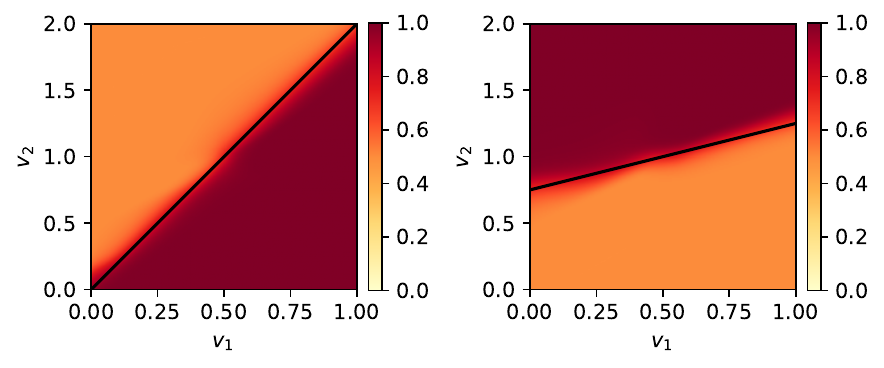}
    \centering \par \scriptsize {$\theta=0.50, \alpha=0.50$ MAE: 0.019}
  \end{subfigure}%
  \begin{subfigure}{.44\linewidth}
    \includegraphics[width=\linewidth, clip,trim=0 20 0 0]{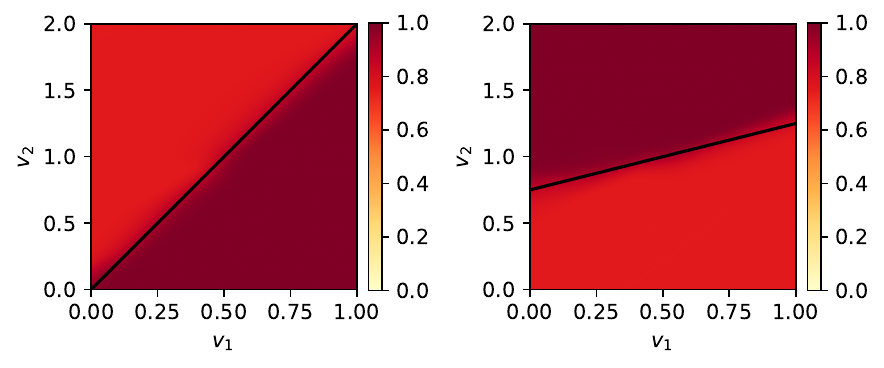}
    \centering \par \scriptsize{$\theta=0.75, \alpha=0.50$, MAE: 0.010}
  \end{subfigure}

  \begin{subfigure}{.44\linewidth}
    \includegraphics[width=\linewidth, clip,trim=0 20 0 0]{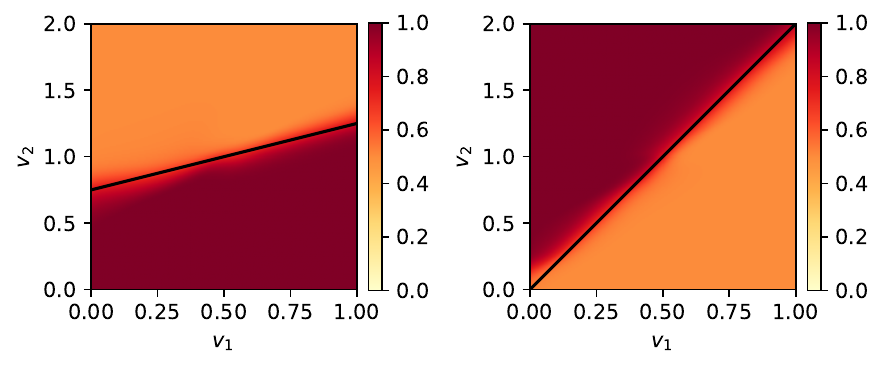}
    \centering \par \scriptsize {$\theta=0.50, \alpha=2.00$, MAE: 0.020}
  \end{subfigure}%
  \begin{subfigure}{.44\linewidth}
    \includegraphics[width=\linewidth, clip,trim=0 20 0 0]{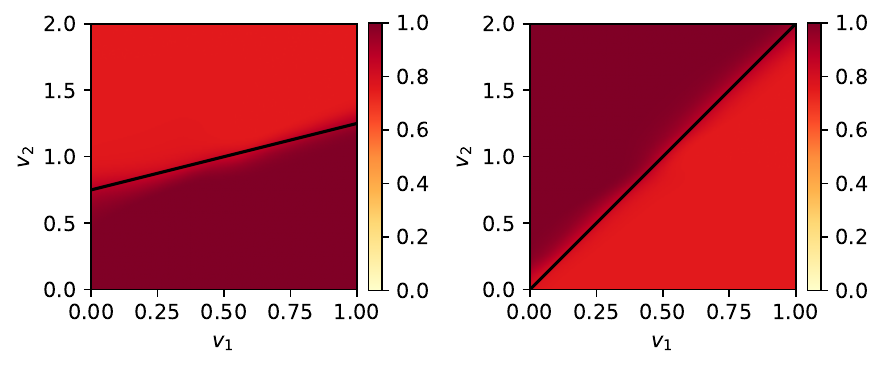}
    \centering \par \scriptsize {$\theta=0.75, \alpha=2.00$, MAE: 0.010}
  \end{subfigure}
  \caption{Setting~\ref{SIX}: Asym UNF}
\end{subfigure}

\caption{\label{fig:multi-ic} Experiments learned for Settings~\ref{SVII} (\textbf{top}), ~\ref{SVIII} (\textbf{middle}), and \ref{SIX} (\textbf{bottom}) for {\em ex post} IC constraints. The plot shows the probability of recommending the correct action to buyer $1$ (left subplot) and buyer $2$ (right subplot) for varying values of $v_1$ (x-axis) and $v_2$ (y-axis). For each setting, the theoretically optimal experiments are represented by solid lines separating the regions. We include in the captions the Mean Absolute Error (MAE) between the optimal recommendation and learned recommendation.}
\end{figure}

\begin{table}[t]
\centering
\small
\begin{tabular}{ccccccc}
\toprule
\multirow{2.5}{*}{Setting} & \multirow{2.5}{*}{$\alpha$} & \multirow{2.5}{*}{$\theta$} & \multicolumn{2}{c}{RegretNet} & Opt \\
\cmidrule(lr){4-5}
& & & {rev} & {rgt} & {rev} \\
\midrule
\multirow{4}{*}{~\ref{SVII}}
    & \multirow{2}{*}{0.5} & (0.50, 0.50) & 0.423 $\pm$ 0.002 & \multirow{4}{*}{$<0.001$} & 0.404 \\
    & & (0.75, 0.25) & 0.213 $\pm$ 0.001 & & 0.202 \\
    & \multirow{2}{*}{2.0} & (0.50, 0.50) & 0.831 $\pm$ 0.003 & & 0.809 \\
    & & (0.75, 0.25) & 0.422 $\pm$ 0.003 & & 0.404 \\
\midrule
\multirow{4}{*}{~\ref{SVIII}}
    & \multirow{2}{*}{0.5} & (0.50, 0.50) & 0.279 $\pm$ 0.000 & \multirow{4}{*}{$<0.001$} & 0.271 \\
    & & (0.75, 0.25) & 0.140 $\pm$ 0.001 & & 0.135 \\
    & \multirow{2}{*}{2.0} & (0.50, 0.50) & 0.553 $\pm$ 0.001 & & 0.542 \\
    & & (0.75, 0.25) & 0.251 $\pm$ 0.062 & & 0.271 \\
\midrule
\multirow{4}{*}{~\ref{SIX}}
    & \multirow{2}{*}{0.5} & (0.50, 0.50) & 0.434 $\pm$ 0.003 & \multirow{4}{*}{$<0.001$} & 0.422 \\
    & & (0.75, 0.25) & 0.219 $\pm$ 0.002 & & 0.211 \\
    & \multirow{2}{*}{2.0} & (0.50, 0.50) &0.865 $\pm$ 0.003 & & 0.845 \\
    & & (0.75, 0.25) & 0.383 $\pm$ 0.117& & 0.422 \\
\bottomrule
\end{tabular}
\caption{Test revenue and expected {\em ex post} regret obtained by RegretNet for Settings~\ref{SVII},~\ref{SVIII}, and~\ref{SIX} under the {\em ex post} IC constraints~\label{tab:ic-rev-rgt}. We report the mean and standard deviation across five independent runs.}
\end{table}

\label{sec:expost62}
We also apply RegretNet to the same settings (\ref{SVII},~\ref{SVIII} and ~\ref{SIX}), but while adopting {\em ex post} IC constraints. This is of interest because there is no known analytical solution for the optimal mechanism for two buyers in the {\em ex post} IC setting.{\footnote{We also provide an illustration here of the analysis in Proposition~\ref{prop:heldout-regret}, applied to an upper confidence bound on regret for some fixed mechanism in the {\em ex post} IC context of Setting~\ref{SVIII}. We have $v_{\max}=1$, and let $\delta=0.05$. With test-set size $L=2^{20}$, as adopted in our experiments, the sampling term in Proposition~\ref{prop:heldout-regret} is $~\approx 0.0072$ when $\alpha=0.5$ and $\approx 0.0143$ when $\alpha=2$. Thus, if the exact sample regret is $0.001$, the corresponding upper confidence bound on expected regret would be $\approx 0.0082$ and $\approx 0.0154$, respectively.%
}}

We visualize the learned mechanisms in Figure~\ref{fig:multi-ic}. From these computational results, we conjecture the structure of the optimal mechanism in the {\em ex post} IC setting. To assess the accuracy of the learned designs, we evaluate revenue and
expected {\em ex post} regret on the test set and present the results
in Table~\ref{tab:ic-rev-rgt}. We also compute mean absolute error (MAE) between the recommendation probabilities produced by RegretNet and the recommendation rule implied by our conjecture and report in the figure captions.

Finally, building on this evidence, we prove the conjecture by adapting Myerson’s framework for single-item auctions. Theorem~\ref{thm:icproof} states the result, with the proof deferred to Appendix~\ref{app:ic-proof}.

\begin{restatable}{theorem}{icproof}
\label{thm:icproof}
Consider the binary-state, binary-action setting in which all buyers share a common interim belief $\theta_i = \theta$. Each buyer $i$ has a value $v_i$ drawn independently from a regular distribution $\mathcal{V}_i$ with continuous density $f_i$ and cumulative distribution $F_i$. Define the virtual value
\[
\phi_i(v_i) = v_i - \frac{1 - F_i(v_i)}{f_i(v_i)}.
\]
Then, the revenue-optimal mechanism satisfying \emph{ex post} incentive compatibility and individual rationality allocates the fully informative experiment to buyer $i$ if
\[
\phi_i(v_i) \geq \frac{\alpha}{n-1}\sum_{j \neq i}\phi_j(v_j).
\]
Otherwise, buyer $i$ receives an uninformative experiment that signals the most likely state under the prior.
\end{restatable}

\subsection{Irregular Distributions}

We present additional results for both BIC and {\em ex post} IC settings for two buyers, binary actions, and interim beliefs $\theta = (0.5, 0.5)$, the same for each buyer. We set $\alpha = 0.5$, and the payoffs are drawn from the irregular distribution whose pdf $f(v)$
is:
\[
f(v) =
\begin{cases}
    2.5, & \text{if } 0 \leq v < 0.3 \\
    0.5, & \text{if } 0.3 \leq v < 0.8
\end{cases}
\]

\begin{figure*}[htpb]
\centering
\begin{subfigure}{.44\linewidth}
     \includegraphics[width=\linewidth, clip,trim=0 20 0 0]{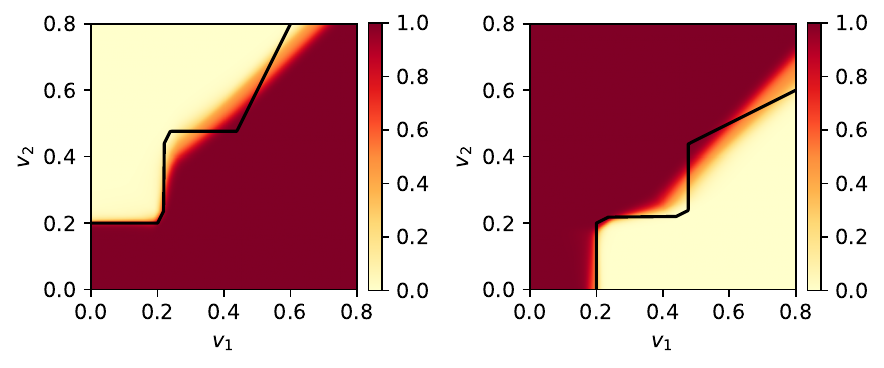}
    \centering \par \scriptsize {$\theta=0.50, \alpha=0.50$, MAE: 0.045}
\end{subfigure}
\begin{subfigure}{.44\linewidth}
     \includegraphics[width=\linewidth, clip,trim=0 20 0 0]{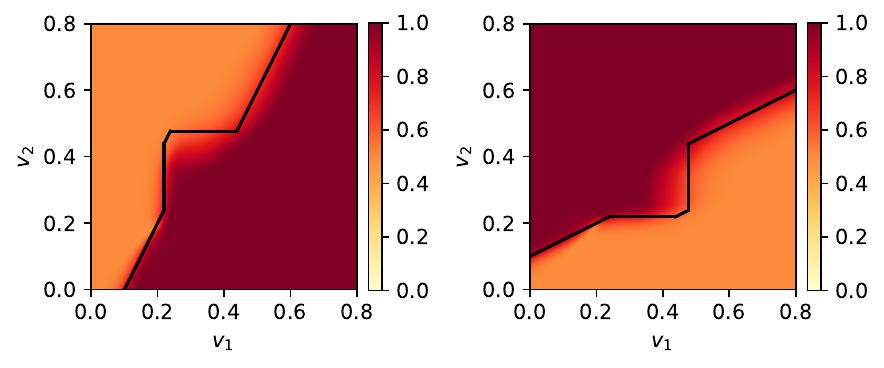}
    \centering \par \scriptsize {$\theta=0.50, \alpha=0.50$, MAE: 0.022}
\end{subfigure}
\caption{Experiments learned for BIC and {\em ex post} IC constraints when the payoffs are drawn from an irregular distribution. For both these settings, we use the ironed virtual value functions to compute the solid line.\label{fig:irr-all}}
\end{figure*}

The solutions for these problems make use of ironed virtual values as the payoff distribution is irregular (see Figure~\ref{fig:irr-all}).

\section{Conclusion}
\label{sec:limit}

We have introduced a new deep neural network architecture and learning framework to study the design of optimal data markets. We have demonstrated through experimental work the flexibility of the framework, showing that it can adapt to relatively complex scenarios and facilitate the discovery of new designs that can be proved to be optimal. Note that while we have only considered matching utility payoffs in this paper, our approach can be extended to non-matching utility payoffs as well.

We also point out some limitations. First, for our approach to continue to provide insights into the theoretically optimal design for larger problems (e.g., with more buyers, more signals, more actions), it will be important to provide interpretability to the mechanisms learned by RegretNet (designs learned by RochetNet, on the other hand, are immediately interpretable).
Second, while our approach scales with number of buyers, or states in the {\em ex post IC} setting, it does not scale as easily with the number of buyers in the BIC setting. The challenge in the BIC setting comes from the ``interim computations" involving conditional expectations over reports of others and scaling beyond what is studied in this paper will require new techniques, for example, exploiting symmetry.
Third, we are making use of gradient-based approaches, which may suffer from local optima in non-convex problems. At the same time, deep learning has shown success in various problem domains despite non-convexity. The experiments reported here align with these observations, with our neural network architectures consistently recovering optimal solutions, when these are known and thus optimality can be verified, and despite non-convex formulations.

Another limitation of our approach concerns incentive alignment. In the multi-buyer setting, our mechanisms are approximately incentive compatible. This raises the question as to how much approximate incentive alignment is sufficient in practice, and whether a low-regret solution is close in design to an exactly incentive-compatible mechanism.
While there is some recent guiding theory~\citep{Conitzer22, daskalakis2012symmetries, COVZ2021} that provides transformations between $\epsilon$-BIC and BIC while controlling revenue loss in the context of auction design, extending these transformations to approximate DSIC settings and problems with both types and actions presents an interesting avenue for future research.

Lastly, we return to where we started, and underline that markets for trading data about individuals raise substantive ethical concerns~\citep{acemoglu2022too, acquisti2016economics, athey2017digital, socialdata, bonatti2022platform, choi2019privacy, fainmesser2022digital}. Our hope is that machine learning frameworks such as those introduced here can be used to strike new kinds of trade-offs, for example allowing individuals to also benefit directly from trades on data about themselves and to embrace privacy and fairness constraints.

\section*{Acknowledgments}
We would like to express our sincere gratitude to Zhe Feng and Alessandro Bonatti for the initial discussions and the anonymous reviewers on earlier versions of this paper for their helpful feedback. The source code for all experiments along with the instructions to run it are available from GitHub at \url{https://github.com/saisrivatsan/deep-data-markets/}

\bibliographystyle{abbrvnat}
\bibliography{references}

\appendix

\section{Omitted Proofs}

In this section, we prove Lemmas~\ref{lem:ir-net}-\ref{lem:bic-net} and Theorem~\ref{thm:icproof}.

\subsection{Proof of Lemma~\ref{lem:ir-net}}

\irnet*
\begin{proof}
For the {\em ex post} IC setting, we first compute a {\em normalized payment},
 $\tilde{t}_i^w(\rho) \in [0, 1]$, for each buyer $i$ by using a sigmoidal unit. We scale this by the difference between utility achieved by the buyer when he is truthful and obedient and the utility of opting out. Thus, the payment $t_i^w(\rho)$ is computed as follows:
 \begin{align}
     t_i^w(\rho) = \tilde{t}_i^w(\rho) \cdot v_i \cdot \left( \sum_{k \in [m]}\pi_{i,k,k}^w(\rho)\theta_{i,k} - \bar{\alpha} \sum_{k \in [m]}\sum_{j \in [n]\setminus i}\pi_{j,k,k}^w(\rho)\theta_{i,k} - (\max_{k} \theta_{i,k} - \alpha)\right)
 \end{align}

To establish ex post IR, we reason about the utility a buyer can obtain when not participating in the mechanism. As per the discussion in Section~\ref{sec:ir}, in an optimal data market the seller will send optimal recommendations to participating buyers in the event that buyer $i$ opts out, thus minimizing the utility of the buyer who opts out.
In particular, the recommendation for any $j\neq i$, for an opting out buyer $i$, is
such that $\mathop{\mathbb{E}}_{\substack{a_{-i} \sim \tilde{\sigma}_{-i}(\omega; \rho_{-i})\\\omega \sim \theta_i}} \left[\mathbf{1}\{a_j = \omega\} \right]= 1$. Given this, the utility of buyer $i$, when opting out, is
\begin{equation}\label{eqn:ir-opt-out}
\begin{aligned}
    \min_{\tilde{\sigma}_{-i}}\max_{\hat{a}_i \in \mathcal{A}_i} &\left(\mathop{\mathbb{E}}_{ \substack{a_{-i} \sim \tilde{\sigma}_{-i}(\omega; \rho_{-i}),\\\omega \sim \theta_i}} \left[u_i( \hat{a}_i, a_{-i}; \omega, \rho_i) \right] \right) \\
   &= \max_{\hat{a}_i \in \mathcal{A}_i} \left( \mathop{\mathbb{E}}_{ \substack{a_{-i} \sim \tilde{\sigma}_{-i}(\omega; \rho_{-i})\\\omega \sim \theta_i}} \left[ v_i \cdot \mathbf{1} \{\hat{a}_i = \omega\} - v_i \bar{\alpha}\sum_{j\in[n]\setminus i} \mathbf{1} \{\hat{a}_j = \omega\} \right] \right) \\
   &= \max_{\hat{a}_i \in \mathcal{A}_i} \left( \mathop{\mathbb{E}}_{ \omega \sim \theta_i} \left[ v_i \cdot \mathbf{1} \{\hat{a}_i = \omega\} - v_i \alpha \right] \right) \\
   &= v_i \cdot \left(\max_k \theta_{i,k} - \alpha\right).
\end{aligned}
\end{equation}

Since the induced experiment $\pi_i^w(\rho)$ is obedient, $\sum_k \pi_{i,k,k}^w(\rho)\theta_{i,k} \ge \max_k \theta_{i,k}$, and since $\pi_{j,k,k}^w(\rho)\le 1$ with $\sum_k \theta_{i,k}=1$, we have $\bar{\alpha}\sum_k\sum_{j\ne i}\pi_{j,k,k}^w(\rho)\theta_{i,k}\le \alpha$, implying the desired inequality. Thus we have $ \sum_{k \in [m]}\pi_{i,k,k}^w(\rho)\theta_{i,k} - \bar{\alpha}\left( \sum_{k \in [m]}\sum_{j \in [n]\setminus i}\pi_{j,k,k}^w(\rho)\theta_{i,k}\right) - \left(\max_{k}\theta_{i,k} - \alpha\right) \geq 0$.

Taking this with the fact that $\tilde{t}_i^w(\rho) \in [0, 1]$, we have
\begin{equation}
\begin{aligned}
    t^w_i(\rho) &= \tilde{t}_i^w(\rho) \cdot v_i \cdot \left( \sum_{k \in [m]}\pi_{i,k,k}^w(\rho)\theta_{i,k} - \bar{\alpha} \sum_{k \in [m]}\sum_{j \in [n]\setminus i}\pi_{j,k,k}^w(\rho)\theta_{i,k} - (\max_{k} \theta_{i,k} - \alpha)\right) \\
    &\leq v_i \cdot \left(\mathop{\mathbb{E}}_{ \substack{a \sim \sigma^w(\omega; \rho),\\ \omega \sim \theta_i}} \left[ \left(\mathbf{1}\{ a_i = \omega\} - \bar{\alpha} \sum_{j\in[n]\setminus i}\mathbf{1}\{ a_j = \omega\}\right)\right] - \left(\max_{k} \theta_{i,k} - \alpha)\right)\right) \\
    &= \mathop{\mathbb{E}}_{ \substack{a \sim \sigma^w(\omega; \rho),\\ \omega \sim \theta_i}} \left[ v_i \cdot \left(\mathbf{1}\{ a_i = \omega\} - \bar{\alpha} \sum_{j\in[n]\setminus i}\mathbf{1}\{ a_j = \omega\}\right)\right] - v_i \cdot \left(\max_{k} \theta_{i,k} - \alpha\right) \\
    &= \mathop{\mathbb{E}}_{ \substack{a \sim \sigma^w(\omega; \rho),\\ \omega \sim \theta_i}} \left[u_i( a; \omega, \rho_i)\right] - v_i \cdot \left(\max_k \theta_{i,k} - \alpha\right)
\end{aligned}
\end{equation}

Rearranging, and substituting $v_i \cdot \left(\max_k \theta_{i,k} - \alpha\right)$ from Eqn~\ref{eqn:ir-opt-out}, we get the IR constraint described in Eqn~\ref{eqn:ir}.
\end{proof}

\subsection{Proof of Lemma~\ref{lem:iir-net}}

\iirnet*
\begin{proof}

For the BIC setting, we work with the {\em interim} payment, and make use of
a normalized interim payment, $\tilde{t}_i(\rho_i)
\in [0, 1]$, for each buyer $i$,
 by using a sigmoidal unit. In this case, we scale this normalized payment by the difference between the interim utility achieved when the buyer is truthful and obedient and the interim utility of opting out. Thus, the {\em interim} payment is
 \begin{align}
     t_i^w(\rho_i) = \tilde{t}_i^w(\rho_i) \cdot v_i \cdot {\mathbb{E}}_{-i}\left[ \sum_{k \in [m]}\pi_{i,k,k}^w(\rho)\theta_{i,k} - \bar{\alpha} \sum_{k \in [m]}\sum_{j \in [n]\setminus i}\pi_{j,k,k}^w(\rho)\theta_{i,k} - \left(\max_k \theta_{i,k} - \alpha\right)\right]
 \end{align}

The utility of the outside option is again computed similarly to the {\em ex post} IC setting, recognizing that the seller in the optimal mechanism will make optimal recommendation to participating buyers in the event that buyer $i$ opts out. In particular, ${\mathbb{E}}_{-i} \left[\mathop{\mathbb{E}}_{\substack{a_{-i} \sim \tilde{\sigma}_{-i}(\omega; \rho_{-i}),\\\omega \sim \theta_i}} \left[\mathbf{1}\{a_j = \omega\}\right]\right] = 1$ to any agent $j \not= i$, in such a case. In particular, the
utility of buyer $i$, when opting out, is
\begin{equation}\label{eqn:iir-opt-out}
\begin{aligned}
    \min_{\tilde{\sigma}_{-i}}\max_{\hat{a}_i \in \mathcal{A}_i} &\left({\mathbb{E}}_{-i} \left[\mathop{\mathbb{E}}_{ \substack{a_{-i} \sim \tilde{\sigma}_{-i}(\omega; \rho_{-i}),\\\omega \sim \theta_i}} \left[u_i( \hat{a}_i, a_{-i}; \omega, \rho_i) \right] \right] \right)\\
   &= \max_{\hat{a}_i \in \mathcal{A}_i} \left( {\mathbb{E}}_{-i} \left[\mathop{\mathbb{E}}_{ \substack{a_{-i} \sim \tilde{\sigma}_{-i}(\omega; \rho_{-i}),\\\omega \sim \theta_i}} \left[ v_i \cdot \mathbf{1} \{\hat{a}_i = \omega\} - v_i \bar{\alpha}\sum_{j\in[n]\setminus i} \mathbf{1} \{a_j = \omega\} \right] \right]\right)\\
   &= \max_{\hat{a}_i \in \mathcal{A}_i} \left( \mathop{\mathbb{E}}_{ \omega \sim \theta_i} \left[ v_i \cdot \mathbf{1} \{\hat{a}_i = \omega\} - v_i \alpha \right] \right)\\
   &= v_i \cdot \left(\max_k \theta_{i,k} - \alpha\right)
\end{aligned}
\end{equation}

Since the induced experiments are obedient, we have $ {\mathbb{E}}_{-i} \left[\sum_{k \in [m]}\pi_{i,k,k}^w(\rho)\theta_{i,k}\right] \geq \max_{k}\theta_{i,k}$. Additionally, ${\mathbb{E}}_{-i} \left[\bar{\alpha}\left( \sum_{k \in [m]}\sum_{j \in [n]\setminus i}\pi_{j,k,k}^w(\rho)\theta_{i,k}\right)\right] \leq \bar{\alpha}\left( \sum_{j \in [n]\setminus i}1\right) = \alpha$. Thus:
\begin{equation*}
    {\mathbb{E}}_{-i} \left[\sum_{k \in [m]}\pi_{i,k,k}^w(\rho)\theta_{i,k} - \bar{\alpha}\sum_{k \in [m]}\sum_{j \in [n]\setminus i}\pi_{j,k,k}^w(\rho)\theta_{i,k} - \left(\max_{k}\theta_{i,k} - \alpha\right)\right] \geq 0\end{equation*}

Taking this with the fact that $\tilde{t}_i^w(\rho_i) \in [0, 1]$, we have
\begin{equation}
\begin{aligned}
t_i^w(\rho_i) &= \tilde{t}_i^w(\rho_i) \cdot v_i \cdot {\mathbb{E}}_{-i}\left[ \sum_{k \in [m]}\pi_{i,k,k}^w(\rho)\theta_{i,k} - \bar{\alpha} \sum_{k \in [m]}\sum_{j \in [n]\setminus i}\pi_{j,k,k}^w(\rho)\theta_{i,k} - \left(\max_k \theta_{i,k} - \alpha\right)\right]\\
    &\leq v_i \cdot {\mathbb{E}}_{-i}\left[ \sum_{k \in [m]}\pi_{i,k,k}^w(\rho)\theta_{i,k} - \bar{\alpha} \sum_{k \in [m]}\sum_{j \in [n]\setminus i}\pi_{j,k,k}^w(\rho)\theta_{i,k}\right] - v_i \cdot \left(\max_k \theta_{i,k} - \alpha\right) \\
    &= {\mathbb{E}}_{-i} \left[\mathop{\mathbb{E}}_{ \substack{a \sim \sigma^w(\omega; \rho),\\ \omega \sim \theta_i} } \left[ v_i \cdot \left(\mathbf{1}\{ a_i = \omega\} - \bar{\alpha} \sum_{j\in[n]\setminus i}\mathbf{1}\{ a_j = \omega\}\right)\right]\right] - v_i \cdot \left(\max_k \theta_{i,k} - \alpha\right) \\
    &= {\mathbb{E}}_{-i} \left[\mathop{\mathbb{E}}_{ \substack{a \sim \sigma^w(\omega; \rho_i),\\ \omega \sim \theta_i}} \left[u_i( a; \omega, \rho_i)\right]\right] - v_i \cdot \left(\max_k \theta_{i,k} - \alpha\right)
\end{aligned}
\end{equation}

Rearranging the above and substituting $v_i \cdot \left(\max_k \theta_{i,k} - \alpha\right)$ from Eqn~\ref{eqn:iir-opt-out}, we get the IIR constraint described in Eqn~\ref{eqn:iir}.

\end{proof}

\subsection{Proof of Lemma~\ref{lem:ic-net}}
\icnet*
\begin{proof}

We first prove the forward direction: if any mechanism $\mathcal{M}=\left(\sigma^w, t^w\right)$ where $\sigma_i^w(\omega, \rho)=\left(\pi_{i, \omega}^w(\rho)\right)$ is ex post IC, then $\text{IC-RGT}_i^w=0, \forall i \in[n]$.

For the \textit{ex post} IC setting, the incentive compatibility constraint requires that for every agent $i$, and for each $\rho \in \mathcal{T}$, and assuming that every other agent reports its type truthfully and follows the recommended action, then for each misreport $\rho_i' \in \mathcal{T}_i$, and each deviation function, $\delta: \mathcal{A}_i \to \mathcal{A}_i$, the following condition:
\begin{equation}\label{lem:icexpost1}
\begin{aligned}
    \mathop{\mathbb{E}}_{ \substack{a \sim \sigma(\omega, \rho), \\\omega \sim \theta_i}} \left[u_i( a, \omega, \rho_i) - t_i(\rho) \right] \geq \mathop{\mathbb{E}}_{\substack{a \sim \sigma(\omega; \rho_i', \rho_{-i}), \\ \omega \sim \theta_i}\ } \left[u_i( \delta (a_i), a_{-i}, \omega, \rho_i)- t_i(\rho_i', \rho_{-i}) \right]
\end{aligned}
\end{equation}
Note that the above holds for any deviation function $\delta$. Therefore, we can write that
\begin{equation}\label{lem:icexpost2}
\begin{aligned}
    \mathop{\mathbb{E}}_{ \substack{a \sim \sigma(\omega, \rho), \\\omega \sim \theta_i}} \left[u_i( a, \omega, \rho_i) - t_i(\rho) \right] \geq \max_\delta \mathop{\mathbb{E}}_{\substack{a \sim \sigma(\omega; \rho_i', \rho_{-i}), \\ \omega \sim \theta_i}\ } \left[u_i( \delta (a_i), a_{-i}, \omega, \rho_i)- t_i(\rho_i', \rho_{-i}) \right]
\end{aligned}
\end{equation}
We can expand to get the following equation:
\begin{equation}
\begin{aligned}
     v_i \cdot \left( \sum_{k \in [m]}\pi_{i,k,k}^w(\rho)\theta_{i,k} - \bar{\alpha} \sum_{k \in [m]}\sum_{j \in [n]\setminus i}\pi_{j,k,k}^w(\rho)\theta_{i,k} \right) - t_i(\rho) \\
     \geq \max_{\delta} v_i \cdot \left( \sum_{k \in [m]}
    \left\{\pi_{i,\delta(k),k}^w(\rho'_i, \rho_{-i})\theta_{i,\delta(k)}\right\} - \bar{\alpha} \sum_{k \in [m]}\sum_{j \in [n]\setminus i}\pi_{j,k,k}^w(\rho'_i, \rho_{-i})\theta_{i,k} \right) - t_i(\rho'_i,\rho_{-i})
\end{aligned}
\end{equation}
Pushing the max inside (since other terms don't involve $\delta(k)$), we note that:
\begin{equation}
    \max_\delta\sum_{k \in [m]}
    \left\{\pi_{i,\delta(k),k}^w(\rho'_i, \rho_{-i})\theta_{i,\delta(k)}\right\} = \sum_{k \in [m]} \max_{k'}
    \left\{\pi_{i,k',k}^w(\rho'_i, \rho_{-i})\theta_{i,k'}\right\}
\end{equation}
Thus we have,
\begin{equation}\label{lem:icexpost3}
\begin{aligned}
     v_i \cdot \left( \sum_{k \in [m]}\pi_{i,k,k}^w(\rho)\theta_{i,k} - \bar{\alpha} \sum_{k \in [m]}\sum_{j \in [n]\setminus i}\pi_{j,k,k}^w(\rho)\theta_{i,k} \right) - t_i(\rho) \\
     \geq v_i \cdot \left( \sum_{k \in [m]} \max_{k' \in [m]}
    \left\{\pi_{i,k',k}^w(\rho'_i, \rho_{-i})\theta_{i,k'}\right\} - \bar{\alpha} \sum_{k \in [m]}\sum_{j \in [n]\setminus i}\pi_{j,k,k}^w(\rho'_i, \rho_{-i})\theta_{i,k} \right) - t_i(\rho'_i,\rho_{-i})
\end{aligned}
\end{equation}

By Equation~\ref{eq::rgtep}, the above is equivalent to $\mathrm{ic{\textit -}rgt}^w_i(\rho'_i, \rho) \leq 0$. Furthermore, this holds for any deviating report $\rho_i'$. Thus, we can write:

\begin{equation}\label{lem:icexpost5}
    \max_{\rho'_i\in \mathcal{T}_i} \mathrm{ic{\textit -}rgt}^w_i(\rho'_i, \rho) \leq 0
\end{equation}
But note that
\begin{equation}
\begin{aligned}\label{lem:icexpost6}
\max _{\rho_i^{\prime} \in \mathcal{T}_i} \mathrm{ic{\textit -}rgt}_i^w\left(\rho_i^{\prime}, \rho\right) &\geq \mathrm{ic{\textit -}rgt}_i^w\left(\rho_i, \rho\right)
    \\&= v_i \cdot \left( \sum_{k \in [m]} \max_{k' \in [m]}
    \left\{\pi_{i,k',k}^w(\rho)\theta_{i,k'}\right\} - \bar{\alpha} \sum_{k \in [m]}\sum_{j \in [n]\setminus i}\pi_{j,k,k}^w(\rho)\theta_{i,k} \right)
    \\&\qquad- v_i \cdot \left( \sum_{k \in [m]}\pi_{i,k,k}^w(\rho)\theta_{i,k} - \bar{\alpha} \sum_{k \in [m]}\sum_{j \in [n]\setminus i}\pi_{j,k,k}^w(\rho)\theta_{i,k} \right)
    \\ &= v_i\cdot \left( \sum_{k \in [m]} \left[\max_{k' \in [m]}
    \left\{\pi_{i,k',k}^w(\rho)\theta_{i,k'}\right\} - \pi_{i,k,k}^w(\rho)\theta_{i,k}\right]\right)
    \\ &\geq 0
\end{aligned}
\end{equation}
where the last step holds since $\max_{k' \in [m]} \left\{\pi_{i,k',k}^w(\rho)\theta_{i,k'}\right\} \geq \pi_{i,k,k}^w(\rho)\theta_{i,k}$

Combining Equations~\ref{lem:icexpost5} and~\ref{lem:icexpost6}, we have $\max_{\rho'_i\in \mathcal{T}_i} \mathrm{ic{\textit -}rgt}^w_i(\rho'_i, \rho) = 0$. Taking the expectation over all profiles $\rho \in \mathcal{T}$, we have $\mathbb{E}_{\rho \sim \mathcal{P}}\left[\max _{\rho_i^{\prime} \in \mathcal{T}_i} \mathrm{ic{\textit -}rgt}_i^w\left(\rho_i^{\prime}, \rho\right)\right] = 0$. Thus, $ \text{IC-RGT}_i^w=0, \forall i \in[n]$. Thus, we've shown that if any mechanism $\mathcal{M}=\left(\sigma^w, t^w\right)$ where $\sigma_i^w(\omega, \rho)=\left(\pi_{i, \omega}^w(\rho)\right)$ is {\em ex post} IC, then $\text{IC-RGT}_i^w=0$, as desired.

Next, we consider the reverse direction: if $\text{IC-RGT}_i^w=0, \forall i \in[n]$, we want to show that any mechanism $\mathcal{M}=\left(\sigma^w, t^w\right)$ where $\sigma_i^w(\omega, \rho)=\left(\pi_{i, \omega}^w(\rho)\right)$ is ex post IC. Starting with $\text{IC-RGT}_i^w=0, \forall i \in[n]$, we have that by definition: $
\mathbb{E}_{\rho \sim \mathcal{P}}\left[\max _{\rho_i^{\prime} \in \mathcal{T}_i} \mathrm{ic{\textit -}rgt}_i^w\left(\rho_i^{\prime}, \rho\right)\right] = 0 $. But Equation~\ref{lem:icexpost6} shows that
$\max_{\rho_i'\in\mathcal T_i}
\mathrm{ic\text{-}rgt}_i^w(\rho_i',\rho)\ge0$.
Since this nonnegative quantity has expectation zero, it equals zero
for $\mathcal P$-almost every $\rho$. Therefore, for each such
profile and every $\rho_i'\in\mathcal T_i$,
$\mathrm{ic\text{-}rgt}_i^w(\rho_i',\rho)\le0$.

Substituting the definition in Equation~\ref{eq::rgtep} gives:
\begin{equation}\label{eq}
\begin{aligned}
    v_i \cdot \Bigg( \sum_{k \in [m]} \max_{k' \in [m]}
    \left\{\pi_{i,k',k}^w(\rho'_i, \rho_{-i})\theta_{i,k'}\right\} &- \bar{\alpha} \sum_{k \in [m]}\sum_{j \in [n]\setminus i}\pi_{j,k,k}^w(\rho'_i, \rho_{-i})\theta_{i,k} \Bigg)
    \\- v_i \cdot \Bigg( \sum_{k \in [m]}\pi_{i,k,k}^w(\rho)\theta_{i,k} - &\bar{\alpha} \sum_{k \in [m]}\sum_{j \in [n]\setminus i}\pi_{j,k,k}^w(\rho)\theta_{i,k} \Bigg)
    - \left(t^w_i(\rho'_i, \rho_{-i}) - t^w_i(\rho)\right) \leq 0
\end{aligned}
\end{equation}
This can be transformed back to:
\begin{equation}
\begin{aligned}
    \mathop{\mathbb{E}}_{ \substack{a \sim \sigma(\omega, \rho) \\\omega \sim \theta_i}} \left[u_i( a, \omega, \rho_i) - t_i(\rho) \right] &\geq \max_\delta \mathop{\mathbb{E}}_{\substack{a \sim \sigma(\omega; \rho_i', \rho_{-i}), \\ \omega \sim \theta_i}\ } \left[u_i( \delta (a_i), a_{-i}, \omega, \rho_i)- t_i(\rho_i', \rho_{-i}) \right]
    \\ &\geq \mathop{\mathbb{E}}_{\substack{a \sim \sigma(\omega; \rho_i', \rho_{-i}) \\ \omega \sim \theta_i}\ } \left[u_i( \delta (a_i), a_{-i}, \omega, \rho_i)- t_i(\rho_i', \rho_{-i}) \right]
\end{aligned}
\end{equation}
This is exactly the {\em ex post} IC constraint in
Equation~\ref{eqn:expost-ic}, except on a set of type profiles of
$\mathcal P$-measure zero.

\end{proof}

\subsection{Proof of Lemma~\ref{lem:bic-net}}
\bicnet*
\begin{proof}

We first prove the forward direction: if any mechanism $\mathcal{M}=\left(\sigma^w, t^w\right)$ where $\sigma_i^w(\omega, \rho)=\left(\pi_{i, \omega}^w(\rho)\right)$ is interim IC, then $\text{BIC-RGT}_i^w=0, \forall i \in[n]$.

For the BIC setting, for each $(\rho_i, \rho_i') \in \mathcal{T}_i^2$ and for each deviation function $\delta: \mathcal{A}_i \to \mathcal{A}_i$, a BIC mechanism satisfies:
\begin{equation}
\begin{aligned}
     {\mathbb{E}}_{-i} \left[\mathop{\mathbb{E}}_{ \substack{a \sim \sigma(\omega, \rho)\\ \omega \sim \theta_i}} \left[u_i( a, \omega, \rho_i) - t_i(\rho) \right]\right] \geq {\mathbb{E}}_{-i} \left[\mathop{\mathbb{E}}_{ \substack{ a\sim \sigma(\omega; \rho_i', \rho_{-i}) \\\omega \sim \theta_i}} \left[u_i( \delta (a_i), a_{-i}, \omega, \rho_i)- t_i(\rho_i', \rho_{-i}) \right] \right]
\end{aligned}
\end{equation}
Note that the above holds for any deviation function $\delta$, therefore we can write that
\begin{equation}
\begin{aligned}
     {\mathbb{E}}_{-i} \left[\mathop{\mathbb{E}}_{ \substack{a \sim \sigma(\omega, \rho)\\ \omega \sim \theta_i}} \left[u_i( a, \omega, \rho_i) - t_i(\rho) \right]\right]& \\\geq \max_\delta {\mathbb{E}}_{-i} &\left[\mathop{\mathbb{E}}_{ \substack{ a\sim \sigma(\omega; \rho_i', \rho_{-i}) \\\omega \sim \theta_i}} \left[u_i( \delta (a_i), a_{-i}, \omega, \rho_i)- t_i(\rho_i', \rho_{-i}) \right] \right]
\end{aligned}
\end{equation}
We can expand to get the following equations:
\begin{equation}
    \begin{aligned}
    v_i &\cdot {\mathbb{E}}_{-i} \left[\sum_{k \in [m]}
   \pi_{i,k,k}^w(\rho_i, \rho_{-i})\theta_{i,k} - \bar{\alpha} \sum_{k \in [m]}\sum_{j \in [n]\setminus i}\pi_{j,k,k}^w(\rho_i, \rho_{-i})\theta_{i,k} \right] - t_i(\rho_i)
   \\ &\geq \max_\delta v_i \cdot {\mathbb{E}}_{-i} \left[ \left( \sum_{k \in [m]}
    \pi_{i,\delta(k),k}^w(\rho'_i, \rho_{-i})\theta_{i,\delta(k)} - \bar{\alpha} \sum_{k \in [m]} \sum_{j \in [n]\setminus i} \pi_{j,k,k}^w(\rho'_i, \rho_{-i})\theta_{i,k} \right) \right]- t_i(\rho'_i)
    \\ &= \max_\delta v_i \cdot \left( \sum_{k \in [m]} {\mathbb{E}}_{-i} \left[
    \pi_{i,\delta(k),k}^w(\rho'_i, \rho_{-i})\theta_{i,\delta(k)} - \bar{\alpha} \sum_{j \in [n]\setminus i} \pi_{j,k,k}^w(\rho'_i, \rho_{-i})\theta_{i,k}\right] \right) - t_i(\rho'_i)
     \\ &= v_i \cdot \left( \sum_{k \in [m]} \max_\delta {\mathbb{E}}_{-i} \left[
    \pi_{i,\delta(k),k}^w(\rho'_i, \rho_{-i})\theta_{i,\delta(k)} - \bar{\alpha} \sum_{j \in [n]\setminus i} \pi_{j,k,k}^w(\rho'_i, \rho_{-i})\theta_{i,k}\right] \right)- t_i(\rho'_i)
     \\ &= v_i \cdot \left( \sum_{k \in [m]} \max_{k' \in [m]} {\mathbb{E}}_{-i} \left[
    \pi_{i,k',k}^w(\rho'_i, \rho_{-i})\theta_{i,k'} - \bar{\alpha} \sum_{j \in [n]\setminus i} \pi_{j,k,k}^w(\rho'_i, \rho_{-i})\theta_{i,k}\right] \right)- t_i(\rho'_i)
    \end{aligned}
\end{equation}
The above equations follow from the linearity of expectations and from pushing the max to the inside (since other terms don't involve $\delta(k)$).

Note that by our definition of \textit{interim} regret in Equation~\ref{eq:bicreg}, the above equation is equivalent to $\mathrm{bic{\textit -}rgt}_i^w\left(\rho_i^{\prime}, \rho_i\right) \leq 0$. Furthermore, this holds for any deviating report $\rho'_i$. Therefore, we can write that

\begin{equation}\label{lem:bicpf1}
\max_{\rho'_i\in \mathcal{T}_i} \mathrm{bic{\textit -}rgt}^w_i(\rho'_i, \rho_i) \leq 0
\end{equation}
But note that
\begin{equation}\label{lem:bicpf2}
\begin{aligned}
\max_{\rho_i'\in\mathcal T_i}
\mathrm{bic\text{-}rgt}_i^w(\rho_i',\rho_i)
&\ge
\mathrm{bic\text{-}rgt}_i^w(\rho_i,\rho_i)
\\
&=
v_i\sum_{k\in[m]}
\left(
\max_{k'\in[m]}
\mathbb E_{-i}
\left[
\pi_{i,k',k}^w(\rho_i,\rho_{-i})\theta_{i,k'}
\right]
-
\mathbb E_{-i}
\left[
\pi_{i,k,k}^w(\rho_i,\rho_{-i})\theta_{i,k}
\right]
\right)
\\
&\ge0.
\end{aligned}
\end{equation}
The equality follows because, at the truthful report, the payment
terms cancel and the externality term is independent of $k'$ and
therefore also cancels. The final inequality holds because $k'=k$ is
feasible in each maximization.

Combining Equations~\ref{lem:bicpf1} and \ref{lem:bicpf2}, we have that $\max_{\rho'_i\in \mathcal{T}_i} \mathrm{bic{\textit -}rgt}^w_i(\rho'_i, \rho_i) = 0$. Taking the expectation over all profiles $\rho$, we have ${\mathbb{E}}_{\rho \sim \mathcal{P}}\left[\max _{\rho_i^{\prime} \in \mathcal{T}_i} \mathrm{bic{\textit -}rgt}_i^w\left(\rho_i^{\prime}, \rho_i\right)\right] = 0 \implies \text{BIC-RGT}_i^w=0, \forall i \in[n]$. Thus, we've shown that if any mechanism $\mathcal{M}=\left(\sigma^w, t^w\right)$ where $\sigma_i^w(\omega, \rho)=\left(\pi_{i, \omega}^w(\rho)\right)$ is BIC, then $\text{BIC-RGT}_i^w=0$, as desired.

Next, we consider the reverse direction: if $\text{BIC-RGT}_i^w=0, \forall i \in[n]$, we want to show that any mechanism $\mathcal{M}=\left(\sigma^w, t^w\right)$ where $\sigma_i^w(\omega, \rho)=\left(\pi_{i, \omega}^w(\rho)\right)$ is BIC. Starting with $\text{BIC-RGT}_i^w=0, \forall i \in[n]$, we have that by definition: ${\mathbb{E}}_{\rho \sim \mathcal{P}}\left[\max _{\rho_i^{\prime} \in \mathcal{T}_i} \mathrm{bic{\textit -}rgt}_i^w\left(\rho'_i, \rho_i\right)\right] = 0 $. But Equation~\ref{lem:bicpf2} shows that
$\max_{\rho_i'\in\mathcal T_i}
\mathrm{bic\text{-}rgt}_i^w(\rho_i',\rho_i)\ge0$.
Since this nonnegative quantity has expectation zero, it equals zero
for $\mathcal P_i$-almost every $\rho_i$. Therefore, for each such
type and every $\rho_i'\in\mathcal T_i$,
$\mathrm{bic\text{-}rgt}_i^w(\rho_i',\rho_i)\le0$.
Plugging in our definition for \textit{interim} regret, we have that:
\begin{equation}
\begin{aligned}
    v_i \cdot \Bigg( \sum_{k \in [m]} \max_{k' \in [m]}
    {\mathbb{E}}_{-i} \Big[\pi_{i,k',k}^w(\rho'_i, \rho_{-i})\theta_{i,k'} &- \bar{\alpha} \sum_{j \in [n]\setminus i} \pi_{j,k,k}^w(\rho'_i, \rho_{-i})\theta_{i,k} \Big]\Bigg)- t_i(\rho'_i)
    \\ \leq v_i \cdot {\mathbb{E}}_{-i} \Bigg[\sum_{k \in [m]}
   \pi_{i,k,k}^w(\rho_i, \rho_{-i})\theta_{i,k} &- \bar{\alpha} \sum_{k \in [m]}\sum_{j \in [n]\setminus i}\pi_{j,k,k}^w(\rho_i, \rho_{-i})\theta_{i,k} \Bigg] - t_i(\rho_i)
\end{aligned}
\end{equation}
This can be transformed back to:

\begin{equation}
\begin{aligned}
     {\mathbb{E}}_{-i} \left[\mathop{\mathbb{E}}_{ \substack{a \sim \sigma(\omega, \rho)\\ \omega \sim \theta_i}} \left[u_i( a, \omega, \rho_i) - t_i(\rho) \right]\right]
     \\&\hspace{-50pt}\geq \max_\delta {\mathbb{E}}_{-i} \left[\mathop{\mathbb{E}}_{ \substack{ a\sim \sigma(\omega; \rho_i', \rho_{-i}) \\\omega \sim \theta_i}} \left[u_i( \delta (a_i), a_{-i}, \omega, \rho_i)- t_i(\rho_i', \rho_{-i}) \right] \right]
     \\&\hspace{-50pt}\geq {\mathbb{E}}_{-i} \left[\mathop{\mathbb{E}}_{ \substack{ a\sim \sigma(\omega; \rho_i', \rho_{-i}) \\\omega \sim \theta_i}} \left[u_i( \delta (a_i), a_{-i}, \omega, \rho_i)- t_i(\rho_i', \rho_{-i}) \right] \right] \quad \forall \delta
\end{aligned}
\end{equation}

This is exactly the BIC constraint in Equation~\ref{eqn:bic}, except
on a set of types of $\mathcal P_i$-measure zero.

\end{proof}

\subsection{Proof of Theorem~\ref{thm:icproof}}

\icproof*

\begin{proof}\label{app:ic-proof}
We consider a setting with binary states, binary actions, and where each buyer has the same interim belief, set to $(\theta, 1 - \theta)$. Without loss of generality, let $\theta \geq (1 - \theta)$. Since the interim beliefs are fixed, we will just represent buyer types with $v_i$ instead of $\rho_i$. Let $x_i(v) = \mathop{\mathbb{E}}_{\substack{a \sim \sigma(\omega;\rho),\\\omega \sim \theta_i}} \left[\mathbf{1} \{ a_i = \omega\} \right]$. Let $\pi_i(v)$ be the matrix representation of the experiment assigned to buyer $i$ i.e. $\pi_{i, \omega}(v) = \sigma_i(\omega, v)$

We first show that a mechanism is obedient if and only if the experiment assigned to the buyer satisfies $x_i(v) \geq \theta$.
If a mechanism is obedient, then $\theta \cdot \pi_{i, 1,1}(v) \geq (1 - \theta) \pi_{i,2,1}(v)$ and $(1 - \theta) \pi_{i,2,2}(v) \geq \theta \cdot \pi_{i,1,2}(v)$. We have $x_i(v) = \theta \cdot \pi_{i,1,1}(v) + (1 - \theta) \pi_{i,2,2}(v) \geq \theta \cdot \pi_{i,1,1}(v) + \theta \cdot \pi_{i,1,2}(v) = \theta (\pi_{i,1,1}(v) + \pi_{i,1,2}(v)) = \theta$.

In the other direction, consider when $x_i(v) \geq \theta$. In this case, we show that both $\theta \cdot \pi_{i, 1,1}(v) \geq (1 - \theta) \pi_{i,2,1}(v) $ and $(1 - \theta) \pi_{i,2,2}(v) \geq \theta \cdot \pi_{i,1,2}(v) $ need to hold for obedience.
Assume to the contrary, one of these fails to hold when $x_i(v) \geq \theta$. We have one of,
\begin{itemize}
    \item $(1 - \theta) \pi_{i,2,2}(v) < \theta \cdot \pi_{i,1,2}(v) $, and we have $x_i(v) = \theta \cdot \pi_{i,1,1}(v) + (1 - \theta) \pi_{i,2,2}(v) < \theta \cdot \pi_{i,1,1}(v) + \theta \cdot \pi_{i,1,2}(v) = \theta (\pi_{i,1,1}(v) + \pi_{i,1,2}(v) ) = \theta$.
    \item $\theta \cdot \pi_{i,1,1}(v) < (1 - \theta) \pi_{i,2,1}(v) $, and we have $x_i(v) = \theta \cdot \pi_{i,1,1}(v) + (1 - \theta) \pi_{i,2,2}(v) < (1 - \theta) \pi_{i,2,1}(v) + (1 - \theta) \pi_{i,2,2}(v) = (1 - \theta) (\pi_{i,2,1}(v) + \pi_{i,2,2}(v) ) = (1 - \theta) \leq \theta$.
\end{itemize}
In either case, we have a contradiction.

For the setting under consideration (fixed beliefs and linear payoffs), we can show that the IC constraints are satisfied if and only if {\em truthfulness} and {\em obedience} constraints are satisfied. This decomposition is closely related to Proposition 3.5 in~\citet{Bonatti2022}, which establishes an analogous result under Bayesian incentive compatibility. We provide a direct ex post argument here for completeness. For the ``only if'' direction, note that ex post incentive compatibility immediately implies truthfulness by restricting attention to deviations that only change the reported type while keeping the recommended action fixed, and it implies obedience by restricting attention to deviations that keep the report fixed and only change the action taken after receiving the recommendation.

For the converse, suppose the mechanism satisfies ex post truthfulness~\eqref{eqn:expost-truth} and ex post obedience~\eqref{eqn:expost-ob}. Fix agent $i$. Since the beliefs are fixed, represent the type profile as $v \in \mathcal{V}$, misreport as $v_i' \in \mathcal{V}_i$, and deviation function $\delta : \mathcal{A}_i \to \mathcal{A}_i$. By ex post truthfulness,
\[
\mathop{\mathbb{E}}_{\substack{a \sim \sigma(\omega, v), \\ \omega \sim \theta_i}}
\left[u_i(a,\omega,v_i) - t_i(v)\right]
\ge
\mathop{\mathbb{E}}_{\substack{a \sim \sigma(\omega, v_i', v_{-i}), \\ \omega \sim \theta_i}}
\left[u_i(a,\omega,v_i) - t_i(v_i', v_{-i})\right].
\]

Because beliefs are fixed and utilities are linear, obedience depends only on the realized signal and not on the reported type. Hence, obedience at $(v_i', v_{-i})$ also holds for the true type $v_i$.

By ex post obedience at $(v_i', v_{-i})$, and since payments do not depend on actions,
\[
\mathop{\mathbb{E}}_{\substack{a \sim \sigma(\omega, v_i', v_{-i}), \\ \omega \sim \theta_i}}
\left[u_i(a,\omega,v_i) - t_i(v_i', v_{-i})\right]
\ge
\mathop{\mathbb{E}}_{\substack{a \sim \sigma(\omega, v_i', v_{-i}), \\ \omega \sim \theta_i}}
\left[u_i(\delta(a_i), a_{-i}, \omega, v_i) - t_i(v_i', v_{-i})\right].
\]
Combining the two inequalities yields~\eqref{eqn:expost-ic}, thus showing that IC constraints are satisfied.

Now denote $\tilde{x}_i(v) = x_i(v) - \frac{\alpha}{n - 1}\sum_{j \in [n]\setminus i} x_j(v)$. Thus, the optimal design problem to solve is:
\begin{align*}
    \texttt{maximize} \quad\mathbb{E}_{v\sim\mathcal{V}} &\left[ \sum_{i \in [n]} t_i(v)\right]
    &\\
    \text{s.t.}\ v_i \tilde{x}_i(v) - t_i(v) &\geq v_i \tilde{x}_i(v'_i; v_{-i}) - t_i(v'_i; v_{-i}) \quad
    &\forall v \in \mathcal{V}, v'_i \in \mathcal{V}_i, i \in [n]\\
    v_i \tilde{x}_i(v) - t_i(v) &\geq v_i (\theta - \alpha) \quad &\forall v \in \mathcal{V}, i \in [n]\\
    x_i(v) &\geq \theta \quad
    &\forall v \in \mathcal{V}, i \in [n]
\end{align*}

The first constraint corresponds to {\em truthfulness} for the IC setting, the second is the IR constraint, and the third is the {\em obedience} constraint.
We have thus reduced the data market design problem to that of Myerson's revenue maximizing single-parameter mechanism design problem. Instead of $\tilde{x}_i(v)$ denoting the probability of allocating an item, it denotes expected payoff after accounting for the negative externalities. Also, rather than allocative constraints, we have obedience constraints on $x_i(v)$ which requires $x_i(v) \geq \theta$. Thus, by Myerson's theory, the total expected revenue is equal to the expected virtual welfare minus some constant $K$ (stemming from IR constraints) for some $\tilde{x}_i(v_i, v_{-i})$ that is non-decreasing in $v_i$. Thus, we have:
\begin{equation}
\begin{aligned}
    \mathbb{E}_{v\sim\mathcal{V}} \left[ \sum_{i \in [n]} t_i(v)\right] &= \mathbb{E}_{v\sim\mathcal{V}} \left[ \sum_{i \in [n]} \phi_i(v_i)\tilde{x}_i(v)\right] - K \\
    &= \mathbb{E}_{v\sim\mathcal{V}} \left[ \sum_{i \in [n]} \phi_i(v_i) \left(x_i(v) - \frac{\alpha}{n - 1} \sum_{j \in [n]\setminus i} x_j(v)\right) \right] - K\\
    &= \mathbb{E}_{v\sim\mathcal{V}} \left[ \sum_{i \in [n]} \left(\phi_i(v_i) - \frac{\alpha}{n - 1}\sum_{j \in [n]\setminus i} \phi_j(v_j)\right) x_i(v)\right] - K
\end{aligned}
\end{equation}
In order to maximize the revenue, we need to maximize the virtual welfare. Thus we can set $x_i(v) = 1$ (a fully informative experiment)
when $\phi_i(v_i) - \frac{\alpha}{n-1}\sum_{j \in [n]\setminus i} \phi_j(v_j) \geq 0$ and we can set $x_i(v) = \theta$ (an uninformative experiment that always sends the same signal regardless of the state) otherwise. This is precisely the mechanism described in the theorem.

Note that such an $\tilde{x}_i(v_i, v_{-i})$ is non-decreasing in $v_i$ (since $x_i(v_i, v_{-i})$ is non-decreasing in $v_i$ and $x_j(v_i, v_{-i})$ for $j \in [n]\setminus i$ is non-increasing in $v_i$ for regular distributions $\mathcal{V}_i$ and $\mathcal{V}_j$). Moreover, $x_i(v) \geq \theta$ is also satisfied.
\end{proof}

{\color{black}
\section{Generalization and Network Capacity Bounds}
\label{app:generalization_bounds}

In this section, we prove Theorem~\ref{thm:gen_bounds} and Theorem~\ref{thm:cover_bound}.

Our analysis follows the statistical-learning framework of \citet[Theorem~2.2 and Appendix~D.2, D.5]{dutting2023jacm}. Relative to the auction setting, the only substantive changes are that buyer-level regret now optimizes jointly over type misreports and action deviations, and that utilities depend on other buyers' experiments through the negative-externality term.

Throughout this discussion, \(n\) denotes the number of buyers, \(m=|\Omega|\) the number of states (and, in the matching-payoff setting, actions), \(v_{\max}\) an upper bound on buyer values, and \(\alpha\) the negative-externality parameter from Section~\ref{sec:prelim}. We work on a compact reported type domain
\[
\mathcal{T}=\prod_{i=1}^n \mathcal{T}_i,
\qquad
\mathcal{T}_i\subseteq [0,v_{\max}]\times\triangle(\Omega),
\]
and all misreports are taken from the corresponding set \(\mathcal{T}_i\). For the neural-network mechanism classes considered below, the induced utility maps are continuous on this compact domain, so the maxima in the regret definitions are attained (equivalently, one may replace every \(\max\) by \(\sup\) without changing the argument).

Let $\mathcal{M}$ be a class of data-market mechanisms $(\sigma,t)$. For the purpose of establishing covering-number bounds, we analyze the architecture under the standard assumption that hidden-layer activations are bounded and $1$-Lipschitz.\footnote{\citet{dutting2023jacm} assume bounded, 1-Lipschitz activations (e.g., tanh) in their covering-number analysis. Our implementation uses Leaky ReLU, which is also 1-Lipschitz but unbounded. Theorem \ref{thm:cover_bound} therefore applies formally to a bounded-activation counterpart of our architecture. In practice, since inputs are bounded and network parameters are norm-constrained, the resulting mappings remain effectively bounded, and the same argument extends.}

Define the $\ell_{\infty,1}$ distance between two mechanisms $(\sigma,t)$ and $(\sigma',t')$ by
\begin{equation}
\label{eq:mech_metric}
d((\sigma,t),(\sigma',t'))
=
\max_{\rho\in\mathcal{T}}
\sum_{i\in[n]}
\left(
\sum_{j,k\in[m]}
|\pi_{i,j,k}(\rho)-\pi'_{i,j,k}(\rho)|
+
|t_i(\rho)-t_i'(\rho)|
\right).
\end{equation}
For any $\epsilon>0$, let $\mathcal{N}_\infty(\mathcal{M},\epsilon)$ denote the minimum number of balls of radius $\epsilon$ needed to cover $\mathcal{M}$ under this distance.

\subsection{Auxiliary Lemmas}

To establish uniform convergence, we rely on empirical Rademacher complexity. Let $\mathcal{F}$ denote a class of bounded functions $f:Z\to[-M,M]$ defined on an input space $Z$, for some $M>0$. Let $D$ be a distribution over $Z$ and let $\mathcal{S}=\{z_1,\ldots,z_L\}$ be a sample drawn i.i.d.\ from $D$. The empirical Rademacher complexity of $\mathcal{F}$ on $\mathcal{S}$ is
\[
\hat{\mathcal{R}}_L(\mathcal{F})
:=
\frac{1}{L}\,
\mathbb{E}_{\sigma}
\left[
\sup_{f\in\mathcal{F}}
\sum_{\ell=1}^L \sigma_\ell f(z_\ell)
\right],
\]
where $\sigma\in\{-1,1\}^L$ and each $\sigma_\ell$ is drawn i.i.d.\ from the uniform distribution on $\{-1,1\}$.

We use the following two standard results.

\begin{lemma}[Uniform Convergence, Lemma~D.1 in \citet{dutting2023jacm}]
\label{lem:rademacher}
Let $\mathcal{S}$ be a sample of size $L$ drawn i.i.d.\ from a distribution $D$ over $Z$. With probability at least $1-\delta$, for all $f\in\mathcal{F}$,
\[
\mathbb{E}_{z\sim D}[f(z)]
\le
\frac{1}{L}\sum_{\ell=1}^L f(z_\ell)
+
2\hat{\mathcal{R}}_L(\mathcal{F})
+
4M\sqrt{\frac{2\log(4/\delta)}{L}}.
\]
\end{lemma}

\begin{lemma}[Massart's Lemma, Lemma~26.8 in \citet{shalev2014understanding}]
\label{lem:massart}
Let $A=\{\mathbf{a}_1,\ldots,\mathbf{a}_N\}\subseteq\mathbb{R}^L$, and let
\[
\overline{\mathbf{a}}
=
\frac{1}{N}\sum_{r=1}^N \mathbf{a}_r.
\]
Then the empirical Rademacher complexity of $A$ is bounded by
\[
\hat{\mathcal{R}}_L(A)
\le
\max_{\mathbf{a}\in A}
\|\mathbf{a}-\overline{\mathbf{a}}\|_2
\cdot
\frac{\sqrt{2\log N}}{L}.
\]
\end{lemma}

\subsection{Proof of Theorem~\ref{thm:gen_bounds}}
\genbound*

\begin{proof}
We follow \citet[Theorem~2.2 and Appendix~D.2]{dutting2023jacm}. The only data-market-specific ingredient is the utility Lipschitz estimate in Step~2 below.

Let
\[
B:=v_{\max}(1+\alpha),
\qquad
c:=\max\{1,\;4B\},
\]
and let
\[
\Delta_i:=\{\delta:\mathcal{A}_i\to\mathcal{A}_i\}.
\]
Since each action space \(\mathcal{A}_i\) is finite, \(\Delta_i\) is finite.

In the BIC setting, we identify the interim payment with its lift to \(\mathcal{T}\):
\[
t_i(\rho):=t_i(\rho_i),
\]
which is constant in \(\rho_{-i}\).

\paragraph{Revenue bound.}
Define
\[
\mathrm{rev}\circ\mathcal{M}
=
\left\{
f:\mathcal{T}\to\mathbb{R}
\ \middle|\
f(\rho)=\sum_{i=1}^n t_i(\rho)
\text{ for some }M=(\sigma,t)\in\mathcal{M}
\right\}.
\]
Since \(|t_i(\rho)|\le B\), every function in \(\mathrm{rev}\circ\mathcal{M}\) takes values in
\[
[-nB,nB]\subseteq[-nc,nc].
\]

Applying Lemma~\ref{lem:rademacher} to the class
\[
-\mathrm{rev}\circ\mathcal{M}
=
\{-f \mid f\in \mathrm{rev}\circ\mathcal{M}\}
\]
with confidence parameter \(\delta/2\), we obtain that with probability at least \(1-\delta/2\), for every \(M=(\sigma,t)\in\mathcal{M}\),
\begin{align}
-\mathbb{E}_{\rho\sim\mathcal{P}}
\!\left[\sum_{i=1}^n t_i(\rho)\right]
&\le
-\frac{1}{L}\sum_{\ell=1}^L \sum_{i=1}^n t_i(\rho^{(\ell)})
+
2\hat{\mathcal{R}}_L(\mathrm{rev}\circ\mathcal{M})
\nonumber\\
&\qquad
+
4nc\sqrt{\frac{2\log(8/\delta)}{L}}.
\label{eq:dm-rev-uc}
\end{align}

Exactly as in \citet[Appendix~D.2.3]{dutting2023jacm}, an \(\eta\)-cover of \(\mathcal{M}\) under \(d(\cdot,\cdot)\) induces an \(\eta\)-cover of \(\mathrm{rev}\circ\mathcal{M}\), because
\[
\sup_{\rho\in\mathcal{T}}
\left|
\sum_{i=1}^n t_i(\rho)-\sum_{i=1}^n \hat t_i(\rho)
\right|
\le
\sup_{\rho\in\mathcal{T}}
\sum_{i=1}^n |t_i(\rho)-\hat t_i(\rho)|
\le
d((\sigma,t),(\hat\sigma,\hat t)).
\]
Hence
\[
\mathcal{N}_\infty(\mathrm{rev}\circ\mathcal{M},\eta)
\le
\mathcal{N}_\infty(\mathcal{M},\eta).
\]

Using the same cover-plus-Massart argument as in \citet[Appendix~D.2.3]{dutting2023jacm},
\[
\hat{\mathcal{R}}_L(\mathrm{rev}\circ\mathcal{M})
\le
\inf_{\eta>0}
\left\{
\eta
+
2nc\sqrt{
\frac{
2\log\!\left(\mathcal{N}_\infty(\mathcal{M},\eta)\right)
}{L}
}
\right\}.
\]
Since \(2nc\ge 1\),
\[
\hat{\mathcal{R}}_L(\mathrm{rev}\circ\mathcal{M})
\le
\inf_{\eta>0}
\left\{
2nc\,\eta
+
2nc\sqrt{
\frac{
2\log\!\left(\mathcal{N}_\infty(\mathcal{M},\eta)\right)
}{L}
}
\right\}.
\]
Setting \(\epsilon=2nc\,\eta\) yields
\[
\hat{\mathcal{R}}_L(\mathrm{rev}\circ\mathcal{M})
\le
\inf_{\epsilon>0}
\left\{
\epsilon
+
2nc\sqrt{
\frac{
2\log\!\left(
\mathcal{N}_\infty\!\left(\mathcal{M},\frac{\epsilon}{2nc}\right)
\right)
}{L}
}
\right\}
=
n\Delta_L.
\]

Substituting this bound into~\eqref{eq:dm-rev-uc} and rearranging gives
\[
\mathbb{E}_{\rho\sim\mathcal{P}}
\!\left[\sum_{i=1}^n t_i(\rho)\right]
\ge
\frac{1}{L}\sum_{\ell=1}^L \sum_{i=1}^n t_i(\rho^{(\ell)})
-
2n\Delta_L
-
4nc\sqrt{\frac{2\log(8/\delta)}{L}}.
\]

\paragraph{Regret bound.}
For a mechanism \(M=(\sigma,t)\in\mathcal{M}\), define buyer \(i\)'s utility maps as follows.

In the {\em ex post} IC setting,
\begin{equation}
U_i^{M,\mathrm{ep}}(\rho_i,\rho',\delta)
:=
\mathop{\mathbb{E}}_{\substack{\omega\sim\theta_i\\ a\sim\sigma(\omega;\rho')}}
\!\left[
u_i(\delta(a_i),a_{-i},\omega,\rho_i)
\right]
-
t_i(\rho').
\label{eq:Ui-ep-dm}
\end{equation}

In the BIC setting,
\begin{equation}
U_i^{M,\mathrm{bic}}(\rho_i,\rho_i',\delta)
:=
\mathbb{E}_{-i}
\left[
\mathop{\mathbb{E}}_{\substack{\omega\sim\theta_i\\ a\sim\sigma(\omega;(\rho_i',\rho_{-i}))}}
\!\left[
u_i(\delta(a_i),a_{-i},\omega,\rho_i)
\right]
-
t_i(\rho_i')
\right].
\label{eq:Ui-bic-dm}
\end{equation}

Let \(\mathcal{U}_i\) denote the class of buyer-\(i\) utility maps induced by \(\mathcal{M}\).
Define buyer-\(i\)'s regret function by
\[
r_i^M(\rho)
=
\begin{cases}
\displaystyle
\max_{\rho_i'\in\mathcal{T}_i,\ \delta\in\Delta_i}
U_i^{M,\mathrm{ep}}(\rho_i,(\rho_i',\rho_{-i}),\delta)
-
U_i^{M,\mathrm{ep}}(\rho_i,\rho,\mathrm{id}),
& \text{{\em ex post} IC,}\\[10pt]
\displaystyle
\max_{\rho_i'\in\mathcal{T}_i,\ \delta\in\Delta_i}
U_i^{M,\mathrm{bic}}(\rho_i,\rho_i',\delta)
-
U_i^{M,\mathrm{bic}}(\rho_i,\rho_i,\mathrm{id}),
& \text{BIC,}
\end{cases}
\]
where in the BIC case we view \(r_i^M\) as a function on \(\mathcal{T}\) via the lift
\[
r_i^M(\rho_i,\rho_{-i}) := r_i^M(\rho_i).
\]

Now define
\[
\mathrm{rgt}\circ\mathcal{U}_i
=
\{r_i^M \mid M\in\mathcal{M}\},
\]
and the sum-regret class
\[
\overline{\mathrm{rgt}}\circ\mathcal{U}
=
\left\{
f:\mathcal{T}\to\mathbb{R}
\ \middle|\
f(\rho)=\sum_{i=1}^n r_i^M(\rho)
\text{ for some common }M\in\mathcal{M}
\right\}.
\]

As in \citet[Appendix~D.2.4]{dutting2023jacm}, the proof proceeds in three steps.

\paragraph{Step 1.}
For each \(i\),
\[
\mathcal{N}_\infty(\mathrm{rgt}\circ\mathcal{U}_i,\epsilon)
\le
\mathcal{N}_\infty(\mathcal{U}_i,\epsilon/2).
\]
This is exactly Step~1 of \citet[Appendix~D.2.4]{dutting2023jacm}. The only difference is that the maximization domain is now
\((\rho_i',\delta)\in \mathcal{T}_i\times \Delta_i\), but the same max-difference argument applies verbatim:

Let \(\widehat{\mathcal U}_i\) be an \(\epsilon/2\)-cover of \(\mathcal U_i\) under \(d_{\mathcal U_i}\).
Fix \(U_i\in\mathcal U_i\) and choose \(\hat U_i\in\widehat{\mathcal U}_i\) such that
\[
d_{\mathcal U_i}(U_i,\hat U_i)\le \epsilon/2.
\]

In the {\em ex post} IC setting, for \(\rho\in\mathcal T\), define
\[
r_i^{U_i}(\rho)
=
\max_{\rho_i'\in\mathcal T_i,\ \delta\in\Delta_i}
U_i(\rho_i,(\rho_i',\rho_{-i}),\delta)
-
U_i(\rho_i,\rho,\mathrm{id}),
\]
and analogously define \(r_i^{\hat U_i}\). Then
\begin{align*}
|r_i^{U_i}(\rho)-r_i^{\hat U_i}(\rho)|
&\le
\left|
\max_{\rho_i',\delta}
U_i(\rho_i,(\rho_i',\rho_{-i}),\delta)
-
\max_{\rho_i',\delta}
\hat U_i(\rho_i,(\rho_i',\rho_{-i}),\delta)
\right|
\\
&\qquad
+
\left|
U_i(\rho_i,\rho,\mathrm{id})
-
\hat U_i(\rho_i,\rho,\mathrm{id})
\right|
\\
&\le
\sup_{\rho_i',\delta}
\left|
U_i(\rho_i,(\rho_i',\rho_{-i}),\delta)
-
\hat U_i(\rho_i,(\rho_i',\rho_{-i}),\delta)
\right|
+
\epsilon/2
\\
&\le
\epsilon.
\end{align*}
Taking the supremum over \(\rho\in\mathcal T\) shows that the regret functions induced by
\(U_i\) and \(\hat U_i\) are within \(\epsilon\) in \(\ell_\infty\).

The BIC case is identical, replacing
\[
r_i^{U_i}(\rho)
\quad\text{by}\quad
r_i^{U_i}(\rho_i)
=
\max_{\rho_i'\in\mathcal T_i,\ \delta\in\Delta_i}
U_i(\rho_i,\rho_i',\delta)
-
U_i(\rho_i,\rho_i,\mathrm{id}),
\]
and then lifting \(r_i^{U_i}\) to \(\mathcal T\) via \(r_i^{U_i}(\rho_i,\rho_{-i})=r_i^{U_i}(\rho_i)\).
Hence
\[
\mathcal{N}_\infty(\mathrm{rgt}\circ\mathcal{U}_i,\epsilon)
\le
\mathcal{N}_\infty(\mathcal{U}_i,\epsilon/2).
\]

\paragraph{Step 2.}
For each \(i\),
\[
\mathcal{N}_\infty(\mathcal{U}_i,\epsilon)
\le
\mathcal{N}_\infty(\mathcal{M},\epsilon/c).
\]
To show this, define the metric on \(\mathcal{U}_i\) by
\[
d_{\mathcal{U}_i}(U_i,\hat U_i)
=
\begin{cases}
\displaystyle
\sup_{\rho\in\mathcal{T},\ \rho_i'\in\mathcal{T}_i,\ \delta\in\Delta_i}
\left|
U_i(\rho_i,(\rho_i',\rho_{-i}),\delta)
-
\hat U_i(\rho_i,(\rho_i',\rho_{-i}),\delta)
\right|,
& \text{{\em ex post} IC,}\\[12pt]
\displaystyle
\sup_{\rho_i\in\mathcal{T}_i,\ \rho_i'\in\mathcal{T}_i,\ \delta\in\Delta_i}
\left|
U_i(\rho_i,\rho_i',\delta)
-
\hat U_i(\rho_i,\rho_i',\delta)
\right|,
& \text{BIC.}
\end{cases}
\]

Fix \(M=(\sigma,t)\) and \(\hat M=(\hat\sigma,\hat t)\).

For the {\em ex post} IC setting, fix \(i\), \(\rho\in\mathcal{T}\), \(\rho_i'\in\mathcal{T}_i\), and \(\delta\in\Delta_i\). Expanding~\eqref{eq:Ui-ep-dm},
\begin{align*}
&
\left|
U_i^{M,\mathrm{ep}}(\rho_i,(\rho_i',\rho_{-i}),\delta)
-
U_i^{\hat M,\mathrm{ep}}(\rho_i,(\rho_i',\rho_{-i}),\delta)
\right|
\\
&\le
v_{\max}
\sum_{k\in[m]}
\left|
\pi_{i,\delta(k),k}(\rho_i',\rho_{-i})
-
\hat\pi_{i,\delta(k),k}(\rho_i',\rho_{-i})
\right|
\\
&\qquad
+
v_{\max}\bar\alpha
\sum_{j\neq i}\sum_{k\in[m]}
\left|
\pi_{j,k,k}(\rho_i',\rho_{-i})
-
\hat\pi_{j,k,k}(\rho_i',\rho_{-i})
\right|
\\
&\qquad
+
\left|
t_i(\rho_i',\rho_{-i})
-
\hat t_i(\rho_i',\rho_{-i})
\right|.
\end{align*}
The three sums on the right are each taken over subsets of the coordinates appearing in
\(d(M,\hat M)\). Moreover,
\[
c\ge 1,
\qquad
c\ge B\ge v_{\max},
\qquad
c\ge B\ge v_{\max}\bar\alpha.
\]
Therefore
\[
\left|
U_i^{M,\mathrm{ep}}(\rho_i,(\rho_i',\rho_{-i}),\delta)
-
U_i^{\hat M,\mathrm{ep}}(\rho_i,(\rho_i',\rho_{-i}),\delta)
\right|
\le
c\,d(M,\hat M).
\]

For the BIC setting, taking expectation over \(\rho_{-i}\sim\mathcal P_{-i}\) and using Jensen's inequality,
\begin{align*}
&
\left|
U_i^{M,\mathrm{bic}}(\rho_i,\rho_i',\delta)
-
U_i^{\hat M,\mathrm{bic}}(\rho_i,\rho_i',\delta)
\right|
\\
&=
\Bigg|
\mathbb{E}_{-i}
\Big[
U_i^{M,\mathrm{ep}}(\rho_i,(\rho_i',\rho_{-i}),\delta)
-
U_i^{\hat M,\mathrm{ep}}(\rho_i,(\rho_i',\rho_{-i}),\delta)
\Big]
\Bigg|
\\
&\le
\mathbb{E}_{-i}
\Big[
\big|
U_i^{M,\mathrm{ep}}(\rho_i,(\rho_i',\rho_{-i}),\delta)
-
U_i^{\hat M,\mathrm{ep}}(\rho_i,(\rho_i',\rho_{-i}),\delta)
\big|
\Big]
\\
&\le
c\,d(M,\hat M).
\end{align*}
Taking the relevant supremum yields
\[
d_{\mathcal{U}_i}(U_i^M,U_i^{\hat M})
\le
c\,d(M,\hat M),
\]
which implies the desired covering-number bound.

\paragraph{Step 3.}
\[
\mathcal{N}_\infty(\overline{\mathrm{rgt}}\circ\mathcal{U},\epsilon)
\le
\mathcal{N}_\infty\!\left(\mathcal{M},\frac{\epsilon}{2nc}\right).
\]
Indeed, let \(\widehat{\mathcal{M}}\) be an \(\epsilon/(2nc)\)-cover of \(\mathcal{M}\). For any \(M\in\mathcal{M}\), choose \(\hat M\in\widehat{\mathcal{M}}\) such that
\[
d(M,\hat M)\le \frac{\epsilon}{2nc}.
\]
By Step~2, for every buyer \(i\),
\[
d_{\mathcal{U}_i}(U_i^M,U_i^{\hat M})
\le
\frac{\epsilon}{2n}.
\]
Then by Step~1,
\[
\sup_{\rho\in\mathcal{T}}
|r_i^M(\rho)-r_i^{\hat M}(\rho)|
\le
\frac{\epsilon}{n}.
\]
Summing over buyers gives
\[
\sup_{\rho\in\mathcal{T}}
\left|
\sum_{i=1}^n r_i^M(\rho)
-
\sum_{i=1}^n r_i^{\hat M}(\rho)
\right|
\le
\epsilon,
\]
which proves the claim.

\paragraph{Range of the regret class.}
For every report and deviation rule,
\[
u_i(\delta(a_i),a_{-i},\omega,\rho_i)
=
v_i
\left(
\mathbf{1}\{\delta(a_i)=\omega\}
-
\bar\alpha\sum_{j\neq i}\mathbf{1}\{a_j=\omega\}
\right)
\in[-\alpha v_i,\;v_i].
\]
Using also \(|t_i(\rho)|\le B\), we obtain
\[
|U_i^{M,\mathrm{ep}}(\rho_i,\rho',\delta)|\le 2B,
\qquad
|U_i^{M,\mathrm{bic}}(\rho_i,\rho_i',\delta)|\le 2B.
\]
Since the truthful-obedient pair \((\rho_i,\mathrm{id})\) is feasible in the maximization domain,
\[
0\le r_i^M(\rho)\le 4B\le c.
\]
Hence every function in \(\overline{\mathrm{rgt}}\circ\mathcal{U}\) takes values in \([0,nc]\).

Applying Lemma~\ref{lem:rademacher} to \(\overline{\mathrm{rgt}}\circ\mathcal{U}\) with confidence parameter \(\delta/2\), we obtain that with probability at least \(1-\delta/2\),
\begin{align}
\mathbb{E}_{\rho\sim\mathcal{P}}
\left[
\sum_{i=1}^n r_i^M(\rho)
\right]
&\le
\frac{1}{L}\sum_{\ell=1}^L \sum_{i=1}^n r_i^M(\rho^{(\ell)})
+
2\hat{\mathcal{R}}_L(\overline{\mathrm{rgt}}\circ\mathcal{U})
\nonumber\\
&\qquad
+
4nc\sqrt{\frac{2\log(8/\delta)}{L}}.
\label{eq:dm-rgt-uc}
\end{align}

Using the same cover-plus-Massart argument as in \citet[Appendix~D.2.4]{dutting2023jacm}, together with Step~3,
\[
\hat{\mathcal{R}}_L(\overline{\mathrm{rgt}}\circ\mathcal{U})
\le
\inf_{\epsilon>0}
\left\{
\epsilon
+
2nc\sqrt{
\frac{
2\log\!\left(
\mathcal{N}_\infty\!\left(
\mathcal{M},\frac{\epsilon}{2nc}
\right)
\right)
}{L}
}
\right\}
=
n\Delta_L.
\]

Substituting this into~\eqref{eq:dm-rgt-uc} and dividing by \(n\) yields
\[
\frac{1}{n}\sum_{i=1}^n \mathrm{RGT}_i
\le
\frac{1}{n}\sum_{i=1}^n \widehat{\mathrm{rgt}}_i
+
2\Delta_L
+
4c\sqrt{\frac{2\log(8/\delta)}{L}}.
\]

Finally, the revenue and regret bounds each hold with probability at
least $1-\delta/2$. A union bound therefore gives their simultaneous
validity with probability at least $1-\delta$, with the concentration
terms displayed in the theorem.
\end{proof}

\subsection{Proof of Theorem~\ref{thm:cover_bound}}
\coverbound*

\begin{proof}
We adapt \citet[Lemma~D.2, Lemma~D.3, and Theorem~D.1]{dutting2023jacm}. The experiment network plays the role of the auction-allocation network, while the normalized-payment network plays the role of the fractional-payment network.

Let \(\mathcal{G}\) denote the class of experiment networks, and let \(\widetilde{\mathcal{P}}\) denote the class of lifted normalized-payment maps. In the BIC setting, each coordinate \(\tilde t_i\) is lifted from \(\mathcal T_i\) to \(\mathcal T\) by
\[
\tilde t_i(\rho):=\tilde t_i(\rho_i).
\]

Because the full parameter vector satisfies \(\|w\|_1\le W\), every layer matrix has induced \(\ell_1\) norm at most \(W\). We therefore apply \citet[Lemma~D.2]{dutting2023jacm} to the feed-forward part of each network and then account separately for the final normalization layer. For the normalized-payment network, the final sigmoid is bounded and \(1\)-Lipschitz. For the experiment network, the final row-wise normalization is the temperature-\(\tau_\pi\) softmax, which is \(1/\tau_\pi\)-Lipschitz in \(\ell_1\); since \(\tau_\pi\) is fixed, this only changes the constant in the resulting covering-number bound.

\paragraph{Step 1: cover \(\mathcal{G}\) and \(\widetilde{\mathcal{P}}\).}
By \citet[Lemma~D.2]{dutting2023jacm}, there exists a universal constant \(C>0\) such that any feed-forward network with width at most \(T\), \(R\) hidden layers, bounded \(1\)-Lipschitz activations, and \(d\) trainable parameters satisfies
\[
\mathcal{N}_\infty(\mathcal{F},\epsilon)
\le
\left\lceil
\frac{C\,T^2(2W)^{R+1}}{\epsilon}
\right\rceil^d.
\]

For the experiment network, the pre-normalization output dimension is \(nm^2\), so we may take
\[
T_\pi=\max\{K,nm^2\}.
\]
By the same calculation as Eq.~(23) in \citet{dutting2023jacm}, applied blockwise to the \(nm\) rows of length \(m\), we obtain
\begin{equation}
\mathcal{N}_\infty(\mathcal{G},\epsilon)
\le
\left\lceil
\frac{C\,\max\{K,nm^2\}^2(2W)^{R+1}}{\epsilon}
\right\rceil^{d_\pi}.
\label{eq:dm-G-cover}
\end{equation}

For the normalized-payment map, the final coordinate-wise sigmoid is bounded and \(1\)-Lipschitz in \(\ell_1\). If a single shared payment network with \(n\) outputs is used, we may take
\[
T_t=\max\{K,n\},
\]
and obtain
\[
\mathcal{N}_\infty(\widetilde{\mathcal{P}},\epsilon)
\le
\left\lceil
\frac{C\,\max\{K,n\}^2(2W)^{R+1}}{\epsilon}
\right\rceil^{d_t}.
\]
If, in the BIC setting, different buyers use different payment
networks, we cover each coordinate class to accuracy $\epsilon/n$ and
multiply the covers. Thus, in either the shared- or separate-network
case, we may use the bound
\begin{equation}
\mathcal{N}_\infty(\widetilde{\mathcal{P}},\epsilon)
\le
\left\lceil
\frac{C\,n\max\{K,n\}^2(2W)^{R+1}}{\epsilon}
\right\rceil^{d_t},
\label{eq:dm-P-cover}
\end{equation}
where $d_t$ denotes the total number of trainable payment parameters
across all payment networks.

\paragraph{Step 2: combine the two covers.}
Define
\[
d_{\mathcal{G}}(\pi,\hat\pi)
=
\sup_{\rho\in\mathcal{T}}
\sum_{i\in[n]}\sum_{j,k\in[m]}
|\pi_{i,j,k}(\rho)-\hat\pi_{i,j,k}(\rho)|,
\]
and
\[
d_{\widetilde{\mathcal{P}}}(\tilde t,\hat{\tilde t})
=
\sup_{\rho\in\mathcal{T}}
\sum_{i\in[n]}
|\tilde t_i(\rho)-\hat{\tilde t}_i(\rho)|.
\]

Let
\[
B:=v_{\max}(1+\alpha).
\]

In the {\em ex post} setting, write
\[
t_i(\rho)=\tilde t_i(\rho)\,s_i^{\mathrm{ep}}(\rho),
\]
where
\[
s_i^{\mathrm{ep}}(\rho)
=
v_i\Bigg(
\sum_{k\in[m]}\pi_{i,k,k}(\rho)\theta_{i,k}
-
\bar\alpha\sum_{j\neq i}\sum_{k\in[m]}\pi_{j,k,k}(\rho)\theta_{i,k}
-
(\max\theta_i-\alpha)
\Bigg).
\]

In the BIC setting, after lifting to \(\mathcal T\),
\[
t_i(\rho)=\tilde t_i(\rho_i)\,s_i^{\mathrm{bic}}(\rho_i),
\]
where
\begin{align*}
s_i^{\mathrm{bic}}&(\rho_i)
=\\
&v_i\,
\mathbb{E}_{-i}
\Bigg[
\sum_{k\in[m]}\pi_{i,k,k}(\rho_i,\rho_{-i})\theta_{i,k}
-
\bar\alpha\sum_{j\neq i}\sum_{k\in[m]}\pi_{j,k,k}(\rho_i,\rho_{-i})\theta_{i,k}
-
(\max\theta_i-\alpha)
\Bigg].
\end{align*}

We only need a uniform absolute bound on these scaling terms. Since
\[
0\le \sum_{k\in[m]}\pi_{i,k,k}(\rho)\theta_{i,k}\le 1,
\qquad
0\le \bar\alpha\sum_{j\neq i}\sum_{k\in[m]}\pi_{j,k,k}(\rho)\theta_{i,k}\le \alpha,
\]
it follows that
\[
|s_i^{\mathrm{ep}}(\rho)|\le B,
\qquad
|s_i^{\mathrm{bic}}(\rho_i)|\le B.
\]

Hence, in either setting,
\[
|t_i-\hat t_i|
=
|\tilde t_i s_i-\hat{\tilde t}_i\hat s_i|
\le
|(\tilde t_i-\hat{\tilde t}_i)s_i|
+
|\hat{\tilde t}_i(s_i-\hat s_i)|
\le
B\,|\tilde t_i-\hat{\tilde t}_i|
+
|s_i-\hat s_i|.
\]

For the {\em ex post} setting,
\begin{align*}
\sum_{i=1}^n |s_i^{\mathrm{ep}}(\rho)-\hat s_i^{\mathrm{ep}}(\rho)|
&\le
v_{\max}
\sum_{i=1}^n\sum_{k\in[m]}
|\pi_{i,k,k}(\rho)-\hat\pi_{i,k,k}(\rho)|
\\
&\qquad
+
v_{\max}\bar\alpha
\sum_{i=1}^n\sum_{j\neq i}\sum_{k\in[m]}
|\pi_{j,k,k}(\rho)-\hat\pi_{j,k,k}(\rho)|
\\
&\le
v_{\max} d_{\mathcal G}(\pi,\hat\pi)
+
v_{\max}\alpha\, d_{\mathcal G}(\pi,\hat\pi)
\\
&=
B\,d_{\mathcal G}(\pi,\hat\pi).
\end{align*}
Therefore
\[
d((\pi,t),(\hat\pi,\hat t))
\le
(1+B)\,d_{\mathcal G}(\pi,\hat\pi)
+
B\,d_{\widetilde{\mathcal P}}(\tilde t,\hat{\tilde t}).
\]

For the BIC setting, for every buyer \(i\) and every \(\rho_i\),
\[
|s_i^{\mathrm{bic}}(\rho_i)-\hat s_i^{\mathrm{bic}}(\rho_i)|
\le
\mathbb{E}_{-i}
\big[
|s_i^{\mathrm{ep}}(\rho_i,\rho_{-i})-\hat s_i^{\mathrm{ep}}(\rho_i,\rho_{-i})|
\big]
\le
B\,d_{\mathcal G}(\pi,\hat\pi).
\]
Summing over buyers gives
\[
\sum_{i=1}^n |s_i^{\mathrm{bic}}(\rho_i)-\hat s_i^{\mathrm{bic}}(\rho_i)|
\le
nB\,d_{\mathcal G}(\pi,\hat\pi),
\]
and therefore
\[
d((\pi,t),(\hat\pi,\hat t))
\le
(1+nB)\,d_{\mathcal G}(\pi,\hat\pi)
+
B\,d_{\widetilde{\mathcal P}}(\tilde t,\hat{\tilde t}).
\]

Now let
\[
C_B:=1+(n+1)B.
\]
In the {\em ex post} setting, the two coefficients sum to \(1+2B\le C_B\); in the BIC setting, they sum to \(1+(n+1)B=C_B\). Hence, in either setting, any pair of covers of radius \(\epsilon/C_B\) for \(\mathcal G\) and \(\widetilde{\mathcal P}\) induces an \(\epsilon\)-cover of \(\mathcal M_{\mathrm{NN}}\). Therefore
\begin{equation}
\mathcal N_\infty(\mathcal M_{\mathrm{NN}},\epsilon)
\le
\mathcal N_\infty\!\left(\mathcal G,\frac{\epsilon}{C_B}\right)
\cdot
\mathcal N_\infty\!\left(\widetilde{\mathcal P},\frac{\epsilon}{C_B}\right).
\label{eq:dm-product-cover}
\end{equation}

\paragraph{Step 3: simplify the logarithm.}

Combining \eqref{eq:dm-G-cover}, \eqref{eq:dm-P-cover}, and \eqref{eq:dm-product-cover} introduces additional factors \(n\) and \(C_B\) inside the logarithm. Since \(B=v_{\max}(1+\alpha)\) is fixed, there exists a constant \(c_B>0\), depending only on \((v_{\max},\alpha)\), such that
\[
\log C_B \le c_B + \log n.
\]
Because \(n\le nm^2\), the resulting \(\log n\) terms are
absorbed into \(\log\max\{K,nm^2\}\). Using also
\[
\max\{K,n\}\le \max\{K,nm^2\},
\]
we conclude that there exists a constant \(C_0>0\), depending only on
universal constants, fixed architecture hyperparameters (including
\(\tau_\pi\)), and on \((v_{\max},\alpha)\), such that for every
\(\epsilon\in(0,1]\),

\[
\log \mathcal N_\infty(\mathcal M_{\mathrm{NN}},\epsilon)
\le
C_0(d_\pi+d_t)\Big(
(R+1)\log(2W)+\log\max\{K,nm^2\}+\log(1/\epsilon)
\Big).
\]

\paragraph{Step 4: plug into \(\Delta_L\).}
Since every subclass has smaller covering number than the ambient class,
\[
\mathcal N_\infty(\mathcal M,\epsilon)
\le
\mathcal N_\infty(\mathcal M_{\mathrm{NN}},\epsilon)
\]
for every \(\mathcal M\subseteq \mathcal M_{\mathrm{NN}}\). Therefore
\[
\Delta_L
=
\inf_{\epsilon>0}
\left\{
\frac{\epsilon}{n}
+
2c\sqrt{
\frac{
2\log\!\left(
\mathcal N_\infty\!\left(\mathcal M,\frac{\epsilon}{2nc}\right)
\right)
}{L}
}
\right\}
\]
\[
\le
\inf_{\epsilon>0}
\left\{
\frac{\epsilon}{n}
+
2c\sqrt{
\frac{
2\log\!\left(
\mathcal N_\infty\!\left(\mathcal M_{\mathrm{NN}},\frac{\epsilon}{2nc}\right)
\right)
}{L}
}
\right\}.
\]
Substituting the logarithmic bound above and choosing \(\epsilon=L^{-1/2}\) yields
\[
\Delta_L
\le
\frac{1}{n\sqrt L}
+
C_2\sqrt{
\frac{
(d_\pi+d_t)\big((R+1)\log(2W)+\log(L\max\{K,nm^2\})\big)
}{L}
}
\]
for some constant \(C_2>0\), depending only on universal constants, fixed architecture hyperparameters (including \(\tau_\pi\)), and on \((v_{\max},\alpha)\). In particular, \(\Delta_L\to 0\) as \(L\to\infty\). This proves the theorem.
\end{proof}

}
{\color{black}
\section{Discrete two-type benchmarks}
\label{app:discrete-two-type}

For completeness, we also evaluate RochetNet on discrete two-type benchmarks corresponding to the comparative-statics exercises in Settings~\ref{SIII} and~\ref{SIV}. These experiments are closer to the original discrete framework studied by \citet{Berg2018}, whereas our main text focuses on continuous analogs.

In the first benchmark, we consider a two-point distribution over beliefs with
\[
\theta^L = \left(\frac{7}{36}, \frac{29}{36}\right),
\qquad
\theta^H = \left(\frac{59}{88}, \frac{29}{88}\right),
\]
and vary the mixture weight \(\gamma \in [0,1]\) on the high type. We compute the informativeness \(I\) of the experiment assigned to the low type. The learned mechanism exhibits the expected comparative static: as the frequency of the high type increases, the informativeness of the low-type experiment decreases.

In the second benchmark, we again consider a two-point distribution over beliefs, keeping the high type fixed at
\[
\theta^H = \left(\frac{59}{88}, \frac{29}{88}\right),
\]
while varying the low type \(\theta^L = (q,1-q)\) so as to increase the precision of the low type's prior belief \(|q-0.5|\). Once again, the learned mechanism exhibits the expected comparative static: as the low type becomes more precise, the informativeness of the experiment assigned to that type decreases.

These results are shown in Figure~\ref{fig:discrete-two-type}. They provide an additional sanity check that RochetNet recovers the same qualitative structure in the discrete benchmarks as in the known theory, while our main contribution remains the continuous-type and multi-buyer settings where analytical solutions are harder to obtain.

\begin{figure}[t]
    \centering
    \includegraphics[width=0.42\textwidth]{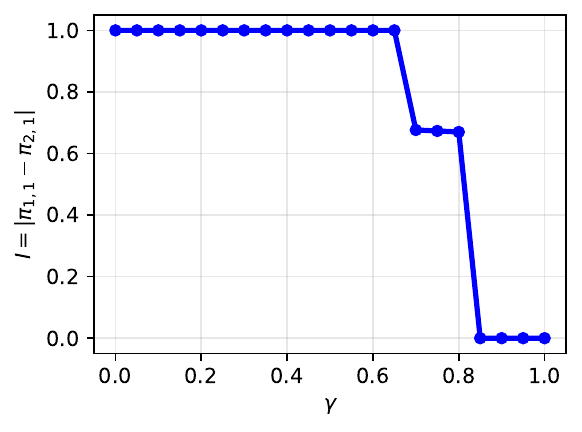}
    \includegraphics[width=0.42\textwidth]{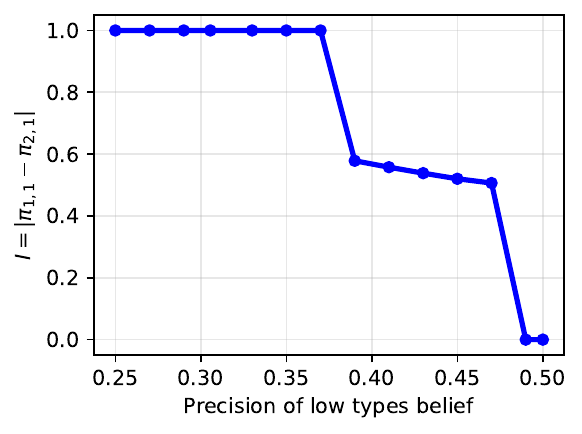}
    \caption{Discrete two-type benchmarks corresponding to the comparative statics in Settings~\ref{SIII} and~\ref{SIV}. \textbf{Left:} informativeness \(I\) of the low-type experiment as a function of the frequency \(\gamma\) of the high type. \textbf{Right:} informativeness \(I\) of the low-type experiment as a function of the precision of the low type's prior belief. In both cases, RochetNet recovers the expected qualitative monotonicity.}
    \label{fig:discrete-two-type}
\end{figure}

}

\section{Setup and Hyper-parameters} \label{app:setup}

We present the hyperparameters for the Single Buyer Setting in Table~\ref{tab:dm-single-hyperparams}, Multi-Buyer Setting with BIC constraints and IC constraints in Table~\ref{tab:bic-hyperparams} and \ref{tab:ic-hyperparams} respectively.

\begin{table}[ht]
\centering
\small
\renewcommand{\arraystretch}{1.3}
\begin{tabular}{lcc}
\toprule
\multicolumn{3}{c}{\textbf{Network Architecture}} \\ \midrule
Number of initialized menus & 1000 & \\
$\tau_\pi$ & 0.2 & \\
$\tau_u$ & 0.005 & \\\midrule
\multicolumn{3}{c}{\textbf{Training Parameters}} \\ \midrule
Max iterations & 20,000 &\\
Batch size & $2^{15}$ & {sampled online} \\
Learning rate & 0.001 & RMSprop Optimizer \\
Maintenance iteration frequency & 1000 & {remove duplicates/unused menus} \\\midrule
\multicolumn{3}{c}{\textbf{Testing Parameters}} \\ \midrule
Test set size & $2^{30}$ samples \\
\bottomrule
\end{tabular}
\caption{Hyperparameters for RochetNet experiments.}\label{tab:dm-single-hyperparams}
\end{table}

\begin{table}[ht]
\centering
\small
\renewcommand{\arraystretch}{1.3}
\begin{tabular}{lcl}
\toprule
\multicolumn{3}{c}{\textbf{Network Architecture}} \\ \midrule
Hidden layers & 3 & \\
Hidden units & 200 & \\
Activation Function & Leaky ReLU & \\
$\tau_\pi$ & 0.1 & \\
Number of samples & \multirow{2}{*}{512} &\\
for computing interim values &&\\
\midrule
\multicolumn{3}{c}{\textbf{Training / Optimizer}} \\ \midrule
Max iterations & 20,000 & \\
Batch size & 128 & {Sampled Online}\\
Learning rate & 0.0005 & AdamW Optimizer\\
\multicolumn{3}{c}{\textbf{Lagrangian Parameters}} \\ \midrule
$\lambda_i^\mathrm{obv}$ for $i \in N$ & 10.0 \\
$\lambda_i^\mathrm{trv}$ for $i \in N$ & 10.0 \\
Lagrangian Update frequency & 100 \\
$\zeta_i^\mathrm{obv}$ for $i \in N$ & 10.0 \\
$\zeta_i^\mathrm{trv}$ for $i \in N$ & 10.0 \\
Penalty increment & 10.0 & {$\zeta_i^\mathrm{obv}$, $\zeta_i^\mathrm{trv}$ is increased by 10} \\
Penalty update frequency & 100 & every 100 iterations \\ \midrule
\multicolumn{3}{c}{\textbf{Testing Parameters}} \\ \midrule
Test Set Size & $2^{20}$ & \\
Evaluation Misreports & 100 & \\
Gradient Ascent Iterations & \multirow{2}{*}{100} & \\
(performed on all misreports) & & \\
Gradient Ascent Learning rate & 0.005 &\\
\bottomrule
\end{tabular}
\caption{Hyperparameters for the multi-buyer setting with BIC constraints.}\label{tab:bic-hyperparams}
\end{table}

\begin{table}[ht]
\centering
\small
\renewcommand{\arraystretch}{1.3}
\begin{tabular}{lcl}
\toprule
\multicolumn{3}{c}{\textbf{Network Architecture}} \\ \midrule
Hidden layers & 3 & \\
Hidden units & 200 & \\
Activation Function & Leaky ReLU & \\
$\tau_\pi$ & 0.1 & \\ \midrule
\multicolumn{3}{c}{\textbf{Training / Optimizer}} \\ \midrule
Max iterations & 20,000 & \\
Batch size & 1024 & {Sampled Online}\\
Learning rate & 0.0005 & AdamW Optimizer\\
Misreports initialization& 100& \\ \midrule
\multicolumn{3}{c}{\textbf{Lagrangian Parameters}} \\ \midrule
$\lambda_i^\mathrm{rgt}$ for $i \in N$ & 0.5 &\\
Lagrangian Update frequency & 100 &\\
$\zeta_i^\mathrm{rgt}$ for $i \in N$ & 1.0 & \\
Penalty increment & 10.0 & {$\zeta_i$ is increased by 10} \\
Penalty update frequency & 1000 & every 1000 iterations \\ \midrule
\multicolumn{3}{c}{\textbf{Testing Parameters}} \\ \midrule
Test Set Size & $2^{20}$ & \\
Evaluation Misreports & 100 & \\
Gradient Ascent Iterations & \multirow{2}{*}{100} & \\
(performed on all misreports) & & \\
Gradient Ascent Learning rate & 0.005 &\\
\bottomrule
\end{tabular}
\caption{Hyperparameters for the multi-buyer setting with IC constraints.}\label{tab:ic-hyperparams}
\end{table}

\end{document}